%% file: main1.tex
\documentclass[journal]{IEEEtran}

\usepackage{packages}

\newtheorem{theorem}{Theorem}

\newtheorem{corollary}[theorem]{Corollary}

\newcounter{MYtempeqncnt}

\begin{document}
\title{The Memory-Enhanced Gaussian Noise (MEGN) Model for Fiber-Optic Channels}

\author{Kaiquan~Wu,~\IEEEmembership{Member, IEEE}, Gabriele~Liga,~\IEEEmembership{Senior Member, IEEE}, Marco~Secondini,~\IEEEmembership{Senior Member, IEEE},
Stella~Civelli,~\IEEEmembership{Member, IEEE},
Hussam~Batshon,~\IEEEmembership{Senior Member, IEEE},
Greg~Raybon,~\IEEEmembership{Fellow, IEEE},
Xi~Chen,~\IEEEmembership{Senior Member, IEEE},
and Alex~Alvarado,~\IEEEmembership{Senior Member, IEEE}
\thanks{This work is part of the project FOCAL with file number NGF.1609.242.054 of the National Growth Fund (NGF) AiNed programme which is financed by the Dutch Research Council (NWO). An earlier version of this paper was presented at Optical Fiber Communication 2025 in San Francisco \cite{wu2025FEGN}. \textit{(Corresponding author: Kaiquan Wu.)}}
\thanks{This work has been submitted to the IEEE for possible publication. Copyright may be transferred without notice, after which this version may no longer be accessible.}
\thanks{Kaiquan~Wu, Gabriele~Liga, and Alex~Alvarado are with the Information and Communication Theory Lab, Signal Processing Systems Group, Department of Electrical Engineering, Eindhoven University of Technology, Eindhoven 5600 MB, The Netherlands (e-mails: \{k.wu, g.liga, a.alvarado\}@tue.nl).}
\thanks{Marco~Secondini and Stella~Civelli are with TeCIP Institute, Scuola Superiore Sant'Anna, Pisa, Italy. Stella Civelli is also with the CNR-IEIIT, Pisa, Italy (e-mail:  marco.secondini@sssup.it, stella.civelli@cnr.it).}
\thanks{Hussam~Batshon, Greg~Raybon and Xi~Chen are with Nokia Bell Labs, Murray Hill, NJ 07974, United States (e-mail:  \{hussam.batshon, gregory.raybon, xi.v.chen\}@nokia-bell-labs.com).}
}

\markboth{Preprint, \today}%
{Shell \MakeLowercase{\textit{et al.}}: Bare Demo of IEEEtran.cls for IEEE Journals}

\maketitle

\begin{abstract}
The enhanced Gaussian noise (EGN) model is widely used for estimating the nonlinear interference (NLI) power accumulated in coherent fiber-optic transmission systems. Given a fixed fiber link, under the assumption that transmitted symbols are independent and identically distributed (i.i.d.), the EGN model establishes that the NLI power depends on time-invariant signal statistics, i.e., the second-, fourth-, and sixth-order moments of the symbols, which are determined by the modulation format and its probability distribution. However, recent advances in coded modulation have sought to mitigate NLI by introducing controlled temporal correlations among transmitted symbols, thereby violating the i.i.d. assumption underlying the EGN model. Among these correlations, symbol energy correlations are believed to exert the most significant influence on NLI. This work presents a rigorous mathematical derivation of a memory extension of the EGN model that explicitly accounts for symbol energy correlations, referred to as the \textit{MEGN} model. The proposed MEGN model is validated through both numerical simulations and transmission experiments. Normalized average NLI power estimations with less than $5\%$ errors across a wide range of symbol rates and transmission distances are reported. The model also provides a theoretical framework for analyzing and optimizing optical transmission systems employing temporally correlated modulation schemes.
\end{abstract}


\input{Tex/Ch_Intro}

\input{Tex/Ch_Model}

\input{Tex/Ch_CCDM_Deri}

\input{Tex/Ch_Results}

\input{Tex/Ch_Conclusions}

\appendices

\input{Tex/Ch_Sym_Corr}

\input{Tex/Ch_Mom_Deri.tex}

\input{Tex/Ch_PSD_Deri.tex}

\ifCLASSOPTIONcaptionsoff
  \newpage
\fi

\bibliographystyle{IEEEtran}
\bibliography{references}

\end{document}

%% file: Tex/Ch_Intro.tex
\section{Introduction}

\IEEEPARstart{T}{he} fiber channel degrades the transmitted signal through signal-dependent nonlinear interference (NLI). The generation of NLI is a complex process arising from the joint effects of fiber impairments and the transmitted signal. An NLI power model that characterizes the relationship between signal statistics and the resulting NLI power is therefore valuable for various applications. For example, an accurate NLI power model provides important insights for the design of NLI-tolerant coded modulation (CM) schemes and enables offline system configuration optimization with respect to parameters such as the symbol rate.

An NLI power spectral density (PSD) model typically follows a common structure that can be expressed as
\begin{align}\label{Eq:NLI_PSD_Intution}
\text{NLI PSD} = \sum  \text{Signal Statistics} \times  \text{Channel Functions}   .
\end{align}
This structure highlights that NLI arises from the interplay between the statistical properties of the transmitted symbols (signal statistics) and the physical characteristics of the fiber channel (channel functions).

\begin{figure*}[!t]
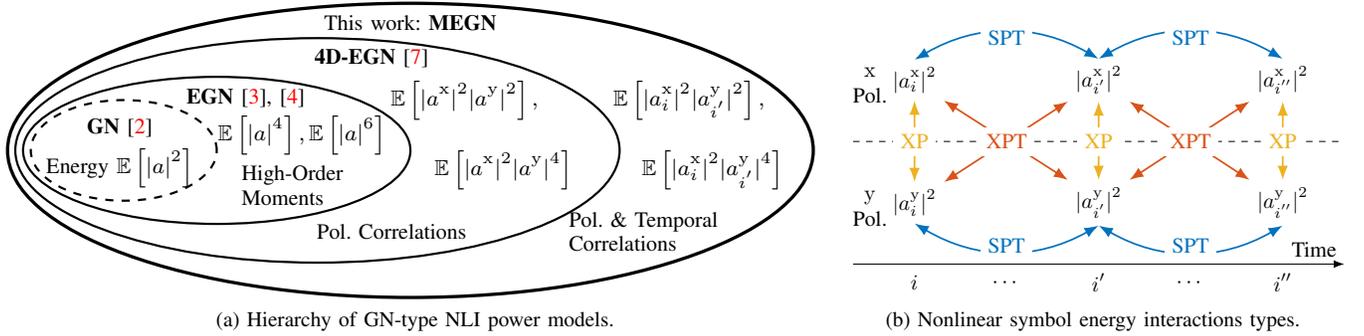

    \centering
    \def\SC{.6}
    \subfloat[Hierarchy of GN-type NLI power models. ]{\resizebox{\SC\linewidth}{!}{\subimport{Figures}{Fig_FEGN_Statistics_Comp.tex}}}
    \def\SC{.38}
    \subfloat[Nonlinear symbol energy interactions types.]{\resizebox{\SC\linewidth}{!}{\subimport{Figures}{symCorr.tex}}}
    \caption{(a): Hierarchy of GN-type NLI power models according to the symbol-energy statistics they capture. The proposed MEGN model further captures temporal symbol-energy correlations. (b): Illustration of nonlinear symbol-energy interaction types considered in the proposed MEGN model, namely self-polarization temporal interactions (SPT), purely cross-polarization interactions (XP), and cross-polarization temporal interactions (XPT).}
    \label{Fig:Models_Comparison}
\end{figure*}

Fig.~\ref{Fig:Models_Comparison}(a) summarizes the symbol-level signal statistics captured by state-of-the-art NLI PSD models. The pioneering Gaussian noise (GN) model \cite{poggiolini2012gn,poggiolini2013gn} reveals that the NLI power simply depends on the average symbol energy (a.k.a. the second-order moment) that is equivalent to the signal power. More precisely, the GN model predicts that the NLI power is proportional to the cube of the signal power. However, as its name suggests, the GN model oversimplifies the signal waveform by treating it as a Gaussian process. In reality, coherent systems employ CM to transmit quadrature amplitude modulation (QAM) symbols, and the modulated pulse train generally violates the Gaussian-process assumption~\cite{secondini2017Scope}. As a consequence, the GN model fails to capture the dependence of NLI on the underlying modulation format.

The enhanced Gaussian noise (EGN) model~\cite{dar2013properties,carena2014egn} is an alternative to the GN model that has been very successful in various applications. The EGN model is derived under the assumption of independent and identically distributed (i.i.d.) symbols. This assumption typically holds when CM schemes (\textit{i}) employ a random interleaver or (\textit{ii}) use sufficiently long blocklengths such that the generated symbols asymptotically mimic an i.i.d. process. The EGN model shows that accounting for the fourth- and sixth-order moments of the transmitted symbols, determined by the 2-D modulation format, improves the accuracy of NLI power prediction compared to the GN model \cite{dar2013properties,carena2014egn}, as illustrated in Fig.~\ref{Fig:Models_Comparison}(a). The \emph{kurtosis} of the transmitted symbol, a metric incorporating the second- and fourth-order moments, has been widely used to improve the design of CM to reduce NLI power. Particularly in the context of constellation shaping~\cite{fehenberger2016probabilistic,gultekin2021kurtosis}, the EGN model can be used to improve the geometry or the probability mass function (PMF) of the QAM symbols for better NLI tolerance.

For advanced CM schemes that generate nonstationary and correlated symbols, the EGN model becomes inadequate. Relaxing the i.i.d. symbol assumption across the four optical field dimensions and the time dimension has therefore become an important research direction. Several refined extensions of the EGN model have been proposed. For example, the 4-D-EGN model~\cite{liga2020extending} considers various correlations of symbols across the polarizations and shows that symbol-energy correlations play a dominant role in the resulting NLI power (see Fig.~\ref{Fig:Models_Comparison}(a)). These symbol statistics partially explain the effectiveness of heuristically designed high-dimensional modulation formats for NLI mitigation \cite{sillekens2022high}.

Due to chromatic dispersion and fiber nonlinearity, the optical fiber can be modeled as a channel with memory, in which symbols interact in a nonlinear manner. Ideally, the signal statistics should include the correlations among the interacting symbols, while the channel functions in \eqref{Eq:NLI_PSD_Intution} should account for the interplay between the memory channel and the correlated symbols. In fact, it has been shown that by varying the strength and length of the temporal symbol-energy correlations, it is possible to control the generated NLI strength~\cite{dar2014shaping,amari2019introducing}. Some advanced CM schemes, such as temporal probabilistic shaping~\cite{yankov2017temporal}, short-blocklength probabilistic amplitude shaping (PAS)~\cite{bocherer2015bandwidth,fehenberger2019analysis,wu2021edi}, and PAS with sequence selection~\cite{civelli2021sequence,wu2022list,civelli2024sequence,wu2026sign}, have recently emerged to exploit temporal correlations of symbols (energy-wise or phase-wise) for NLI reduction. In particular, PAS, although originally designed for additive white Gaussian noise (AWGN) channels, can mitigate NLI simply by decreasing the shaping blocklength, in which case the transmitted symbols possess strong temporal symbol-energy correlations. However, only limited progress has been made toward incorporating such correlations in NLI power models, aside from a few attempts using \emph{heuristic models}~\cite{agrell2014capacity,cho2022shaping,civelli2023nonlinear,borujeny2023constant,deng2025NLI4D}. The additional nonlinear shaping gains brought by temporal symbol-energy correlations currently lack rigorous support from an underlying theoretical model. Without an effective channel model, these techniques must rely on computationally intensive split-step Fourier method (SSFM) simulations~\cite{yankov2017temporal,civelli2021sequence} or heuristic NLI metrics~\cite{wu2022list,cho2022shaping,askari2023probabilistic,deng2025NLI4D} to reflect the NLI power induced by the correlated symbols, which limits interpretability and generalization.

In this paper, to the best of our knowledge, we present the first EGN-based \emph{analytical model} that incorporates the temporal symbol-energy correlations of the transmitted symbols. We call this model the memory-EGN (MEGN) model. The proposed model can be viewed as an extension of the aforementioned GN-type models to account for additional temporal symbol-energy correlations such as between the second- and fourth-order moments, as illustrated in Fig.~\ref{Fig:Models_Comparison}(a). Specifically, as illustrated in Fig.~\ref{Fig:Models_Comparison}(b), the MEGN model accounts for three types of symbol-energy interactions spanning the time and polarization domains:
(\textit{i}) self-polarization temporal interactions (SPT),
(\textit{ii}) cross-polarization temporal interactions (XPT), and
(\textit{iii}) purely cross-polarization interactions (XP).

In this work, we restrict our analysis of NLI to \textit{single-channel interference} (SCI). This model is evaluated for PAS schemes with constant-composition distribution matcher (CCDM)-shaped amplitudes~\cite{bocherer2015bandwidth,schulte2015constant} and independent and uniformly distributed (i.u.d.) signs. The transmitted symbol energies are therefore (\textit{i}) temporally correlated and (\textit{ii}) exhibit nonstationary higher-order statistics. We also examine different signaling scenarios where such correlations may extend to the in-phase/quadrature (I/Q) components and polarizations through the amplitude-to-symbol mapping strategies proposed in~\cite{skvortcov2020huffman}. Moreover, analytical expressions for the energy correlations of the considered signaling schemes are derived and presented. Our results show that the proposed model exhibits excellent accuracy, achieving an estimation error for the normalized average NLI power below $5\%$ for various symbol rates and distances. A minimal MATLAB implementation of the proposed MEGN model is publicly available in~\cite{wu2026megncode}.

\begin{figure*}
    \centering
    \def\SC{1}
    \resizebox{\SC\linewidth}{!}{\subimport{Figures}{paperStructure.tex}}
    \caption{Organization of the paper. The flowchart from left to right corresponds to the analytical derivations (in Appendices), the formulas for the MEGN model, and the numerical validation in this work.}
    \label{Fig:paper_structure}
\end{figure*}

The remainder of this paper is organized as follows and summarized in Fig.~\ref{Fig:paper_structure}. In the main body of the paper, shown as the blue blocks in Fig.~\ref{Fig:paper_structure}, Section~\ref{Sec:FEGN} presents the proposed MEGN model, including the full formulation in Theorem~\ref{Thm:MEGN_Full} and its approximations in Corollaries~\ref{Thm:MEGN_Approx} and~\ref{Thm:MEGN_Approx_PM}. Section~\ref{Sec:CCDM_Corr_Deri} presents the energy correlations of the CCDM symbol sequences and provides the analytical expressions. Section~\ref{Sec:Results} reports numerical validation, whereas Section~\ref{Sec:BellLabResults} presents experimental validation. Finally, Section~\ref{Sec:Conclusions} concludes the paper and outlines possible extensions.

The derivations of the MEGN model are relegated to appendices (shown as red blocks in Fig.~\ref{Fig:paper_structure}). Appendix~\ref{App:symCorr} analyzes the statistical properties of the temporal symbol-energy correlations arising from the Manakov equations. Appendix~\ref{App:Moments} calculates the high-order frequency moments by incorporating these energy correlations into the Manakov framework. Appendix~\ref{App:Kernels_Derivation} uses the resulting frequency moments to derive the corresponding NLI PSD expressions and to decouple them into energy correlations and channel functions. Finally, Appendix~\ref{App:DP_NLI_PSD} extends these results from the single-polarization setting to the dual-polarization case, yielding the main PSD formula for the MEGN model. The primary notation adopted throughout the MEGN model is summarized in Table~\ref{Tab:Notation}.

\begin{table}[t]
\centering
\caption{Notation convention used throughout this paper.}
\label{Tab:Notation}
\renewcommand{\arraystretch}{1.2}
\begin{tabular}{ll}
\hline
\hline
Symbol & Description \\
\hline
$\jmath$ & Imaginary unit \\
$\delta$ & Kronecker delta \\
$\Exp{\cdot}$ & Expectation operator \\
$f$ & Frequency \\
$m,n,k$ & Discrete frequency component index \\
$\mathbb{Z}$ & Integers \\
$\w$ & Discrete time index of symbols \\
$\tau$ & Delay between two time indices \\
$\ppol$ & Superscript polarization indicator, $\ppol \in \{\px,\py\}$ \\
$u^{\ppol}_{\w}$ & Amplitude at time $\w$ in polarization $\ppol$ \\
$a^{\ppol}_{\w}$ & Complex symbol at time $\w$ in polarization $\ppol$ \\
$v_{n}$ & Frequency component with index $n$ \\
$R(\cdot)$ & Correlation function \\
$K(\cdot)$ & Covariance function \\
$M_s$ & Correlation length of transmitted symbols \\
$M_c$ & Fiber channel memory length \\
$M$ & Memory length used in the MEGN model \\
$N$ & Shaping blocklength \\
$s(t),\,S(f)$ & Pulse shape in time and frequency domains \\
$G(f)$ & NLI PSD \\
$z$ & Longitudinal spatial coordinate along the fiber \\
$\alpha$ & Optical field attenuation coefficient \\
$\beta_2$ & Chromatic dispersion coefficient \\
$D$ & Dispersion parameter \\
$\gamma$ & Fiber nonlinear coefficient \\
$L_s$ & Fiber span length \\
$N_s$ & Total number of spans in the link \\
$R_s$ & Symbol rate \\
\hline
\hline
\end{tabular}
\end{table}

%% file: Tex/Ch_Model.tex
\section{The Memory-EGN (MEGN) Model}\label{Sec:FEGN}

This section presents the MEGN model. Starting from the NLI PSD characterized by the EGN model, denoted as $G_{\text{EGN}}(f)$, the proposed MEGN model accounts for the PSD deviation from $G_{\text{EGN}}(f)$ caused by the temporal correlations of symbol energies, denoted as $G_{C}(f)$. Therefore, the PSD corrected by MEGN is
\begin{align}\label{Eq:FEGN_Model_0}
    G_{\text{MEGN}}(f) = G_{\text{EGN}}(f) + G_{C}(f).
\end{align}
As shown in Fig.~\ref{Fig:Models_Comparison}(b), the symbol-energy interactions fall into three categories. Accordingly, the correction term $G_{C}(f)$ in \eqref{Eq:FEGN_Model_0} is decomposed as
\begin{align}\label{Eq:FEGN_Model}
    G_{C}(f) = G_{\text{SPT}}(f) + G_{\text{XPT}}(f) + G_{\text{XP}}(f),
\end{align}
where the term $G_{\text{XP}}(f)$ has been investigated in the 4-D-EGN model \cite{liga2020extending}. The main contribution of the present model is therefore the derivation of the terms $G_{\text{SPT}}(f)$ and $G_{\text{XPT}}(f)$, which manifest themselves as the \textit{memory} effects of the fiber channel. Following the structure in \eqref{Eq:NLI_PSD_Intution}, these terms are expressed as the product of symbol-energy correlations and channel functions. 

Regarding the system setup, we consider single-channel transmission over identical spans of the same fiber type, with each span loss exactly compensated by lumped amplification. In what follows, Sec.~\ref{Sec:FEGN_Pre} first reviews the general NLI PSD expression derived from the Manakov equation and highlights the symbol-correlation-dependent terms that motivate the proposed MEGN model. The rest of this section presents the model formulation, with detailed derivations deferred to the appendices. According to \eqref{Eq:NLI_PSD_Intution}, the required signal statistics, namely energy correlations, and channel functions are defined in Secs.~\ref{Sec:FEGN_symbolCorr} and~\ref{Sec:FEGN_channelFunctions}, respectively. We briefly review the EGN model in Sec.~\ref{Sec:EGN}. Sec.~\ref{Sec:SXPT} then presents the NLI PSD contributions arising from the different symbol-energy interactions illustrated in Fig.~\ref{Fig:paper_structure}(b) and \eqref{Eq:FEGN_Model}, thereby completing the MEGN model formulation in Theorem~\ref{Thm:MEGN_Full}. Finally, Sec.~\ref{Sec:FEGN_Approximations} discusses the complexity of the MEGN model and introduces two approximations, which are detailed in Corollaries~\ref{Thm:MEGN_Approx} and~\ref{Thm:MEGN_Approx_PM}.

\subsection{Preliminaries}\label{Sec:FEGN_Pre}

We analyze the impact of the transmitted signal on the NLI in the frequency domain. The transmitted symbols on the four-dimensional optical carrier are represented by the symbol vector $(a^{\px}_{\w}, a^{\py}_{\w})$. We consider a \textit{discrete-time} signal with an arbitrarily large period $W \to \infty$. After pulse shaping, the signal is represented by a set of discrete frequency components with spacing $f_0 \to 0$, where $R_s = W f_0$. The frequency component at frequency $n f_0$ and polarization $\ppol\in\{\px,\py\}$ is then given by
\begin{align}
    v^{\ppol}_{n} = & \sqrt{f_0}S^{\ppol} {(n f_0)}\sum_{\w=0}^{W-1}a^{\ppol}_{\w}\exp{\left(-\jmath\frac{2\pi\w}{W} n\right)} ,
\end{align}
and we assume that same pulse shaping is used for $\px$ and $\py$ polarizations that is
\begin{align}
    S (f) = S^{\px}(f) = S^{\py}(f).
\end{align}

After transmission through one homogeneous fiber span of length $L_s$, based on the Manakov equation, the general expression for the NLI PSD is
\begin{align}
    G(f) = \GSCIx(f) + \GSCIy(f),
\end{align}
where $\GSCIx(f)$ and $\GSCIy(f)$ denote the NLI PSD in the $\px$- and $\py$-polarizations, respectively. Their expressions are equivalent and can be obtained from one another by interchanging the polarization indices. For $\GSCIx(f)$, the NLI PSD is given by~\cite{Poggiolini2012}\cite[Appendix~D]{Carena2014}

\begin{subequations}\label{Eq:G_SCI}
\begin{align}
     \GSCIx(f)  = &\frac{64}{81} f_0^3 \exp{(-2\alpha L_s)}\sum_{i=-\infty}^{+\infty} \delta(f-if_0)  \nonumber \\
    & \sum_{m,n,k\in \mathcal{S}_i} \sum_{m',n',k'\in \mathcal{S}_i} \zeta(k,m,n) \zeta^*(k',m',n') \nonumber \\
    &    \Bigl( \Exp{v^{\px}_{m} v^{\px *}_{n} v^{\px}_{k} v^{\px *}_{m'} v^{\px}_{n'} v^{\px *}_{k'} }
    \label{Eq:6FreqProduct_1}  \\
    & +\Exp{v^{\px}_{m} v^{\px *}_{n} v^{\px}_{k} v^{\px *}_{m'} v^{\py}_{n'} v^{\py *}_{k'} }
    \label{Eq:6FreqProduct_2}  \\
    & + \Exp{v^{\px}_{m} v^{\px *}_{m'} v^{\px}_{n'} v^{\px *}_{k'}  v^{\py *}_{n} v^{\py}_{k} }
    \label{Eq:6FreqProduct_3}  \\
    & + \Exp{v^{\px}_{m} v^{\px *}_{m'}  v^{\py *}_{n} v^{\py}_{k}  v^{\py}_{n'} v^{\py *}_{k'}}
    \Bigr),\label{Eq:6FreqProduct_4}
\end{align}
\end{subequations}
where $\zeta(k,m,n)$ represents the NLI generation efficiency in discrete-time given as
\begin{align}\label{Eq:NLI_effc_Discrete}
  & \zeta(k,m,n) \nonumber \\
  & = \gamma\frac{1-\exp{( -2\alpha Ls+\jmath 4\pi^2\beta_2 f_0^2 (k-n)(m-n)L_s)}}{2\alpha-\jmath 4\pi^2\beta_2  f_0^2 (k-n)(m-n)}.
\end{align}

As shown in \eqref{Eq:G_SCI}, the signal dependence of the NLI arises from four sixth-order moments of the frequency components. For example, \eqref{Eq:6FreqProduct_1} can be expanded in the time domain as

\begin{align} \label{Eq:FreqMom}
    &\Exp{v^{\px}_{m} v^{\px *}_{n} v^{\px}_{k} v^{\px *}_{m'} v^{\px}_{n'} v^{\px *}_{k'}}\nonumber  \\
    &\quad = f_0^3 \Pss_{mnkm'n'k'}
    \sum_{\mL} \Exp{\axw{1}{}\axw{2}{*}\axw{3}{}\axw{4}{*}\axw{5}{}\axw{6}{*}} \nonumber \\
    &\qquad \exp\!\left(
    -\jmath\frac{2\pi}{W}
    \left(
    \w_1 m - \w_2 n + \w_3 k - \w_4 m' + \w_5 n' - \w_6 k'
    \right)
    \right),
\end{align}
where $\Pss_{mnkm'n'k'}$ represents the beating of six pulse frequency components given as
\begin{align*}
  \Pss_{mnkm'n'k'}  =& S(mf_0) S^{ *} (nf_0) S(kf_0) \nonumber \\
  &  S^{ *} (m'f_0) S(n'f_0) S^{ *} (k'f_0),
\end{align*}
and $\mL$ represents the set of six time indices
\begin{align} \label{Eq:TotalTimeSpace}
    \mL  \!\triangleq \! \left\{ (\w_1,\w_2,\ldots,\w_6) \in \mathbb{Z}^6  \!:\!   0 \!\leq\! \w_1,\w_2,\ldots,\w_6 \!\leq\! W\!-\!1  \right\} .
\end{align}

The expression in \eqref{Eq:G_SCI} shows that, at a discrete-time level, the NLI power is fundamentally determined by symbol correlations across time and polarization. However, the size of the time-domain index space $\mL$ in \eqref{Eq:TotalTimeSpace} is prohibitively large for directly evaluating the impact of symbol correlations on the NLI. Fortunately, under reasonable assumptions on the symbol statistics, most subspaces of $\mL$ give rise to symbol-correlation terms that contribute negligibly to the NLI in \eqref{Eq:G_SCI}. As a result, only the subspaces associated with energy correlations remain relevant to the NLI, reducing the six-dimensional time-domain index space to at most three dimensions. In the next section, we introduce the relevant symbol-energy correlations and refer the reader to Appendix~\ref{App:symCorr} for further details.

\subsection{Symbol-Energy Correlation Statistics}\label{Sec:FEGN_symbolCorr}

From now on, we treat the sequence of symbol energies $[\ldots,|a^{\px}_{\w}|^2,|a^{\py}_{\w}|^2,|a^{\px}_{\w+1}|^2,|a^{\py}_{\w+1}|^2,\ldots ]$ as a random process. The following assumptions on their statistics are made:
\begin{enumerate}
   \item first-order stationarity;
   \item symmetric with respect to both polarizations;
   \item wide-sense cyclostationarity with \textit{finite} fundamental period equal to the symbol correlation length $M_s$.
\end{enumerate}

Assumption 1) indicates that the symbol statistics at individual times are time invariant\cite{wu2021edi}. Assumption 2) further implies identical statistics for $\px$ and $\py$ polarizations. With these two assumptions, we have
\begin{align}
    P_{\text{ch}}\triangleq \Exp{|a^{\ppol}_{\w}|^2} & =\mathbb{E}\{|a|^2 \}, \quad
    \Exp{|a^{\ppol}_{\w}|^4} =   \mathbb{E}\{|a|^4 \}.
\end{align}
where $\ppol \in\{\px,\py\}$ and $P_{\text{ch}}$ denotes the average symbol energy per polarization. The energy correlations of interest are
\begin{subequations}\label{Eq:SymCorr_Intro}
\begin{align}
    \Rc{\w,\wtt}{\ppol \ppol' }  &\triangleq \Exp{|a^{\ppol}_{\w}|^2 |a^{\ppol'}_{\wtt}|^2} , \\
    \Ra{\w,\wtt}{\ppol \ppol' }  &\triangleq  \Exp{|a_{p\!, \w }|^2 |a_{p^\prime\!, \wtt}|^4}, \\
    \Rb{\w,\wtt,\wttt}{\ppol \ppol' \ppol''}   &\triangleq   \Exp{|a^{\ppol}_{\w}|^2 |a^{\ppol'}_{\wtt}|^2 |a^{\ppol''}_{\wttt}|^2},
\end{align}
\end{subequations}
where $\ppol,\ppol',\ppol''\in\{\px,\py\}$ are polarization indices. The corresponding covariance functions are thus defined as
\begin{subequations}\label{Eq:SymCov_Intro}
\begin{align}
    &\Kc{\w,\wtt}{\ppol \ppol' }
    \triangleq
    \Rc{\w,\wtt}{\ppol \ppol' }
    - \Exp{|a|^2}^2,
    \label{Eq:SymCov1_Intro}  \\
    &\Ka{\w,\wtt}{\ppol \ppol' }
    \triangleq
    \Ra{\w,\wtt}{\ppol \ppol' }
    - \Exp{|a|^2}\Exp{|a|^4},
    \label{Eq:SymCov2_Intro} \\
    &\Kb{\w,\wtt,\wttt}{\ppol \ppol' \ppol''}
    \triangleq
    \Rb{\w,\wtt,\wttt}{\ppol \ppol' \ppol''}
    - \Exp{|a|^2}^3 .
    \label{Eq:SymCov3_Intro}
\end{align}
\end{subequations}
Depending on the time indices $\w,\wtt,\wttt$ and polarization indices $\ppol,\ppol',\ppol''$, these correlation functions capture different interaction types in Fig.~\ref{Fig:Models_Comparison}(b).

Due to assumption 2), if cross-polarization energy correlations are present, they are symmetric. For instance,
\begin{align*}
     \Kc{\w ,\wtt}{\px\py} = \Kc{\w ,\wtt}{\py\px} .
\end{align*}

Assumption 3) further implies that although the correlations may vary with time, they repeat periodically with period $M_s$. For instance,
\begin{align*}
    \Kc{\w,\wtt}{\px\py} = \Kc{\w +M_s,\wtt+M_s}{\px\py}.
\end{align*}
Since $M_s$ is finite, two symbols become independent, once their temporal separation exceeds $M_s$, leading to zero-covariance. For instance,
\begin{align*}
    \Kc{\w,\wtt}{\px\py} = 0, \quad \forall |\wtt-\w| \geq M_s.
\end{align*}

Additionally, cyclo-stationarity implies that the time-averaged correlation functions are symmetric with respect to the time delay, which significantly simplifies the derivation of the frequency moments (see Appendix~\ref{App:Moments}).

As we are interested in the average NLI power over time, the time-dependent symbol functions in \eqref{Eq:SymCov1_Intro}--\eqref{Eq:SymCov3_Intro} can be equivalently replaced by their time-averaged versions. Given time delay $\tau=\wtt-\w,\taup=\wttt-\w$, they are defined as
\begin{subequations}\label{Eq:Cov_General}
\begin{align}
    \bKc{ \tau}{\ppol \ppol' }    &\triangleq  \frac{1}{M_s} \sum_{\w=1}^{M_s}  \Kc{\w,\w+\tau}{\ppol \ppol' }  ,   \label{Eq:SymCovAvg1}  \\
    \bKa{\tau}{\ppol \ppol' }    &\triangleq \frac{1}{M_s} \sum_{\w=1}^{M_s} \Ka{\w,\w+\tau}{\ppol \ppol' }  ,   \label{Eq:SymCovAvg2} \\
    \bKb{ \tau,\taup}{\ppol \ppol' \ppol'' }    & \triangleq  \frac{1}{M_s} \sum_{\w=1}^{M_s}  \Kb{\w,\w+\tau,\w+\taup}{\ppol \ppol' \ppol'' } .    \label{Eq:SymCovAvg3}
\end{align}
\end{subequations}

\subsection{Channel Functions with Memory}\label{Sec:FEGN_channelFunctions}
The channel functions essentially reflect the effective beating among triplets of frequency components. Moreover, depending on the groups of symbols experiencing the nonlinear interactions, which is characterized by different subspaces of the time-domain index set $\mL$, the channel functions can be further decomposed into \textit{kernels}. We refer the reader to Appendix~\ref{App:Moments}-\ref{App:Kernels_Derivation} for further details.

The beating among a triplet of pulse frequency components is given by
\begin{align}
    \btF = S(f_1) S^*(f_1+f_2-f) S(f_2) \; \mu(f_1,f_2,f),
\end{align}
where  $\mu(f_1,f_2,f)$ is the link function.
For identical spans of the same fiber type, with lumped amplifiers exactly compensating for the loss of each span, it is defined as
\begin{align}
\mu(f_1,f_2,f)&=\zeta(f_1,f_2,f)\nu(f_1,f_2,f),
\end{align}
where $\zeta(f_1,f_2,f)$ is the continuous version of \eqref{Eq:NLI_effc_Discrete}
\begin{align}
  & \zeta(f_1,f_2,f) \nonumber \\
  & = \gamma\frac{1-\exp{( -2\alpha Ls+\jmath 4\pi^2\beta_2 (f_1-f)(f_2-f)L_s)}}{2\alpha-\jmath 4\pi^2\beta_2 (f_1-f)(f_2-f)},
\end{align}
and $\nu(f_1,f_2,f)$ represents the coherent interference of the NLI field contributions produced in different spans, i.e.,
\begin{align}
\nu(f_1,f_2,f) &  =  \frac{\sin\left(2\beta_2\pi^2 (f_1-f)(f_2-f)N_s L_s\right)}{\sin\left(2\beta_2\pi^2 (f_1-f)(f_2-f) L_s\right)} \cdot \nonumber  \\
 & \exp{[\jmath2\beta_2\pi^2(f_1-f)(f_2-f)(N_s-1)L_s]}.
\end{align}

Then, the following kernels are known and investigated in the EGN model \cite[Eqs.~(7)--(9), (107)--(109)]{Carena2014}
\begin{subequations}\label{Eq:Kernel_EGN}
\begin{align}
  \phi_1(f)\!  & =  \frac{16}{27} R_s^2 \iint_{\mathcal{B}_2}  \Bigl|   S(f_1)  \Bigl|^2  \Bigl|   S(f_2)  \Bigl|^2 \nonumber \\
  & \quad \Bigl|  S(f_1+f_2-f)  \Bigl|^2   \Bigl| \mu (f_1,f_2,f) \Bigr|^2, \\
  \phi_2(f)\! & =  \frac{16}{81} R_s^2     \int_{\mathcal{B}_1} df_1  \Bigl|   S(f_1)  \Bigl|^2    \Bigl| \int_{\mathcal{B}_1} df_2  S( f_2 ) \nonumber \\
  & \quad  S^*(f_1+f_2-f) \mu (f_1,f_2,f) \Bigr|^2, \\
  \phi_3(f)\! & = \frac{16}{81} R_s^2 \int_{\mathcal{B}_1} df_1 |S(f_1)|^2 \Bigl|  \int_{\mathcal{B}_1} \!\! df_2   S( f_2 ) \nonumber \\
  &  \quad  S^*(f_1\!-\!f_2\!+\!f)  \mu (f_1\!-\!f_2\!+\!f,f_2,f) \Bigr|^2 \!,\\
  \phi_4(f)\! &  = \frac{16}{81} R_s    \Bigl| \iint_{\mathcal{B}_2}  df_1  df_2   \btF  \Bigr|^2 ,
\end{align}
\end{subequations}
where the integral domains $ \mathcal{B}_1$ and $ \mathcal{B}_2$ are
\begin{subequations}
\begin{align*}
    \mathcal{B}_1 = & \{\, f : f  \in [-R_s/2,\, R_s/2] \,\}, \\
    \mathcal{B}_2 = &  \{\, (f_1, f_2) : f_1, f_2 \in [-R_s/2,\, R_s/2] \,\}.
\end{align*}
\end{subequations}

\input{Tex/Eq_Kernels}

The new kernels accounting for energy correlations are given in \eqref{Eq:Kn_chi1}--\eqref{Eq:Kn_psi3}, which include the additional delay arguments $\tau$ and $\taup$. We also note that all the kernels in \eqref{Eq:Kernel_EGN} and \eqref{Eq:Kn_chi1}-\eqref{Eq:Kn_psi3} are strictly positive.

\subsection{EGN: NLI for I.I.D. Symbols}\label{Sec:EGN}

Ideally, the correlations in \eqref{Eq:SymCorr_Intro} must be explicitly retained as genuine components to compute the NLI PSD in \eqref{Eq:G_SCI}. However, the EGN model simplifies \eqref{Eq:SymCorr_Intro} under symmetric polarization and the i.i.d. symbol assumptions, i.e.,
\begin{subequations}
\begin{align}
    \Rc{\w,\wtt}{\ppol \ppol'}  &=  \Exp{|a|^2} ^2 , \\
    \Ra{\w,\wtt}{\ppol \ppol'} &=  \Exp{|a|^2} \Exp{|a|^4}, \\
    \Rb{\w,\wtt,\wttt}{\ppol \ppol' \ppol''}  &=  \Exp{|a|^2} ^3 .
\end{align}
\end{subequations}
Consequently, the covariance functions in \eqref{Eq:Cov_General} become zero. Hence, $G_{\text{EGN}}(f)$ is computed \cite[Eq.~(5)]{Carena2014} as
\begin{align}\label{Eq:G_EGN}
    G_{\text{EGN}}(f) & = \Exp{|a|^2} ^3 \kappa_1 \nonumber \\
    &  +\left(\Exp{|a|^2} \Exp{|a|^4}-2\Exp{|a|^2} ^3\right) \kappa_2 \nonumber \\
    & + \left( \Exp{|a|^6}-9\Exp{|a|^2} \Exp{|a|^4}+12\Exp{|a|^2} ^3\right) \kappa_3,
\end{align}
where the channel functions $\kappa_1$, $\kappa_2$, and $\kappa_3$ are computed from the kernels in \eqref{Eq:Kernel_EGN} as
\begin{subequations}\label{Eq:Kn_EGN}
\begin{align}
   \kappa_1  \triangleq & \; \phi_1 , \label{Eq:Kn_EGN_1}\\
   \kappa_2   \triangleq & \; 5\phi_2 +  \phi_3 \label{Eq:Kn_EGN_2},\\
   \kappa_3   \triangleq & \;  \phi_4. \label{Eq:Kn_EGN_3}
\end{align}
\end{subequations}
Furthermore, the symbol statistic $\Exp{|a|^2} \Exp{|a|^4}$ in the second term of \eqref{Eq:G_EGN}, when normalized by $\Exp{|a|^2}^3$, reduces to the well-known NLI metric \textit{kurtosis}.

Therefore, the EGN model predicts the NLI accurately only for transmissions of i.i.d. symbols. To compensate for the NLI neglected by the EGN model when transmitting non-i.i.d. symbols that are correlated at the energy level, the proposed MEGN model requires the covariance functions in \eqref{Eq:Cov_General}.

\subsection{MEGN: NLI for Symbol Energy Interactions}\label{Sec:SXPT}
The NLI PSD due to the SPT interactions of symbol energies in Fig.~\ref{Fig:paper_structure}(b) is
\begin{align}\label{Eq:G_SPT_total}
    G_{\text{SPT}}(f) = & G^{(1)}_{\text{SPT}}(f)+ G^{(2)}_{\text{SPT}}(f),
\end{align}
where $G^{(1)}_{\text{SPT}}(f)$ involves a single summation, while $G^{(2)}_{\text{SPT}}(f)$ involves a double summation and they are given by
\begin{subequations}\label{Eq:G_SPT}
\begin{align}
    G^{(1)}_{\text{SPT}}(f) = &\sum_{\tau=1}^{M}  \Exp{|a|^2}\bKSc{\tau}  \KnSc{\tau,f}   \nonumber   \\
    & +  \bKSa{ \tau} \KnSa{\tau,f} , \label{Eq:G_SPT_1} \\
   G^{(2)}_{\text{SPT}}(f) = & \sum_{\tau=1}^{M} \sum_{\taup=\tau+1}^{\tau+M} \left[ \bKSb{\tau,\taup}   -\Exp{|a|^2}\right. \left. \left( \bKSc{\tau} \right. \right. \nonumber \\
    &  \left.  \left. +\bKSc{\taup}  +\bKSc{\taup-\tau}\right) \right] \KnSb{\tau,f}. \label{Eq:G_SPT_2}
\end{align}
\end{subequations}
The $M$ indicates the memory length of the NLI interactions, and its role on the model accuracy and complexity will be addressed in the next section.

The required SPT-type covariance functions $\bKSc{\tau}$, $\bKSa{ \tau}$ and $\bKSb{\tau,\taup}$ in \eqref{Eq:G_SPT} are represented in the $\px$ polarization as
\begin{subequations}\label{Ch:Cov_SPT}
\begin{align}
    \bKSc{ \tau}  & \triangleq  \left.\bKc{ \tau}{\px\px}\right. ,  \label{Ch:Cov_SPT_1}  \\
    \bKSa{\tau} & \triangleq  \bKa{ \tau}{\px\px},  \label{Ch:Cov_SPT_2}  \\
    \bKSb{ \tau,\taup} & \triangleq  \left.\bKb{ \tau,\taup}{\px\px\px}\right . . \label{Ch:Cov_SPT_3}
\end{align}
\end{subequations}
The corresponding channel functions $\KnSc{\tau,f}, \KnSa{\tau,f}$ and $ \KnSb{\tau,\taup,f}$ in \eqref{Eq:G_SPT} are
\begin{subequations}\label{Ch:Kn_SPT}
\begin{align}
   \KnSc{\tau,f}  \triangleq & \;5\xi_1(\tau,f) + 5\psi_1(\tau,f) + 2\psi_2(\tau,f)   \nonumber  \\
   & -\!22\chi_1(\tau,f)\!-\!21\chi_2(\tau,f) \!-\!5  \chi_3(\tau,f) , \label{Ch:Kn_SPT_1}\\
  \KnSa{\tau,f}  \triangleq & \;4\chi_1(\tau,f) + 4\chi_2(\tau,f)  +  \chi_3(\tau,f)\label{Ch:Kn_SPT_2},\\
   \KnSb{\tau,\taup,f}  \triangleq & \;4\xi_2(\tau,\taup,f) + 2\psi_3(\tau,\taup,f).\label{Ch:Kn_SPT_3}
\end{align}
\end{subequations}

The NLI PSD due to the XPT symbol energy interactions is written as
 \begin{align} \label{Eq:G_XPT_total}
    G_{\text{XPT}}(f) & =  G^{(1)}_{\text{XPT}}(f)+ G^{(2)}_{\text{XPT}}(f),
\end{align}
where
\begin{subequations}\label{Eq:G_XPT}
\begin{align}
    G^{(1)}_{\text{XPT}}(f) = &  \sum_{\tau=1}^{M}  \Exp{|a|^2}\bKXc{\tau} \KnXc{\tau,f}  \nonumber \\
   & +   \bKXa{ \tau} \KnXa{\tau,f}   \nonumber\\
    & +  \left[\bKXb{0,\tau}-\Exp{|a|^2} \bKXc{ 0} \right]  \KnXbi{\tau,f}, \label{Eq:G_XPT_2D} \\
  G^{(2)}_{\text{XPT}}(f) =  &   \sum_{\tau=1}^{M} \sum_{\taup=\tau+1}^{\tau+M} \left[ \bKXb{\tau,\taup}   -\Exp{|a|^2}\right. \left. \left( \bKXc{\tau} \right. \right.  \nonumber \\
    &  \left.  \left.  +\bKSc{\taup} +\bKXc{\taup-\tau}\right) \right] \KnXbii{\tau,\taup,f} . \label{Eq:G_XPT_3D}
\end{align}
\end{subequations}
The XPT-type covariance functions in \eqref{Eq:G_XPT} are represented as
\begin{subequations}\label{Ch:Cov_XPT}
\begin{align}
    \bKXc{ \tau}  & \triangleq  \left.\bKc{ \tau}{\px \py}\right., \label{Ch:Cov_XPT_1}   \\
    \bKXa{\tau} & \triangleq  \left.\bKa{ \tau}{\px \py}\right.,  \label{Ch:Cov_XPT_2}   \\
    \bKXb{ \tau,\taup} & \triangleq  \left.\bKb{ \tau,\taup}{\px \py \px}\right. .  \label{Ch:Cov_XPT_3}
\end{align}
\end{subequations}
Note that the term $\bKSc{\tau}$ in \eqref{Eq:G_XPT_3D} appears to account for the SPT interaction inherent in $\bKXb{ \tau,\taup}$. The associated channel functions in \eqref{Eq:G_XPT} are
\begin{subequations}\label{Ch:Kn_XPT}
\begin{align}
   \kappa^{(1)}_{\text{X}0}(f)  & \triangleq  3\phi_4(f), \label{Ch:Kn_XP_1}\\
   \kappa^{(2)}_{\text{X}0}(f)  &\triangleq  - 12 \phi_4(f)+5\phi_2(f) + \phi_3(f) ,\label{Ch:Kn_XP_2}\\
   \KnXc{\tau,f}  &\triangleq    4\xi_1(\tau,f) + \psi_1(\tau,f) +  \psi_2(\tau,f)   \nonumber  \\
     &   \quad -14\chi_1(\tau,f) -9\chi_2(\tau,f) - \chi_3(\tau,f), \label{Ch:Kn_XPT_1}  \\
   \KnXa{\tau,f}   & \triangleq  2\chi_1(\tau,f) +  \chi_2(\tau,f),\label{Ch:Kn_XPT_2}  \\
   \KnXbi{\tau,f} &\triangleq   6\chi_1(\tau,f) +  5\chi_2(\tau,f) +\chi_3(\tau,f), \label{Ch:Kn_XPT_31} \\
   \KnXbii{\tau,\taup,f} &\triangleq   5\xi_2(\tau,\taup,f) + \psi_3(\tau,\taup,f).\label{Ch:Kn_XPT_32}
\end{align}
\end{subequations}

For XP-type energy interactions, the signal statistics are characterized by $\bKXc{0}$ and $\bKXa{0}$ and are given as
\begin{align} \label{Eq:G_XP_total}
    & G_{\text{XP}}(f) =  \bKXa{ 0} \kappa^{(1)}_{\text{X}0}(f)  +\Exp{|a|^2} \bKXc{ 0}\kappa^{(2)}_{\text{X}0}(f).
\end{align}

Finally, in view of \eqref{Eq:FEGN_Model_0}, \eqref{Eq:FEGN_Model}, \eqref{Eq:G_SPT_total}, \eqref{Eq:G_XPT_total}, and \eqref{Eq:G_XP_total}, the MEGN model is presented in the following theorem.

\begin{theorem}\label{Thm:MEGN_Full}
For the transmission of symbols exhibiting temporal and cross-polarization energy correlations over the optical fiber channel, the NLI PSD is expressed as
\begin{align} \label{Eq:MEGN_Full}
    G_{\text{MEGN}}(f) & =  G_{\text{EGN}}(f) + G^{(1)}_{\text{SPT}}(f) + G^{(2)}_{\text{SPT}}(f) \nonumber \\
    & \quad  + G^{(1)}_{\text{XPT}}(f) + G^{(2)}_{\text{XPT}}(f) + G_{\text{XP}}(f) ,
\end{align}
where $G_{\text{EGN}}(f)$ is computed by \eqref{Eq:Kernel_EGN}, \eqref{Eq:G_EGN} and \eqref{Eq:Kn_EGN}. Based on the kernels in \eqref{Eq:Kn_chi1}-\eqref{Eq:Kn_psi3}, $G^{(1)}_{\text{SPT}}(f)$ and $G^{(2)}_{\text{SPT}}(f)$ are computed by \eqref{Eq:G_SPT}, \eqref{Ch:Cov_SPT}, and \eqref{Ch:Kn_SPT}; $G^{(1)}_{\text{XPT}}(f)$ and $G^{(2)}_{\text{XPT}}(f)$ by \eqref{Eq:G_XPT}, \eqref{Ch:Cov_XPT}, and \eqref{Ch:Kn_XPT_1}--\eqref{Ch:Kn_XPT_32}; and $G_{\text{XP}}(f)$ by \eqref{Ch:Cov_XPT_1}, \eqref{Ch:Cov_XPT_2}, \eqref{Ch:Kn_XP_1}, and \eqref{Ch:Kn_XP_2}.
\end{theorem}

Moreover, the MEGN model provides theoretical justification for our previously proposed NLI metric, the \textit{energy dispersion index} (EDI) \cite{wu2021edi,wu2021exponentially}, which reflects the statistics in \eqref{Ch:Cov_SPT_1} and \eqref{Ch:Cov_XPT_1}.

\subsection{Complexity Analysis and Approximations}\label{Sec:FEGN_Approximations}

\def\WD{1.1*\linewidth}
\def\HT{0.6*\linewidth}
\begin{figure}
\centering
\setkeys{Gin}{width=0.01\textwidth}
\resizebox{.9\linewidth}{!}{\subimport{Figures}{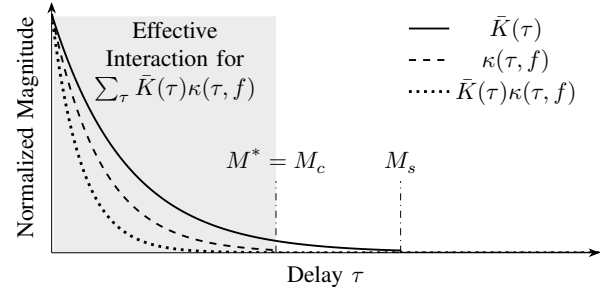}}
\caption{An illustration of the relationship between symbol correlation length $M_s$, channel memory $M_c$ and optimal memory $M$ to be used in the MEGN model, when $M_c<M_s$.}
\label{Fig:lengthComp}
\end{figure}

As seen from \eqref{Eq:G_SPT} and \eqref{Eq:G_XPT}, the complexity of the MEGN model increases with the memory length $M$ due to the summations, with each term in the sum involving a double integral. More specifically, \eqref{Eq:G_SPT_1} and \eqref{Eq:G_XPT_2D} show that the computational complexity scales linearly for the 1-D summations in $G^{(1)}_{\text{SPT}}(f)$ and $G^{(1)}_{\text{XPT}}(f)$, whereas \eqref{Eq:G_SPT_2} and \eqref{Eq:G_XPT_3D} show quadratic scaling for the 2-D summations in $G^{(2)}_{\text{SPT}}(f)$ and $G^{(2)}_{\text{XPT}}(f)$. Therefore, the model can be approximated by properly choosing the memory length $M$ and simplifying the corresponding summation terms.

Regarding the memory length $M$, in the extreme case, one may make it infinitely long to account for everything, even though in reality, zero interactions occur between two distant symbols further away from each other. Ideally, it is sufficient to use the memory length to account for the effective interaction arising from the channel memory effects and the signaling scheme. At one time instance, only the neighboring symbols of the symbol of interest, within both the correlation length $M_s$ and the channel memory $M_c$, interact to alter the NLI with respect to i.i.d. symbols. Hence, the minimum length to be used in the MEGN model without losing accuracy is
\begin{align*}
    M^* =  \min (M_s,M_c )
\end{align*}
An example is shown in Fig.~\ref{Fig:lengthComp} to illustrate the relationship between this ideal MEGN model memory $M^*$, $M_s$ and $M_c$ when $M_c<M_s$. For further simplicity, we use a truncated memory with a constant value so that the majority of the contributions from the symbol correlations and the channel functions combined have been included.

Regarding the 2-D summations in $G^{(2)}_{\text{SPT}}(f)$ \eqref{Eq:G_SPT_2} and $G^{(2)}_{\text{XPT}}(f)$ \eqref{Eq:G_XPT_3D}, we propose discarding them. As we will show in Sec.~\ref{Sec:Results}, these two terms make negligible contributions to the NLI. The resulting simplified MEGN model is then given in the following corollary
\begin{corollary}\label{Thm:MEGN_Approx}
For the transmission of symbols exhibiting temporal and cross-polarization energy correlations over the optical fiber channel, the NLI PSD is approximated as
    \begin{align} \label{Eq:FEGN_Approx}
    G_{\text{MEGN}}(f)  & \approx  G_{\text{EGN}}(f) + G^{(1)}_{\text{SPT}}(f) \nonumber \\
    & + G^{(1)}_{\text{XPT}}(f) + G_{\text{XP}}(f),
\end{align}
where $G_{\text{EGN}}(f)$ is computed by \eqref{Eq:Kernel_EGN}, \eqref{Eq:G_EGN} and \eqref{Eq:Kn_EGN}. Based on the kernels in \eqref{Eq:Kn_chi1}--\eqref{Eq:Kn_xi1}, \eqref{Eq:Kn_psi1} and \eqref{Eq:Kn_psi2}, $G^{(1)}_{\text{SPT}}(f)$ is determined by \eqref{Eq:G_SPT_1}, \eqref{Ch:Cov_SPT_1}, \eqref{Ch:Cov_SPT_2}, \eqref{Ch:Kn_SPT_1}, and  \eqref{Ch:Kn_SPT_2}; $G^{(1)}_{\text{XPT}}(f)$ by \eqref{Eq:G_XPT_2D}, \eqref{Ch:Cov_XPT_1}, \eqref{Ch:Cov_XPT_2}, \eqref{Ch:Kn_XPT_1}--\eqref{Ch:Kn_XPT_31}; and $G_{\text{XP}}(f)$ by \eqref{Ch:Cov_XPT_1}, \eqref{Ch:Cov_XPT_2}, \eqref{Ch:Kn_XP_1}, and \eqref{Ch:Kn_XP_2}.
\end{corollary}


Moreover, when the two polarizations are independent, as in polarization-multiplexed (PM) signaling, the MEGN model admits further simplifications. In this case, for $\forall \tau \neq 0$, the XPT correlations reduce to
\begin{align*}
    \bKXc{\tau} &= 0,\\
    \bKXa{\tau} & = 0 , \\
    \bKXb{\tau,\taup} &  = \Exp{|a|^2}\bKSc{\taup }.
\end{align*}
Hence, the $G_{\text{XPT}}(f)$ in \eqref{Eq:G_XPT} is greatly simplified. In this special PM case, the \textit{approximated} MEGN model in \eqref{Eq:FEGN_Approx} is given in the following corollary.

\begin{corollary}\label{Thm:MEGN_Approx_PM}
For the transmission of symbols exhibiting only temporal energy correlations over the optical fiber channel, the NLI PSD is approximated as
\begin{align}\label{Eq:Corr_SCI_1}
   G_{\text{MEGN}}(f)  & \approx  G_{\text{EGN}}(f)  + \sum_{\tau=1}^{M}  \Exp{|a|^2}\bKSc{\tau}    \kappa^{\text{PM}}_{\text{S}1}(\tau,f)\nonumber \\
   & +   \bKSa{ \tau}  \kappa^{\text{PM}}_{\text{S}2} (\tau,f) ,
\end{align}
where $G_{\text{EGN}}(f)$ is computed by \eqref{Eq:Kernel_EGN}, \eqref{Eq:G_EGN} and \eqref{Eq:Kn_EGN}, and the energy correlations $\bKSc{\tau}$ and $\bKSa{ \tau}$ are given in \eqref{Ch:Cov_SPT_1} and \eqref{Ch:Cov_SPT_2}, while the channel functions $\kappa^{\text{PM}}_{\text{S}1}(\tau,f)$ and $\kappa^{\text{PM}}_{\text{S}2}(\tau,f)$ are defined from the kernels in \eqref{Eq:Kn_chi1}--\eqref{Eq:Kn_xi1}, \eqref{Eq:Kn_psi1} and \eqref{Eq:Kn_psi2}, as
\begin{subequations}
\begin{align}
    \kappa^{ \text{PM}}_{\text{S}1} (\tau,f)\triangleq & \;  5\xi_1(\tau,f) + 5\psi_1(\tau,f) + 2\psi_2(\tau,f) \nonumber \\
    & - 16\chi_1(\tau,f) - 16 \chi_2(\tau,f) -4 \chi_3(\tau,f), \\
    \kappa^{ \text{PM}}_{\text{S}2}(\tau,f) \triangleq & \; 4\chi_1(\tau,f) + 4\chi_2(\tau,f) +  \chi_3(\tau,f).
\end{align}
\end{subequations}

\end{corollary}

%% file: Tex/Eq_Kernels.tex
\begin{figure*}[!t]\label{Eq:Kns} 
\normalsize
\setcounter{MYtempeqncnt}{\value{equation}}
\setcounter{equation}{19}
\begin{subequations}
\begin{align} 
    & \chi_1(\tau,f) \!=\! \frac{32}{81} R_s  \left[   \Bigl| \iint_{\mathcal{B}_2}  df_1   df_2   \btF \cos^2{ \frac{\pi\tau(f-f_2)}{R_s} } \Bigr|^2   -   \Bigl| \iint_{\mathcal{B}_2}  df_1 df_2   \btF  \sin^2{\frac{\pi\tau(f-f_2)}{R_s} } \Bigr|^2 \right] \label{Eq:Kn_chi1}\\
    & \chi_2(\tau,f) \!=\! \frac{32}{81} R_s \left[    \Bigl| \iint_{\mathcal{B}_2}  df_1   df_2    \btF  \cos{ \frac{2\pi\tau f_1}{R_s}  }   \Bigr|^2   + \Bigl| \iint_{\mathcal{B}_2}  df_1   df_2   \btF  \sin{ \frac{2\pi\tau f_1 }{R_s}  } \Bigr|^2 \right] \label{Eq:Kn_chi2} \\
    &  \chi_3(\tau,f) \!=\! \frac{32}{81} R_s   \left[   \Bigl| \iint_{\mathcal{B}_2}  df_1   df_2  \btF   \cos{ \frac{2\pi\tau (f_1+f_2-f)}{R_s}  }   \Bigr|^2  \! \!+\!  \Bigl| \iint_{\mathcal{B}_2}  df_1   df_2    \btF \sin{ \frac{2\pi\tau (f_1+f_2-f)}{R_s}  } \Bigr|^2 \right] \label{Eq:Kn_chi3} \\
    & \xi_1(\tau,f)\!=\!  \frac{32}{81} R_s^2    \left[  \int_{\mathcal{B}_1} df_1  \Bigl| \cos{ \frac{\pi\tau(f-f_1)}{R_s} } S(f_1)  \Bigl|^2  \right. \Bigl| \int_{\mathcal{B}_1} df_2  S( f_2 ) S^*(f_1+f_2-f)  \mu (f_1,f_2,f) \Bigr|^2 \nonumber  \\  
    & - \int_{\mathcal{B}_1} df_1  \Bigl| \sin{ \frac{\pi\tau(f-f_1)}{R_s} } S(f_1)  \Bigl|^2  \left. \Bigl|\int_{\mathcal{B}_1} df_2  S( f_2 ) S^*(f_1+f_2-f)  \mu (f_1,f_2,f) \Bigr|^2 \right] \label{Eq:Kn_xi1} \\
    & \xi_2(\tau,\taup,f) \!=\! \frac{16}{81} R_s  \left[   \Bigl| \iint_{\mathcal{B}_2}  df_1   df_2    \btF  \right. \left[ \exp{\left(\jmath\frac{2\pi\tau(f_1-f)}{R_s}\right)}+\exp{\left(\jmath\frac{2\pi\taup(f_1-f)}{R_s}\right)} \right] \Bigr|^2  \nonumber  \\  
    &-\left.   \Bigl| \iint_{\mathcal{B}_2}  df_1   df_2    \btF  \exp{\left(\jmath\frac{2\pi\tau (f_1-f)}{R_s} \right)} \Bigr|^2 \right. -\left.   \Bigl| \iint_{\mathcal{B}_2}  df_1   df_2    \btF  \exp{\left(\jmath\frac{2\pi\taup (f_1-f)}{R_s} \right)} \Bigr|^2 \right.\nonumber  \\  
    &+\left.   \Bigl| \iint_{\mathcal{B}_2}  df_1   df_2    \btF  \right.   \left[ \exp{\left(\jmath\frac{2\pi(\tau-\taup)(f-f_1)}{R_s}\right)}+\exp{\left(\jmath\frac{2\pi\tau(f-f_1)}{R_s}\right)} \right] \Bigr|^2 \nonumber  \\  
    &-\left.  \Bigl| \iint_{\mathcal{B}_2}  df_1   df_2    \btF \exp{\left(\jmath\frac{2\pi(\tau-\taup)(f-f_1)}{R_s} \right)}\Bigr|^2 \right.  -\left.  \Bigl| \iint_{\mathcal{B}_2}  df_1   df_2    \btF \exp{\left(\jmath\frac{2\pi\tau (f-f_1)}{R_s} \right)}\Bigr|^2 \right.\nonumber  \\  
    &+    \Bigl| \iint_{\mathcal{B}_2}  df_1   df_2    \btF  \left[ \exp{\left(\jmath\frac{2\pi(\taup-\tau)(f-f_1)}{R_s}\right)}+\exp{\left(\jmath\frac{2\pi\taup(f-f_1)}{R_s}\right)} \right]\Bigr|^2   \nonumber \\  
    &-\left.   \Bigl| \iint_{\mathcal{B}_2}  df_1   df_2    \btF  \exp{\left(\jmath\frac{2\pi(\tau-\taup) (f_1-f)}{R_s} \right)}  \Bigr|^2 \right.  -\left. \Bigl| \iint_{\mathcal{B}_2}  df_1   df_2    \btF  \exp{\left(\jmath\frac{2\pi\taup (f-f_1)}{R_s} \right)} \Bigr|^2 \right]  \label{Eq:Kn_xi2} \\
    & \psi_1(\tau,f) \!=\! \frac{32}{81} R_s^2   \left[   \int_{\mathcal{B}_1} df_1 |S(f_1)|^2 \Bigl| \int_{\mathcal{B}_1} df_2   S( f_2 )\right.   S^*(f_1+f_2-f)  \mu (f_1,f_2,f) \cos { \frac{2\pi\tau f_2}{R_s} } \Bigr|^2  \nonumber  \\  
    & +  \int_{\mathcal{B}_1} df_1 |S(f_1)|^2 \Bigl| \int_{\mathcal{B}_1} df_2   S( f_2 )   \left.  S^*(f_1+f_2-f) \mu (f_1,f_2,f) \sin { \frac{2\pi\tau f_2 }{R_s} } \Bigr|^2 \right] \label{Eq:Kn_psi1} \\
    & \psi_2(\tau,f) \!=\!  \frac{16}{81} R_s^2   \left[  \int_{\mathcal{B}_1} df_1 |S(f_1)|^2 \Bigl|  \int_{\mathcal{B}_1} df_2   S( f_2 ) \right. S^*(f_1-f_2+f)  \mu (f_1-f_2+f,f_2,f) \cos { \frac{2\pi\tau(f_1-f_2+f)}{R_s} } \Bigr|^2   \nonumber  \\
    & +   \int_{\mathcal{B}_1} df_1 |S(f_1)|^2 \Bigl|  \int_{\mathcal{B}_1} df_2   S( f_2 ) S^*(f_1-f_2+f) \left.   \mu (f_1-f_2+f,f_2,f) \sin { \frac{2\pi\tau(f_1-f_2+f)}{R_s} } \Bigr|^2 \right] \label{Eq:Kn_psi2} \\
    & \psi_3(\tau,\taup\!,f)  \!=\!   \frac{16}{81} \!R_s\!  \left[   \Bigl| \iint_{\mathcal{B}_2}  df_1   df_2    \btF \right.  \exp{\left(\jmath\frac{2\pi (-\tau f_1 \!-\! \taup\! f_2)}{R_s}\right)} \Bigr|^2 \!\! +\! \left.   \Bigl| \iint_{\mathcal{B}_2}  df_1   df_2   \btF  \exp{\left(\jmath\frac{2\pi(-\taup\! f_1 \!-\! \tau f_2)}{R_s} \right)}  \Bigr|^2 \right. \nonumber  \\  
   &+\left.   \Bigl| \iint_{\mathcal{B}_2}  df_1   df_2   \btF  \exp{\left(\jmath\frac{2\pi (\tau f_1 - (\taup-\tau) f_2)}{R_s}\right)} \Bigr|^2 \right. \!\!+\! \left.   \Bigl| \iint_{\mathcal{B}_2}  df_1   df_2   \btF  \exp{\left(\jmath\frac{2\pi((\tau-\taup) f_1+\tau f_2  )}{R_s} \right)} \Bigr|^2 \right.\nonumber  \\  
   &+\!\left.   \Bigl| \iint_{\mathcal{B}_2}  df_1   df_2    \btF  \exp{\left(\jmath\frac{2\pi((\taup\!-\tau) f_1\!+\!\taup\!f_2  )}{R_s} \right)} \Bigr|^2 \right.  \!\!\!\!+\!\!\!\left.   \Bigl| \iint_{\mathcal{B}_2}  df_1   df_2    \btF  \exp{\left(\jmath\frac{2\pi (\taup\! f_1 \!-\!  (\tau-\taup) f_2)}{R_s}\right)} \Bigr|^2 \right]\label{Eq:Kn_psi3} 
\end{align}
\end{subequations}
\setcounter{equation}{20}
\hrulefill
\vspace*{4pt}
\end{figure*}

%% file: Tex/Ch_CCDM_Deri.tex
\section{Case Study: Energy Correlation of Symbols with Constant Composition Amplitudes} \label{Sec:CCDM_Corr_Deri}

The symbol-energy correlations (or covariances) required in \eqref{Ch:Cov_SPT} and \eqref{Ch:Cov_XPT} are generic quantities that can be computed numerically from the transmitted data, either by brute-force evaluation or more efficiently using progressive binning with logarithmic density \cite{laurence2006fast,urquidi2024binning}, or by an FFT-based approach via the Wiener--Khinchin theorem \cite{box2015time}. Alternatively, analytical expressions might be derived for specific shaping algorithms with known structural properties.

CCDM is one of the most widely used distribution-matching algorithms for PAS. Due to the block-wise energy constraint imposed by CCDM, the symbol energies exhibit an inherent negative correlation~\cite{wu2021edi}, yielding negative covariance in \eqref{Eq:Cov_General}. 

In this section, we present analytical expressions for QAM symbols generated by \textit{ideal} CCDM, meaning that all possible amplitude blocks with the same constant composition are used as CCDM codewords with equal probability. First, we present analytical expressions for the correlations between amplitudes within the same CCDM shaping block in Sec.~\ref{Sec:CCDM_Intra_Amp}. These correlated amplitudes are then mapped onto the I/Q components of the $\px$ and $\py$ polarizations to form blocks of correlated QAM symbols, where the resulting correlation structure depends on the chosen mapping strategy. In this work, only the 1-D, 2-D, and 4-D mapping strategies are considered \cite{skvortcov2020huffman}. Next, we show analytical expressions for the correlations between QAM symbols within a shaped block in Sec.~\ref{Sec:CCDM_Intra_Sym}. Finally, inter-block effects are considered, and analytical expressions for the energy covariance functions in \eqref{Ch:Cov_SPT} and \eqref{Ch:Cov_XPT} are presented in Sec.~\ref{Sec:CCDM_Inter_Sym}. In particular, $\bKSc{\tau}$ in \eqref{Ch:Cov_SPT_1} was derived in our previous work \cite[Appendix A]{wu2021edi}, and for brevity the derivations of other energy correlations are omitted here.

\subsection{Intra-Block Amplitude Correlation} \label{Sec:CCDM_Intra_Amp}

For amplitude $\Ampu$ in the alphabet $\mathcal{U}$, consider the generation of a constant-composition codeword as drawing amplitudes in series from $N$ amplitudes without replacement. The number of each possible amplitude is a constant number denoted by $N_{\Ampu}$ in every sequence, and thus $N = \sum_{\Ampu\in\mathcal{U}} N_{\Ampu}$. Although the composition is predetermined, the ordering of amplitudes is uncertain. We can easily obtain its $k$-th order moments
\begin{align}
\Exp{\Ampu^k} & =\sum_{\Ampu\in\mathcal{U}} \Ampu^k \frac{N_{\Ampu}}{N}. \label{EA2}
\end{align}
The amplitudes within each codeword are inherently correlated due to the constant composition. For the relevant energy-level statistics when $k$ is an even integer, we consider the energy correlations given in the following theorems.

\begin{theorem}\label{Thm:CCAc}
Consider a constant-composition amplitude block $(\Ampu_{1}, \Ampu_{2}, \ldots, \Ampu_{N})$ with amplitudes drawn from the alphabet $\mathcal{U}$. For any distinct amplitude-position indices $\w, \wtt \in \{1,2,\ldots,N\}$ with $\w \neq \wtt$, the correlation of the second-order moments is given by
\begin{align}
    \rho_{1}
\triangleq \Exp{\Ampu_{\w}^2\Ampu_{\wtt}^2} & = \frac{N\Exp{\Ampu^2 }^2-\Exp{\Ampu^4}}{N-1} ,  \\
& <  \Exp{\Ampu^2}^2.
\end{align}
Furthermore, the correlation between the second- and fourth-order moments is given by
\begin{align}
    \rho_{2}  \triangleq  \Exp{\Ampu_{\w}^2\Ampu_{\wtt}^4} & = \frac{N\Exp{\Ampu^2}\Exp{\Ampu^4}-\Exp{\Ampu^6}}{N-1} \label{eq:Amp24Ac} , \\
    & < \Exp{\Ampu^2} \Exp{\Ampu^4} .
\end{align}

\end{theorem}

\begin{theorem}\label{Thm:AmpAc_222}
Consider a constant-composition amplitude block $(\Ampu_{1}, \Ampu_{2}, \ldots, \Ampu_{N})$ with amplitudes drawn from the alphabet $\mathcal{U}$. For any distinct amplitude-position indices $\w, \wtt,\wttt \in \{1,2,\ldots,N\}$ with $\w \neq \wtt,\w \neq \wttt,\wtt \neq \wttt$, the correlation among a triplet of the second-order moments is

\begin{align}
    \rho_{3}  \triangleq \Exp{\Ampu_{\w}^2 \Ampu_{\wtt}^2 \Ampu_{\wttt}^2} = & \; \frac{N^2}{(N-1)(N-2)}\Ea{2}^3 \nonumber \\
     & -\frac{ 3N }{(N-1) (N-2)} \Ea{2} \Ea{4}\nonumber  \\
     & +\frac{2 }{(N-1) (N-2)}\Ea{6}, \label{eq:Amp222Ac} \\
     < & \; \Ea{2}^3  .
\end{align}

\end{theorem}

It is readily verified from Theorems~\ref{Thm:CCAc} and~\ref{Thm:AmpAc_222} that these intra-block amplitude-energy correlations vanish as $N \to \infty$. 

\subsection{Intra-Block Symbol-Energy Correlation }\label{Sec:CCDM_Intra_Sym}

\begin{figure}
\centering
\setkeys{Gin}{width=0.01\textwidth}
\subfloat[1-D mapping, $M_s=8$.]{\resizebox{1\linewidth}{!}{\subimport{Figures}{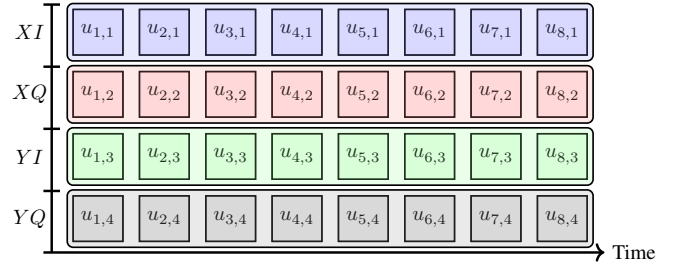}}}\\
\subfloat[2-D mapping, $M_s=4$. ]{\resizebox{1\linewidth}{!}{\subimport{Figures}{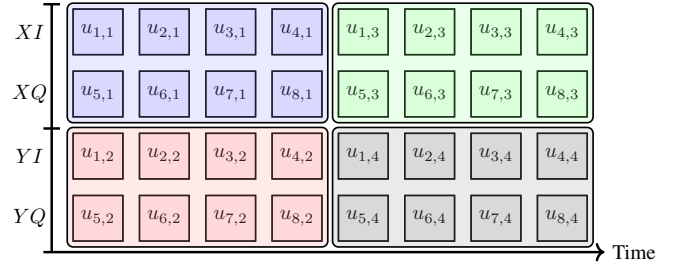}}}\\
\subfloat[4-D mapping, $M_s=2$. ]{\resizebox{1\linewidth}{!}{\subimport{Figures}{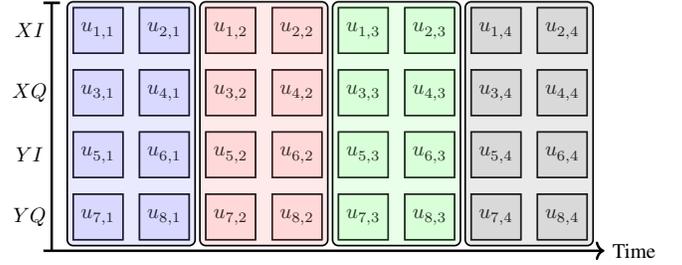}}}
\caption{An illustration of the 1-D, 2-D, and 4-D mapping strategies from four amplitude blocks with length $N=8$ to the optical field dimensions. Different amplitude blocks are distinguished by color and by the second subscript of $u$. }
\label{Fig:124Dmapping}
\end{figure}

Fig.~\ref{Fig:124Dmapping} illustrates the different mapping strategies for four amplitude blocks. In 1-D mapping, each amplitude block is assigned to a single optical-field dimension, resulting in four independent dimensions. In contrast, 2-D and 4-D mapping distribute each amplitude block across multiple dimensions, thereby introducing inter-dimensional correlations in the resulting QAM symbols. In particular, under 1-D and 2-D mapping, the two polarizations remain independent, so only SPT energy interactions arise. By contrast, 4-D mapping introduces both XP and XPT interactions.

Without loss of generality, in the following analysis, the shaping blocklength $N$ is assumed to be an integer multiple of four. Hence, for $H$-D mapping, the time-domain QAM symbol vectors $(a^{\px}_{\w}, a^{\py}_{\w})$ are correlated over each block of length
\begin{align}
    M_s = N/H,
\end{align} which therefore defines the correlation length. Using Theorems~\ref{Thm:CCAc} and~\ref{Thm:AmpAc_222}, the temporal symbol-energy correlations within a block are expressed as follows.

\begin{theorem}\label{Thm:CCSym2D_Intra}
Consider a block of QAM symbols $(a^{\px}_{1}, a^{\py}_{1}, a^{\px}_{2}, a^{\py}_{2}, \ldots, a^{\px}_{M_s}, a^{\py}_{M_s})$, generated by mapping a constant-composition amplitude block $(\Ampu_{1}, \Ampu_{2}, \ldots, \Ampu_{N})$, with amplitudes drawn from the alphabet $\mathcal{U}$, using $H$-D mapping. Then, for any distinct time indices $\w, \wtt \in \{1,2,\ldots,M_s\}$ and any $\ppol, \ppol' \in \{\px,\py\}$, the second-order moment correlation in the SPT case with $\ppol=\ppol'$ is given by
\begin{align}
\setlength{\nulldelimiterspace}{0pt}
\rho_{S1}   & \triangleq \Exp{|a^{\ppol}_{\w}|^2 |a^{\ppol }_{\wtt}|^2}\nonumber \\
&  = \left\{\begin{IEEEeqnarraybox}[\relax][c]{l's}
2\rho_{1}+ 2\Exp{\Ampu^2}^2, & \text{if} $H=1$,\\
4\rho_{1}, & \text{if} $H=2$ \text{or} $4$.
 \end{IEEEeqnarraybox}\right.
\end{align}
The XPT case with $\ppol\neq\ppol'$ is given by
\begin{align}
\setlength{\nulldelimiterspace}{0pt}
\rho_{X1}  & \triangleq \Exp{|a^{\ppol}_{\w}|^2 |a^{\ppol' }_{\wtt}|^2}\nonumber \\
&  =  \left\{\begin{IEEEeqnarraybox}[\relax][c]{l's}
4\Exp{\Ampu^2}^2, & \text{if} $H=1$  \text{or} $2$, \\
4\rho_{1}, & \text{if} $H=4$ .
 \end{IEEEeqnarraybox}\right.
\end{align}

Furthermore, for the correlation between the second- and fourth-order moments, the SPT case with $\ppol=\ppol'$ is given by
\begin{align}
\setlength{\nulldelimiterspace}{0pt}
\rho_{S2} &  \triangleq \Exp{|a^{\ppol}_{\w}|^2 |a^{\ppol}_{\wtt}|^4} \nonumber \\
& = \left\{\begin{IEEEeqnarraybox}[\relax][c]{l's}
2\rho_{2}+ 2\Exp{\Ampu^2}\Exp{\Ampu^4} + 4\rho_{1}\Exp{\Ampu^2} , & \text{if} $H=1$,\\
4\rho_{2} + 4\rho_{3}, & \text{if} $H=2$ \text{or} $4$.
 \end{IEEEeqnarraybox}\right.
\end{align}
the XPT case with $\ppol\neq\ppol'$ is given by
\begin{align}
\setlength{\nulldelimiterspace}{0pt}
\rho_{X2}  &  \triangleq \Exp{|a^{\ppol}_{\w}|^2 |a^{\ppol'}_{\wtt}|^4} \nonumber \\
& =  \left\{\begin{IEEEeqnarraybox}[\relax][c]{l's}
4\Exp{\Ampu^2}\Exp{\Ampu^4} + 4 \Exp{\Ampu^2}^3 , & \text{if} $H=1$,\\
4\Exp{\Ampu^2}\Exp{\Ampu^4} + 4 \rho_{1}\Exp{\Ampu^2}  , & \text{if} $H=2$, \\
4\rho_{2} + 4 \rho_{3} , & \text{if} $H=4$ .
 \end{IEEEeqnarraybox}\right.
\end{align}

\end{theorem}

\begin{theorem}\label{Thm:CCSym3D_Intra}
Consider a block of QAM symbols $(a^{\px}_{1}, a^{\py}_{1}, a^{\px}_{2}, a^{\py}_{2}, \ldots, a^{\px}_{M_s}, a^{\py}_{M_s})$, generated by mapping a constant-composition amplitude block $(\Ampu_{1}, \Ampu_{2}, \ldots, \Ampu_{N})$, with amplitudes drawn from the alphabet $\mathcal{U}$, using $H$-D mapping. Then, for any distinct time indices $\w, \wtt, \wttt \in \{1,2,\ldots,M_s\}$ satisfying $\w \neq \wtt$, $\w \neq \wttt$, and $\wtt \neq \wttt$, and any $\ppol, \ppol', \ppol'' \in \{\px,\py\}$, the correlation among a triplet of second-order moments is characterized as follows. In the SPT case with $\ppol=\ppol'=\ppol''$, it is given by
\begin{align}
\setlength{\nulldelimiterspace}{0pt}
 \rho_{S3}  & \triangleq \Exp{|a^{\ppol}_{\w}|^2 |a^{\ppol  }_{\wtt}|^2 |a^{\ppol }_{\wttt}|^2 }\nonumber \\
&  =  \left\{\begin{IEEEeqnarraybox}[\relax][c]{l's}
2\rho_{3}+ 6\rho_{1}\Exp{\Ampu^2}, & \text{if} $H=1$,\\
8\rho_{3}, & \text{if} $H=2$ \text{or} $4$.
 \end{IEEEeqnarraybox}\right.
\end{align}
In the XPT case with $\ppol=\ppol'',\ppol'\neq\ppol''$, it is given by
\begin{align}
\setlength{\nulldelimiterspace}{0pt}
\rho_{X3} & \triangleq \Exp{|a^{\ppol}_{\w}|^2 |a^{\ppol'  }_{\wtt}|^2 |a^{\ppol'' }_{\wttt}|^2 }\nonumber \\
&  =   \left\{\begin{IEEEeqnarraybox}[\relax][c]{l's}
4\rho_{1}\Exp{\Ampu^2} + 4 \Exp{\Ampu^2}^3 , & \text{if} $H=1$,\\
8\rho_{1}\Exp{\Ampu^2}, & \text{if} $H=2$,\\
8 \rho_{3} , & \text{if} $H=4$.
 \end{IEEEeqnarraybox}\right.
\end{align}

\end{theorem}

\subsection{Inter-Block Symbol-Energy Correlation}\label{Sec:CCDM_Inter_Sym}

The MEGN model requires the time-averaged covariances defined in \eqref{Eq:Cov_General}, which account for inter-block effects, namely that two QAM symbols may be independent when positioned in different shaped blocks. Based on the intra-block correlations in Theorems~\ref{Thm:CCSym2D_Intra} and~\ref{Thm:CCSym3D_Intra}, the resulting time-averaged symbol-energy correlations are given as follows.

\begin{theorem} \label{Thm:CCSym2D}
Consider a sequence of QAM symbols $(a^{\px}_{1}, a^{\py}_{1}, a^{\px}_{2}, a^{\py}_{2}, \ldots)$, obtained by mapping constant-composition amplitude blocks of blocklength $N$, with amplitudes drawn from the alphabet $\mathcal{U}$, using $H$-D mapping. When $N$ is an integer multiple of four, the correlation length is $M_s=N/H$. Then, for any time delay $\tau>0$ and any $\ppol, \ppol' \in \{\px,\py\}$, the time-averaged second-order moment correlation in the SPT case with $\ppol=\ppol'$ is given by
\begin{align} \label{Eq:Rs1}
\setlength{\nulldelimiterspace}{0pt}
\bRSc{\tau} & \triangleq  \frac{1}{M_s} \sum_{\w=1}^{M_s} \Exp{|a^{\ppol}_{\w}|^2 |a^{\ppol}_{\w +\tau}|^2} \nonumber \\
& =  \left\{\begin{IEEEeqnarraybox}[\relax][c]{l's}
\frac{ \tau  \Exp{|a|^2}^2+(M_s- \tau )\rho_{S1} }{M_s}, & \text{if} $ \tau < M_s$, \\
\Exp{|a|^2}^2, &  \text{if} $ \tau \geq M_s$.
 \end{IEEEeqnarraybox}\right.
\end{align}
Furthermore, for the correlation between the second- and fourth-order moments, it is given by
\begin{align} \label{Eq:Rs2}
\setlength{\nulldelimiterspace}{0pt}
\bRSa{\tau} & \triangleq \frac{1}{M_s} \sum_{\w=1}^{M_s} \Exp{|a^{\ppol}_{\w}|^2 |a^{\ppol}_{\w +\tau}|^4} \nonumber \\
& =   \left\{\begin{IEEEeqnarraybox}[\relax][c]{l's}
\frac{ \tau  \Exp{|a|^2}\Exp{|a|^4}+(M_s- \tau )\rho_{S2} }{M_s}, & \text{if} $\tau < M_s$, \\
\Exp{|a|^2}\Exp{|a|^4}, &  \text{if} $ \tau \geq M_s$.
 \end{IEEEeqnarraybox}\right.
\end{align}
In the XPT case with $\ppol\neq\ppol'$, $\bRXc{\tau}$ and $\bRXa{\tau}$ can be obtained in the same manner as in \eqref{Eq:Rs1} and \eqref{Eq:Rs2} by replacing $\rho_{S1} $, $\rho_{S2} $ with $\rho_{X1}$, $\rho_{X2}$.
\end{theorem}

 \begin{theorem}\label{Thm:CCSym3D}
Consider a sequence of QAM symbols $(a^{\px}_{1}, a^{\py}_{1}, a^{\px}_{2}, a^{\py}_{2}, \ldots)$, obtained by mapping constant-composition amplitude blocks of blocklength $N$, with amplitudes drawn from the alphabet $\mathcal{U}$, using $H$-D mapping. When $N$ is an integer multiple of four, the correlation length is $M_s=N/H$. Then, for time delays $\taup>\tau>0$ and any $\ppol, \ppol', \ppol'' \in \{\px,\py\}$, the correlations among a triplet of second-order moments are characterized as follows. In the SPT case with $\ppol=\ppol'=\ppol''$, it is given in \eqref{Eq:RS3}.

\begin{figure*}[!t]
\normalsize
\setcounter{MYtempeqncnt}{\value{equation}}
\setcounter{equation}{52}
\begin{align} \label{Eq:RS3}
\setlength{\nulldelimiterspace}{0pt}
\bRSb{\tau,\taup}  & \triangleq \frac{1}{M_s} \sum_{\w=1}^{M_s} \Exp{|a^{\ppol}_{\w}|^2 |a^{\ppol}_{\w +\tau}|^2 |a^{\ppol}_{\w +\taup}|^2} \nonumber \\
& =   \left\{\begin{IEEEeqnarraybox}[\relax][c]{l's}
\frac{ \taup \Exp{|a|^2}\rho_{S1}+(M_s- \taup ) \rho_{S3} }{M_s}  , &  \text{if\:}  $\taup  < M_s$,  \\
\frac{ \tau \Exp{|a|^2}^3+(M_s- \tau ) \Exp{|a|^2}\rho_{S1} }{M_s}  , & \text{if\:}  $\taup  \geq M_s$ ,  $\tau  < M_s$  \text{\:and\:}    $\taup-\tau  \geq M_s$, \\
\frac{ (\taup-M_s) \Exp{|a|^2}^3+(2M_s- \taup ) \Exp{|a|^2}\rho_{S1} }{M_s}  , &  \text{if\:} $ \taup  \geq M_s $,   $\tau  < M_s$   \text{\:and\:}   $\taup-\tau  < M_s$,   \\
\frac{ (\taup-\tau) \Exp{|a|^2}^3+(M_s- \taup+\tau ) \Exp{|a|^2}\rho_{S1} }{M_s} , &  \text{if\:}    $\tau  \geq M_s$  \text{\:and\:}    $\taup-\tau  < M_s$,    \\
\Exp{|a|^2}^3, &  \text{otherwise.}
 \end{IEEEeqnarraybox}\right.
\end{align}

\setcounter{equation}{53}
\hrulefill
\vspace*{4pt}
\end{figure*}

In the XPT case with $\ppol=\ppol''$ and $\ppol'\neq\ppol''$, the result for 4-D mapping is given as in \eqref{Eq:RS3} by replacing $\rho_{S1} $, $\rho_{S3} $ with $\rho_{X1}$, $\rho_{X3}$. For 1-D and 2-D mapping, since polarizations are independent of each other, we have
\begin{align}
\setlength{\nulldelimiterspace}{0pt}
\bRXb{\tau,\taup}  & \triangleq \frac{1}{M_s} \sum_{\w=1}^{M_s} \Exp{|a^{\ppol}_{\w}|^2 |a^{\ppol'}_{\w +\tau}|^2 |a^{\ppol''}_{\w +\taup}|^2} \nonumber \\
& =  \left\{\begin{IEEEeqnarraybox}[\relax][c]{l's}
\frac{ \tau  \Exp{|a|^2}^3+(M_s- \tau  ) \rho_{1}\Exp{|a|^2} }{M_s} , & \text{if} $ \tau  < M_s$  \\ \Exp{|a|^2}^3 , & \text{if} $ \tau \geq M_s$.
 \end{IEEEeqnarraybox}\right.
\end{align}

\end{theorem}

Finally, the symbol-energy covariances in \eqref{Ch:Cov_SPT} and \eqref{Ch:Cov_XPT} are obtained by subtracting the corresponding products of moments from the correlations in Theorems~\ref{Thm:CCSym2D} and~\ref{Thm:CCSym3D}.



%% file: Tex/Ch_Results.tex
\section{Numerical Validation}\label{Sec:Results}

\input{Tex/Tab_System_Setup}

\subsection{Simulation Setup}
The proposed MEGN model is numerically validated using a multi-span, dual-polarization optical communication system employing standard single-mode fiber (SMF), under various transmission distances $N_s$ and symbol rates $R_s$. The key simulation parameters are summarized in Table~\ref{Tab:FiberParam}. The fiber link is simulated as follows. First, the transmitted symbols are pulse-shaped using a root-raised cosine (RRC) filter. Signal propagation over $N_{s}$ spans is modeled using the split-step Fourier method, where each span is followed by an Erbium-doped fiber amplifier (EDFA) to compensate for fiber attenuation. At the receiver, electronic dispersion compensation (EDC) is applied, after which the channel of interest is extracted using a matched filter. Finally, the residual phase rotation is compensated.

\begin{figure*}[t!]
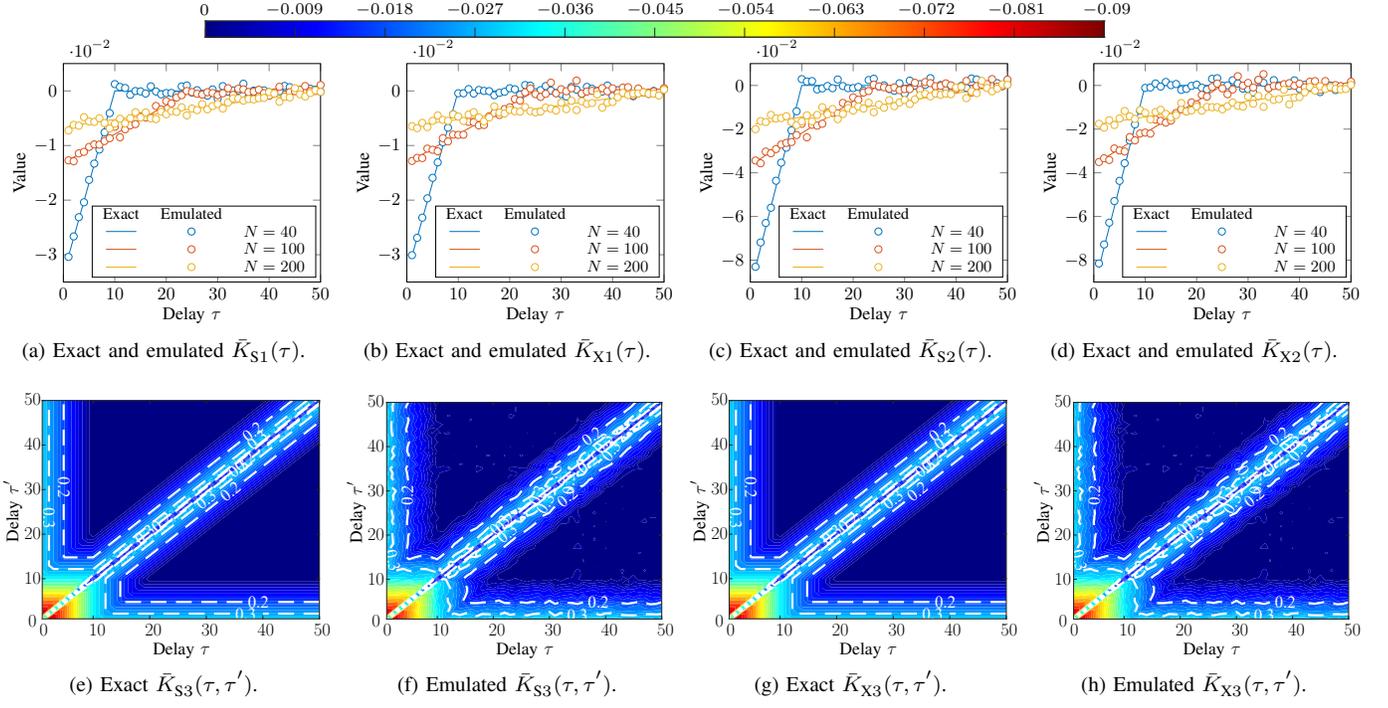

\centering

\resizebox{0.7\linewidth}{!}{\subimport{Figures}{ColorBar3}} \vspace{-1em}
\\
\subfloat[Exact and emulated $\bKSc{\tau}$. ]{
\resizebox{0.245\linewidth}{!}{\subimport{Figures}{SymCorr_KS1}}}
\subfloat[Exact and emulated $\bKXc{\tau}$.]{
\resizebox{0.245\linewidth}{!}{\subimport{Figures}{SymCorr_KX1}}}
\subfloat[Exact and emulated $\bKSa{\tau}$. ]{
\resizebox{0.245\linewidth}{!}{\subimport{Figures}{SymCorr_KS2}}}
\subfloat[Exact and emulated $\bKXa{\tau}$.]{
\resizebox{0.245\linewidth}{!}{\subimport{Figures}{SymCorr_KX2}}}
\\
\subfloat[Exact $\bKSb{\tau,\taup}$. ]{
\resizebox{0.245\linewidth}{!}{\subimport{Figures}{SymCorr_KS3_Ana}}}
\subfloat[Emulated $\bKSb{\tau,\taup}$.]{
\resizebox{0.245\linewidth}{!}{\subimport{Figures}{SymCorr_KS3_Num}}}
\subfloat[Exact $\bKXb{\tau,\taup}$. ]{
\resizebox{0.245\linewidth}{!}{\subimport{Figures}{SymCorr_KX3_Ana}}}
\subfloat[Emulated $\bKXb{\tau,\taup}$. ]{
\resizebox{0.245\linewidth}{!}{\subimport{Figures}{SymCorr_KX3_Num}}}
\caption{Illustration of the symbol-energy covariances in \eqref{Ch:Cov_SPT} and \eqref{Ch:Cov_XPT} for PAS-64QAM with constant-composition amplitudes, 4-D mapping and unitary power. The exact analytical values are obtained from Theorems~\ref{Thm:CCSym2D} and~\ref{Thm:CCSym3D}, and the emulated values are obtained by numerical estimation. (a)--(d): the covariances involving two symbol energies for blocklength $N=40$, $100$, and $200$; (e)--(h): the covariances involving three symbol energies for $N=40$, presented as contour plots.}
\label{Fig:Results_SymCorrl}
\end{figure*}

For the signaling format, we consider PAS-64QAM with amplitudes generated from constant-composition amplitude sequences with 1-D, 2-D, or 4-D mapping. The amplitude probability mass function is $[0.4, 0.3, 0.2, 0.1]$ for the amplitudes in ascending order. As shown in Sec.~\ref{Sec:CCDM_Corr_Deri}, different mapping strategies and blocklengths $N$ lead to distinct energy correlation patterns in the transmitted symbols and, consequently, to different NLI power levels.

The proposed MEGN model is validated in terms of the NLI power coefficient $\eta$, and the effective SNR. Given the NLI PSD $G_{\text{NLI}}(f) $, the NLI power coefficient $\eta$ is defined as
\begin{align}
    \eta = \frac{P_{\text{NLI}}}{P_{\text{ch}}^3} =  P_{\text{ch}}^{-3} \int_{-R_s/2}^{R_s/2} G(f) df.
\end{align}
The $\eta$ obtained from the EGN or MEGN model is denoted as $\eta_{\text{EGN}}$ or $\eta_{\text{MEGN}}$, respectively, and the value obtained from simulation is denoted as $\eta_{\text{Sim}}$. The accuracy is quantified by its deviation from the simulated reference
\begin{align}
   \Delta \eta = \frac{|\eta_{\text{Sim}}-\eta_{\text{Mod}}|}{\eta_{\text{Sim}}}
\end{align}

The effective SNR is then defined as
\begin{align}
    \text{SNR}_\text{eff} = \frac{P_{\text{ch}}}{P_{\text{ASE}}+P_{\text{NLI}}} =\frac{P_{\text{ch}}}{P_{\text{ASE}}+\eta P_{\text{ch}}^3}
\end{align}
The \textit{optimal} effective SNR is achieved when $P_{\text{NLI}}$ reaches half of the $P_{\text{ASE}}$ as the power $P_{\text{ch}}$ increases \cite[Ch.~9.10.2]{mukherjee2020springer}.

\subsection{Symbol-Energy Correlations and Channel Functions }

We first analyze the required signal statistics and the channel functions required in the MEGN model. Fig.~\ref{Fig:Results_SymCorrl} visualizes the temporal symbol-energy correlations in \eqref{Ch:Cov_SPT} and \eqref{Ch:Cov_XPT} for the considered PAS-64QAM signaling. The results are obtained for 4-D mapping such that the transmitted symbols exhibit all types of interactions. The results are obtained either exactly from the analytical expressions in Theorems~\ref{Thm:CCSym2D} and~\ref{Thm:CCSym3D}, or by numerical emulation. All covariances are non-positive. The excellent agreement between the two confirms the validity of Theorems~\ref{Thm:CCSym2D} and~\ref{Thm:CCSym3D}. In Fig.~\ref{Fig:Results_SymCorrl}(a)--(d), for 4-D mapping, the covariances vanish when the delay $\tau$ exceeds the correlation length $M_s=N/4$. In addition, $\bKSc{\tau}$ and $\bKXc{\tau}$ have the same magnitude, and so do $\bKSa{\tau}$ and $\bKXa{\tau}$, because the corresponding SPT and XPT interactions arise from the same amplitude block. Fig.~\ref{Fig:Results_SymCorrl}(e)--(h) presents the two-dimensional covariances for $N=40$. The isolines mark the regions where the covariance magnitude decreases to $30\%$ and $20\%$ of its peak value. It shows that when the three symbols are separated by more than $M_s=10$, they become mutually independent and the covariance becomes zero, corresponding to the two blue triangular regions in each subfigure.

\begin{figure*}[t!]
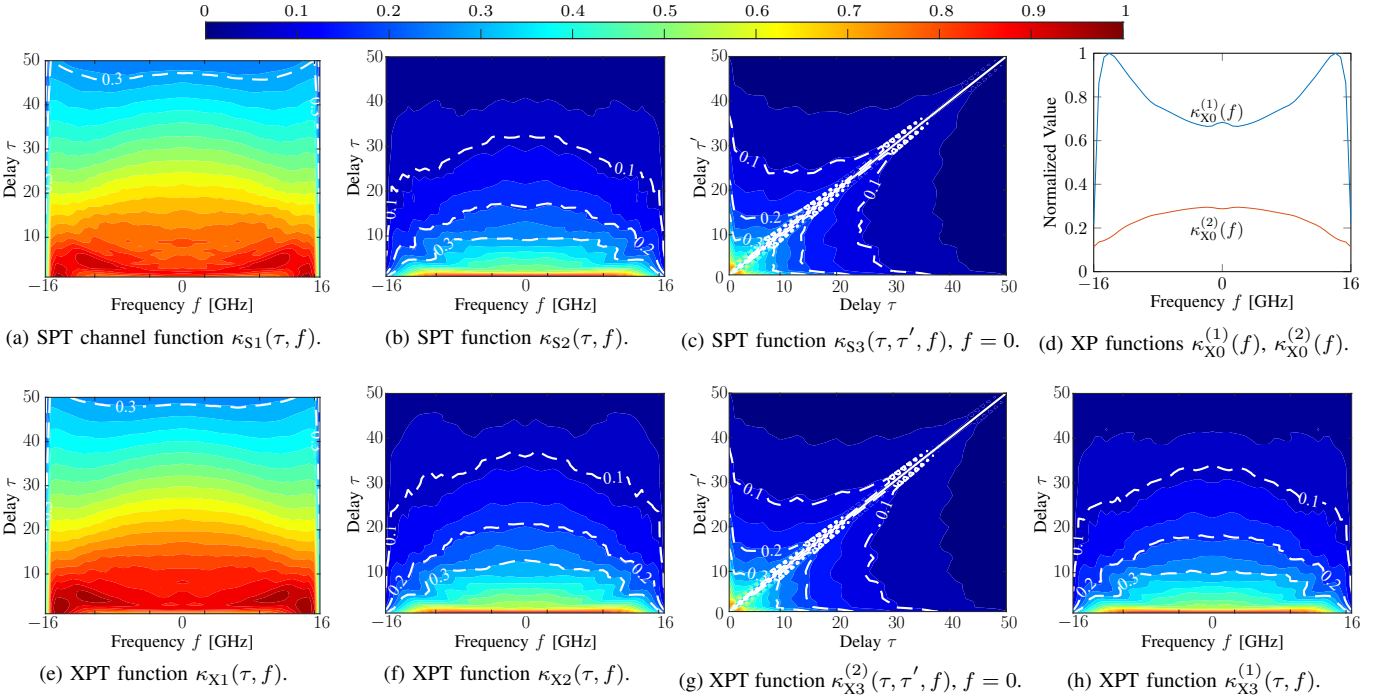

\centering

\resizebox{0.7\linewidth}{!}{\subimport{Figures}{ColorBar1}} \vspace{-1em}
\\
\subfloat[SPT channel function $ \KnSc{ \tau,f }$. ]{
\resizebox{0.245\linewidth}{!}{\subimport{Figures}{FEGN_Kernel_S1}}}
\subfloat[SPT function $ \KnSa{ \tau,f }$.]{
\resizebox{0.245\linewidth}{!}{\subimport{Figures}{FEGN_Kernel_S2}}}
\subfloat[SPT function $ \KnSb{ \tau,\taup,f }$, $f=0$. ]{
\resizebox{0.245\linewidth}{!}{\subimport{Figures}{FEGN_Kernel_S3}}}
\subfloat[XP functions $  \kappa^{(1)}_{\text{X}0}(f) $, $  \kappa^{(2)}_{\text{X}0}(f) $.]{
\resizebox{0.245\linewidth}{!}{\subimport{Figures}{FEGN_Kernel_X0}}}
\\
\subfloat[XPT function $ \KnXc{ \tau,f }$. ]{
\resizebox{0.245\linewidth}{!}{\subimport{Figures}{FEGN_Kernel_X1}}}
\subfloat[XPT function $ \KnXa{ \tau,f }$.]{
\resizebox{0.245\linewidth}{!}{\subimport{Figures}{FEGN_Kernel_X2}}}
\subfloat[XPT function $ \KnXbii{ \tau,\taup,f }$, $f=0$. ]{
\resizebox{0.245\linewidth}{!}{\subimport{Figures}{FEGN_Kernel_X3}}}
\subfloat[XPT function $ \KnXbi{ \tau,f }$. ]{
\resizebox{0.245\linewidth}{!}{\subimport{Figures}{FEGN_Kernel_X3_1tau}}}
\caption{Illustration of the MEGN channel functions for SPT in \eqref{Ch:Kn_SPT_1}--\eqref{Ch:Kn_SPT_3}, XP in \eqref{Ch:Kn_XP_1}--\eqref{Ch:Kn_XP_2}, and XPT in \eqref{Ch:Kn_XPT_1}--\eqref{Ch:Kn_XPT_32}, for transmission at $R_s=32$ GBd and $N_s=10$ spans. In each subfigure, the function is normalized w.r.t. its maximum value. }
\label{Fig:Results_Kernels}
\end{figure*}

Fig.~\ref{Fig:Results_Kernels} visualizes the proposed SPT channel functions in \eqref{Ch:Kn_SPT_1}--\eqref{Ch:Kn_SPT_3}, and the XP and XPT channel functions in \eqref{Ch:Kn_XP_1}--\eqref{Ch:Kn_XPT_32}, for $N_s=10$-span transmission at $R_s=32$ GBd.
Each subfigure illustrates how the normalized strength of each NLI interaction type is distributed across frequency and decays as delay increases. Comparing Fig.~\ref{Fig:Results_Kernels}(a)--(c) with Fig.~\ref{Fig:Results_Kernels}(e)--(g), we observe that they follow similar qualitative patterns regardless of whether the interaction occurs within or across polarizations. In particular, Fig.~\ref{Fig:Results_Kernels}(c) and (g), for visualization purposes, set $f=0$ to show functions of two delay variables. These channel functions show symmetry with respect to $\tau=\taup$, while the remaining channel functions show symmetry with respect to frequency $f=0$. In Fig.~\ref{Fig:Results_Kernels}(a) and (e), the channel functions retain up to $30\%$ of their strength even after the delay reaches $\tau=50$, while in the others, except (d), the strength decays much more rapidly, dropping below $30\%$ at roughly~$\tau=10$. In particular, Fig.~\ref{Fig:Results_Kernels}(d) shows the XP channel function, where the interaction between symbol energies (second-order moments), $\kappa^{(1)}_{\text{X}0}(f)$, exhibits a larger impact than the interaction between the second-order and fourth-order moments, $\kappa^{(2)}_{\text{X}0}(f)$.

\def\WD{0.7*\linewidth}
\def\HT{0.55*\linewidth}
\begin{figure}[t!]
\centering
\setkeys{Gin}{width=0.01\textwidth}
\resizebox{1\linewidth}{!}{\subimport{Figures}{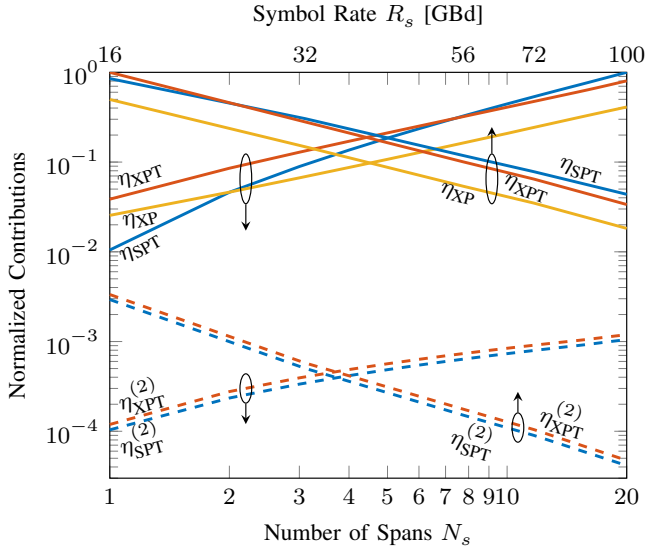}}
\caption{The contributions due to SPT, XP, and XPT-type energy interactions to NLI for blocklength $N=100$ with 4-D mapping vs. (bottom log-scale axis) distance at $R_s=32$ GBd or (top log-scale axis) symbol rate at $N_s=10$ spans. }
\label{Fig:Contribution}
\end{figure}

\def\WD{0.7*\linewidth}
\def\HT{0.55*\linewidth}
\begin{figure}[t!]
\centering
\setkeys{Gin}{width=0.01\textwidth}
\resizebox{1\linewidth}{!}{\subimport{Figures}{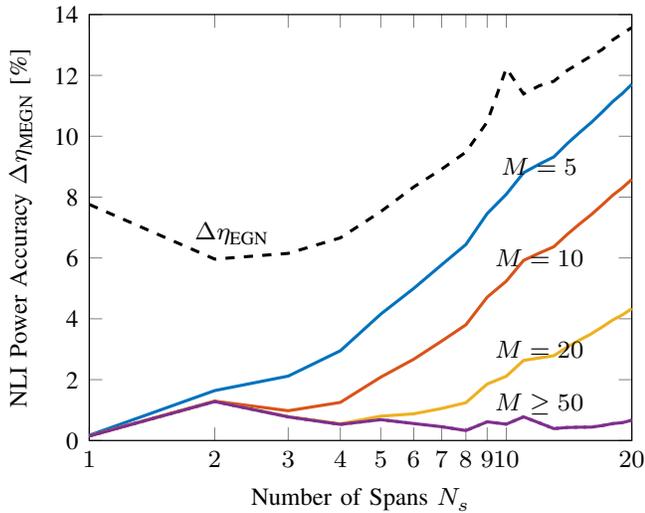}}
\caption{The MEGN model accuracy as the memory $M$ increases for $N=400$ with 4-D mapping, at $R_s=32$ GBd. }
\label{Fig:memoryAccuracy}
\end{figure}

\begin{figure*}
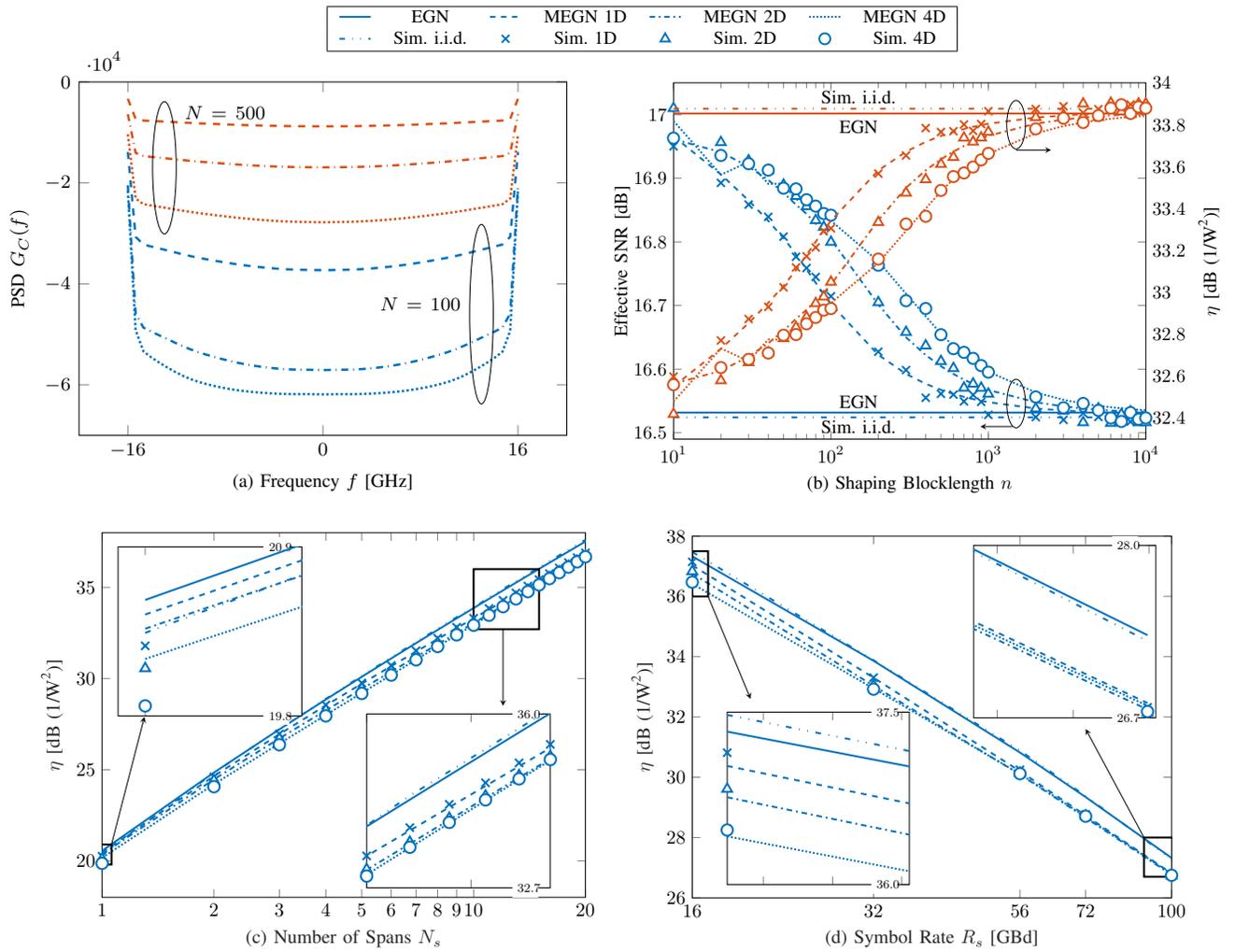

\centering
\setkeys{Gin}{width=0.01\textwidth}

\definecolor{mycolor1}{rgb}{0.00000,0.44700,0.74100}%
\hspace{2em}
\vspace{-1em}
\resizebox{0.5\linewidth}{!}{
\begin{tabular}{ |cccccccc|}
    \hline
    \ref{plot:EGN} & EGN    &\ref{plot:1D_FEGN} &MEGN 1-D   & \ref{plot:2D_FEGN} &MEGN 2-D&  \ref{plot:4D_FEGN}&MEGN 4-D \\
     \ref{plot:iid_Sim}& Sim. i.i.d.& \ref{plot:1D_Sim} & Sim. 1-D  & \ref{plot:2D_Sim}& Sim. 2-D &
       \ref{plot:4D_Sim}& Sim. 4-D\\
 \hline
\end{tabular}}
\def\WD{0.4*\linewidth}
\def\HT{0.29*\linewidth}
\subfloat{
\resizebox{0.45\linewidth}{!}{\subimport{Figures}{NLI_PSD}}
}
\def\WD{0.49*\linewidth}
\def\HT{0.37*\linewidth}
\subfloat{
\resizebox{0.49\linewidth}{!}{\subimport{Figures}{snr_vs_blk}}
} \\
\subfloat{
\resizebox{0.45\linewidth}{!}{\subimport{Figures}{FEGN_Eta_vs_span}}
}
\subfloat{
\resizebox{0.45\linewidth}{!}{\subimport{Figures}{FEGN_Eta_vs_symbolRate}}
}
\caption{(a): NLI PSD reduction $G_{C}(f)$ in \eqref{Eq:FEGN_Model} with shaping blocklength $n=100,500$ and 1-D/2-D/4-D mappings for transmission at $R_s=32$ GBd and $N_s=10$ spans. (b): The optimal effective SNR (left axis) and the NLI power coefficient $\eta$ (right axis) vs. the CCDM shaping blocklength $n$ after transmission of $10$ spans, for transmission at $R_s=32$ GBd and $N_s=10$ spans. The NLI power efficiency obtained from the simulation and EGN/MEGN models vs. (c) the number of spans $N_s$ at $R_s=32$ GBd and (d) the symbol rate $R_s$ at $N_s=10$ spans. }
\label{Fig:Results_combined}
\end{figure*}

\begin{figure*}
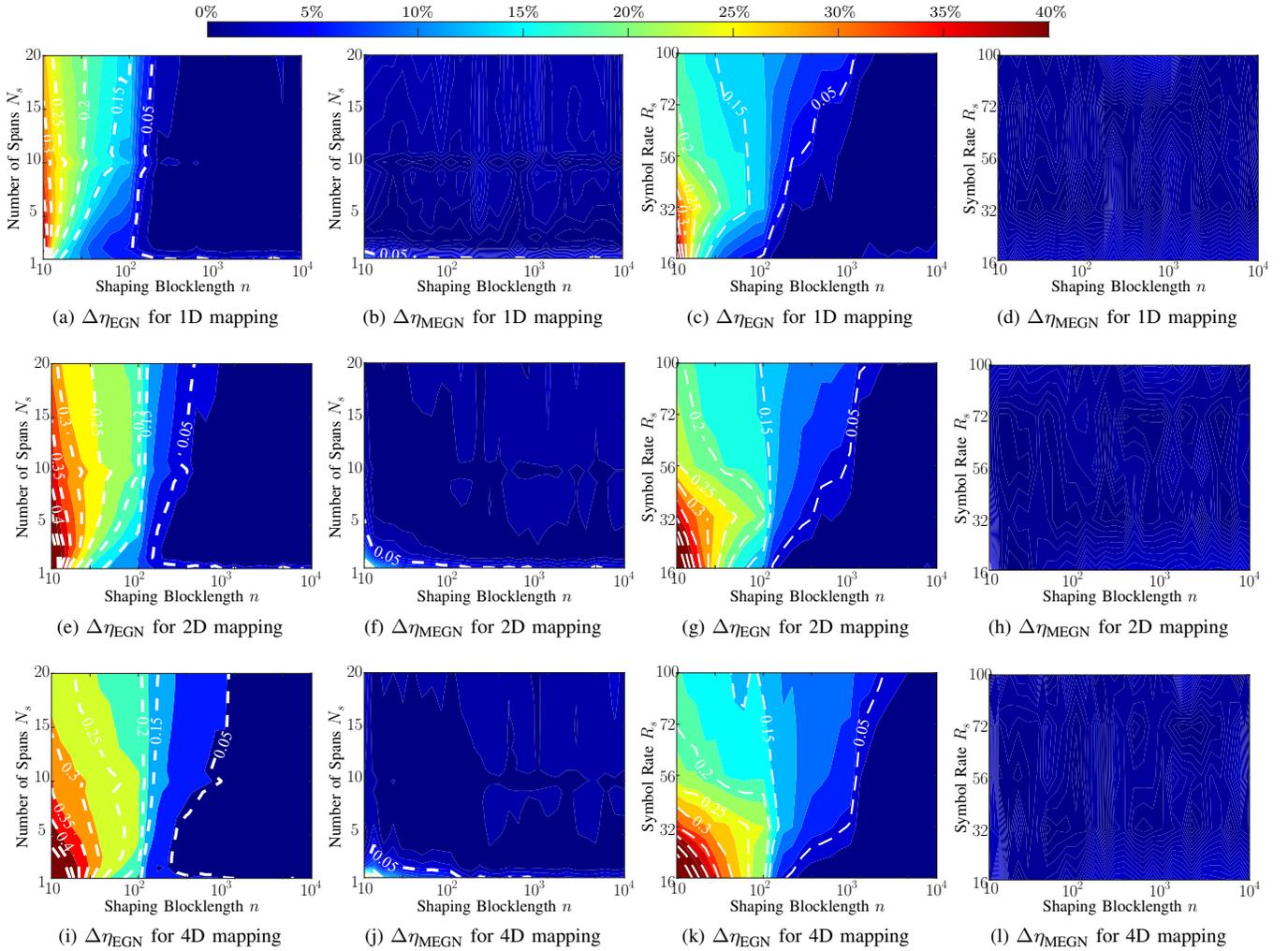

\centering
\setkeys{Gin}{width=0.01\textwidth}   \resizebox{0.7\linewidth}{!}{\subimport{Figures}{ColorBar2}} \vspace{-0.8em}
\\
\subfloat[$\Delta\eta_{\text{EGN}}$ for 1-D mapping  ]{\resizebox{0.245\linewidth}{!}{\subimport{Figures}{EGN_Err_1D_span}} }
\subfloat[$\Delta\eta_{\text{MEGN}}$ for 1-D mapping   ]{\resizebox{0.245\linewidth}{!}{\subimport{Figures}{FEGN_Err_1D_span}}}
\subfloat[$\Delta\eta_{\text{EGN}}$ for 1-D mapping  ]{\resizebox{0.245\linewidth}{!}{\subimport{Figures}{EGN_Err_1D_symbolRate}} }
\subfloat[$\Delta\eta_{\text{MEGN}}$ for 1-D mapping   ]{\resizebox{0.245\linewidth}{!}{\subimport{Figures}{FEGN_Err_1D_symbolRate}}}
\\
\subfloat[$\Delta\eta_{\text{EGN}}$ for 2-D mapping    ]{\resizebox{0.245\linewidth}{!}{\subimport{Figures}{EGN_Err_2D_span}}}
\subfloat[$\Delta\eta_{\text{MEGN}}$ for 2-D mapping   ]{\resizebox{0.245\linewidth}{!}{\subimport{Figures}{FEGN_Err_2D_span}}}
\subfloat[$\Delta\eta_{\text{EGN}}$ for 2-D mapping    ]{\resizebox{0.245\linewidth}{!}{\subimport{Figures}{EGN_Err_2D_symbolRate}}}
\subfloat[$\Delta\eta_{\text{MEGN}}$ for 2-D mapping   ]{\resizebox{0.245\linewidth}{!}{\subimport{Figures}{FEGN_Err_2D_symbolRate}}}
\\
\subfloat[$\Delta\eta_{\text{EGN}}$ for 4-D mapping    ]{\resizebox{0.245\linewidth}{!}{\subimport{Figures}{EGN_Err_4D_span}}}
\subfloat[$\Delta\eta_{\text{MEGN}}$ for 4-D mapping   ]{\resizebox{0.245\linewidth}{!}{\subimport{Figures}{FEGN_Err_4D_span}}}
\subfloat[$\Delta\eta_{\text{EGN}}$ for 4-D mapping    ]{\resizebox{0.245\linewidth}{!}{\subimport{Figures}{EGN_Err_4D_symbolRate}}}
\subfloat[$\Delta\eta_{\text{MEGN}}$ for 4-D mapping   ]{\resizebox{0.245\linewidth}{!}{\subimport{Figures}{FEGN_Err_4D_symbolRate}}}

\caption{Comparison of the estimation errors $\Delta \eta$ for EGN and MEGN for various shaping blocklengths and 1-D (first row), 2-D (second row), and 4-D (third row) mappings. The first two columns present the accuracy for various numbers of spans $N_s$ at symbol rate $R_s=32$ GBd, while the last two columns present the accuracy for various symbol rates $R_s$ at $N_s=10$ spans. }
\label{Fig:Results_Error_symbolRate}
\end{figure*}



\subsection{MEGN Model Breakdown and Approximation}

In this section, we evaluate the NLI contributions of the SPT, XP, and XPT symbol-energy interactions, together with the effect of the memory length in the MEGN model. Based on these results, we validate the approximations proposed for the MEGN model in Sec.~\ref{Sec:FEGN_Approximations}.

The overall contributions from the different interaction types are summarized in Fig.~\ref{Fig:Contribution}.
We consider the representative case of shaping blocklength $N=100$ with 4-D mapping.
First, both XP and XPT contributions increase almost linearly with distance, with XPT contributing slightly more than XP.
This trend highlights the importance of temporal cross-polarization interactions when designing high-dimensional modulation formats.
Second, the SPT contribution is initially lower than XP and XPT in the first few spans, but it grows more rapidly with distance and eventually surpasses both.
This indicates that, for reducing NLI through high-dimensional modulation formats, the dominant interaction to target (intra-pol. vs.\ inter-pol.) depends strongly on the intended transmission distance. Finally, the contribution from SPT/XPT interactions involving three symbols is negligible, which is more than two orders of magnitude smaller than the other terms. This observation justifies the approximation in~\eqref{Eq:FEGN_Approx}.

Fig.~\ref{Fig:memoryAccuracy} shows the prediction accuracy, quantified by the NLI power-coefficient error $\Delta\eta_\text{MEGN}$. The considered signaling format is $N=400$ with 4-D mapping, which gives a correlation length of $M_s=100$. The symbol rate is fixed at $R_s=32$~GBd, while the transmission distance is varied, leading to different channel memories $M_c$. The accuracy is evaluated by applying the MEGN model with the same fixed memory length to all transmission distances. As discussed in Sec.~\ref{Sec:FEGN_Approximations}, the optimal choice of $M$ should ideally be determined jointly by the correlation length $M_s$ and the channel memory $M_c$. As seen in Fig.~\ref{Fig:memoryAccuracy}, the EGN model has limited accuracy, with $\Delta\eta_\text{MEGN}$ exceeding $6\%$. For the MEGN model, the accuracy improves with increasing $M$, as more symbol-energy interactions are included, until approximately $M=50$. Beyond this point, the improvement becomes negligible, even for transmission distances up to 20 spans. This suggests that, for the considered systems, the dominant bottleneck is the channel memory $M_c$ rather than the correlation length $M_s$, and we may infer that for $N_s=20$ span transmission, the channel memory $M_c \approx 50$.

We note that using the full formula or a longer memory length $M$ yields only negligible performance improvement for the MEGN model. Therefore, for simplicity, the following accuracy analysis is carried out using the approximated MEGN model in \eqref{Eq:FEGN_Approx} with $M=50$.

\subsection{MEGN Model Accuracy}
In this section, we assess the accuracy of the MEGN model for various transmission distances and symbol rates. Fig.~\ref{Fig:Results_combined}(a) shows the NLI reduction due to the short blocklength obtained from the MEGN model. It can be seen that the shorter shaping blocklength $N=100$ generally enables more NLI reduction than the longer one. Frequencies closer to zero benefit more from the NLI reduction. Moreover, extending the symbol-energy interactions to XPT through 2-D and 4-D mapping enables a further reduction in NLI, which is more pronounced for $N=100$ than for $500$.

Fig.~\ref{Fig:Results_combined}(b) presents the prediction of the proposed MEGN model in terms of the NLI power coefficient $\eta$ and the optimal effective SNR, at a fixed transmission distance ($10$ spans) and symbol rate ($32$ GBd) but for various shaping blocklengths $N$.
The MEGN model, evaluated for the 10-span transmission, shows excellent agreement with the simulation results. Here the MEGN model uses $M=50$, which is an approximate window that, as shown in Fig.~\ref{Fig:Results_Kernels}, captures the kernel functions down to at least $30\%$ of their normalized magnitude.
Because the temporal symbol-energy correlations vary with $N$ and the mapping strategy, the resulting SNR also depends on these signaling parameters.
When $N$ increases beyond $10^3$, the energy correlations become negligible, and the SNR converges to the value predicted by the EGN model. Since the 2-D and 4-D mapping strategies extend the energy correlations into more dimensions, they feature stronger energy correlations at shorter correlation lengths, causing $\bKSc{\tau}$, $\bKSa{\tau}$, and $\bKSb{\tau,\taup}$ to contribute more in the small-delay region of the channel functions, as illustrated in Fig.~\ref{Fig:Contribution}. This leads to a more pronounced NLI reduction. The overall errors remain small, and the largest discrepancies, still negligible, appear for very short blocklengths $N=10,20$ under 2-D and 4-D mapping.

Fig.~\ref{Fig:Results_combined}(c)--(d) examines the model prediction for $N=100$ for (c) various transmission distances at a fixed symbol rate and (d) various symbol rates at a fixed distance. Overall, the NLI increases with transmission distance and decreasing symbol rate. Under these conditions, energy correlations have a more pronounced impact on the NLI, as indicated by the larger separation between the curves of different signaling schemes. As for the first inset in Fig.~\ref{Fig:Results_combined}(c), the discrepancy at one span is due to the limited accuracy of the EGN model over the first few spans. Apart from this, the other insets show that the MEGN model predicts the NLI accurately, with the corresponding curves passing almost exactly through the simulation points.

Fig.~\ref{Fig:Results_Error_symbolRate} further examines the model accuracy across various signaling schemes, transmission distances and symbol rates, where the MEGN model accuracy is compared side by side against the EGN model.
The EGN model does not accurately predict the NLI in the short-blocklength regime, and the prediction error increases as the shaping blocklength decreases.
Moreover, when moving from 1-D to 2-D and then to 4-D mapping, the area in which the EGN model achieves an error below $5\%$ increases.
Regarding the transmission distances and symbol rates, the estimation errors of the EGN model become distinct as distance and symbol rate decrease.
In contrast, the proposed MEGN model closely matches the simulation results, maintaining errors within $5\%$ over almost the entire region, with the exception of the first span where small deviations remain.

\section{Experimental Validation}\label{Sec:BellLabResults}

\begin{figure}
    \centering
    \def\SC{1}
    \def\WD{0.8*\linewidth}
    \def\HT{0.6*\linewidth}
    \resizebox{\SC\linewidth}{!}{\subimport{Figures}{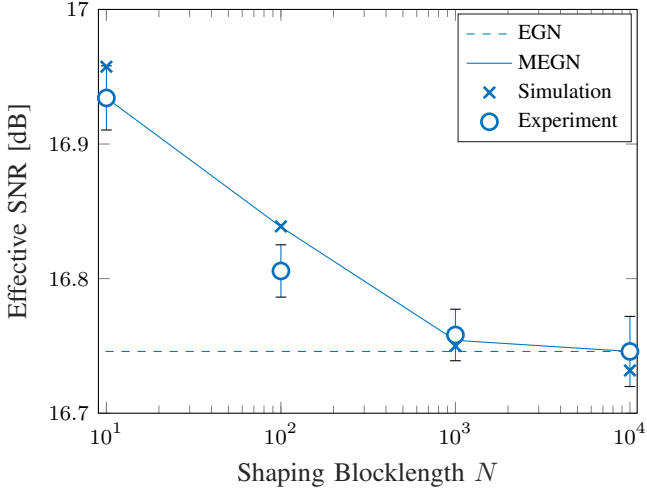}  }
    \caption{The effective SNR vs. the CCDM shaping blocklength $N$ at fixed launch power obtained by experiment and predicted by simulation and the model. The error bar of the experimental data represents the standard error of the mean. }
    \label{Fig:BellResults}
\end{figure}

The proposed MEGN model was further validated experimentally. The PAS-64QAM signaling at $R_s=32$~GBd was generated using RRC pulse shaping with $5\%$ roll-off, consistent with the parameters adopted in the numerical simulations. The transmitter employed two $128$~GSa/s digital-to-analog converters (DACs), while the receiver used a $256$~GSa/s real-time analog-to-digital converter (ADC). The transmission link consisted of six spans of standard single-mode fiber (SSMF), with a total length of $372$~km ($62$~km per span). The fiber parameters were $0.15$~dB/km attenuation, chromatic dispersion below $22.4$~ps/(nm$\cdot$km), and an effective area of $153~\mu\text{m}^2$. Each span exhibited approximately $11$~dB loss, which was fully compensated by an EDFA with a $5$~dB noise figure. The experiment was conducted in a single-channel configuration, and the launch power was set to $-5$~dBm.

For the receiver digital signal processing (DSP) chain, I/Q deskewing was first applied to compensate fixed I/Q delays introduced by the transmitter and optical frontend, followed by frequency-offset compensation. Chromatic dispersion was compensated digitally using EDC, after which a matched filter was applied. An adaptive multiple-input multiple-output (MIMO) equalizer incorporating least-mean square (LMS) adaptation and phase recovery was used to decouple the signal per dimension and mitigate residual phase noise. Finally, we measured the effective SNR from the recovered symbols.

In this experimental validation, we predict the SNR at a \textit{fixed launch power} via simulation and the model, as shown in Fig.~\ref{Fig:BellResults}. In the experiment, a penalty exists due to imperfect DSP and inherent background noise. To compensate for this penalty, we apply additional AWGN noise in the simulation and model, such that the saturation SNR at $N=10^4$ matches the experimental one. Fig.~\ref{Fig:BellResults} demonstrates that both the simulations and the MEGN model accurately predict the blocklength-dependent SNR behavior, again confirming the validity of the proposed MEGN model.

%% file: Tex/Tab_System_Setup.tex
\begin{table}[]
    \centering
    \caption{Simulation Parameters.}
    \label{Tab:FiberParam}
    \begin{tabular}{c|c} 
    \hline\hline\textbf{Parameter} & \textbf{Value} \\
    \hline Modulation & $64$QAM \\
    Polarization & Dual \\
    Center wavelength ($\lambda$) &  1550 nm \\
    Symbol rate ($R_s$) & $\{16,32,56,72,100\}$ GBd \\
    Pulse shape & Root-raised cosine \\
    Pulse roll-off & $5\% $ \\
    \hline Span length ($L_s$) & $100$ km\\
    \# Spans ($N_s$) & $[1,20]$ \\
    Fiber loss ($\alpha$) & $0.22$ dB/km \\
    Dispersion parameter ($D$) & $16.7$ ps/nm/km \\
    Nonlinear parameter ($\gamma$) & $1.3$ 1/W/km \\
    EDFA noise figure & $6$ dB \\
    \hline Oversampling factor & $2$  \\
    \hline\hline
    \end{tabular}
\end{table}

%% file: Tex/Ch_Conclusions.tex
\section{Conclusions}\label{Sec:Conclusions}

We have presented a \emph{novel analytical model} for predicting NLI power in optical systems that utilize symbol correlations across the four dimensions of the optical field. Our key finding demonstrates that the proposed model achieves high accuracy, which was rigorously validated against simulations and experiments. Crucially, the model's performance excels where established methods such as the EGN model lose accuracy, particularly in the short-blocklength regime, which is a critical area for practical PAS operation.

While this work focused on PAS, the fundamental principles allow for a straightforward extension to other advanced high-dimensional modulation formats. The use of this model is significant for several key applications: \textit{(i)} capacity analysis: our model quantitatively reveals how correlation across optical field dimensions can be strategically exploited to suppress NLI, opening a new avenue for refining capacity limits and optimizing spectral efficiency in fiber-optic channels. \textit{(ii)} DSP for NLI mitigation: the model serves as a powerful tool for DSP optimization, for example, to improve sequence selection in PAS or guide the design of multidimensional modulation formats. \textit{(iii)} system optimization: the model enables the selection of optimal signaling and channel parameters, such as identifying the best combination of shaping blocklength, mapping strategy, and symbol rate for a given system configuration.

Finally, the current framework focuses on SCI and can be directly applied to evaluate SCI in wavelength-division multiplexing (WDM) transmission scenarios. Moreover, the present MEGN framework can be naturally extended to model cross-channel interference (XCI) and multi-channel interference (MCI). Since XCI and MCI are the dominant impairments in practical WDM systems, extending the proposed short-blocklength-compatible model to account for these effects constitutes a natural and impactful direction for future work.

%% file: Tex/Ch_Sym_Corr.tex
\section{Symbol-Energy Correlation Analysis}\label{App:symCorr}

In this appendix, we analyze the temporal symbol-energy correlations embedded in \eqref{Eq:G_SCI}, which are essentially captured by the frequency moments \eqref{Eq:6FreqProduct_1}--\eqref{Eq:6FreqProduct_4}. Under the assumptions in Sec.~\ref{Sec:FEGN_symbolCorr}, any subspace of $\mL$ in \eqref{Eq:TotalTimeSpace} that gives rise to a correlation term containing a first-order symbol at a distinct time index does not contribute to the frequency moments in \eqref{Eq:FreqMom}. For instance, consider a subspace of $\mL$ as
\begin{align*}
    \mL_0 \triangleq & \left\{ (\w,\wtt) \in \mathbb{Z}^2 \,:\, 0 \leq \w ,\wtt \leq W-1, \, \w=\w_1, \right.\nonumber \\
    &\left. \wtt=\w_2=\w_3=\cdots=\w_6, \,
    \w\neq\wtt \right \}.
\end{align*}

For the frequency moment in \eqref{Eq:FreqMom} conditioned on this subset, due to the fact that $\Exp{ a^{\px}_{ \w} a^{\px*}_{ \wtt} |a^{\px}_{ \wtt}|^4}=0$ for $\w\neq\wtt$, consequently, \eqref{Eq:FreqMom} evaluates to zero in the subspace $\mL_0$, i.e.,
\begin{align*}
    &\left.\Exp{v^{\px}_{m} v^{\px *}_{n} v^{\px}_{k} v^{\px *}_{m'} v^{\px}_{n'} v^{\px *}_{k'}}\right|_{\mL_0} \\
    & = f_0^3 \Pss_{mnkm'n'k'}
    \sum_{\mL_0} \Exp{ a^{\px}_{ \w} a^{\px*}_{ \wtt} |a^{\px}_{ \wtt}|^4}  \\
    &\qquad \exp\!\left(
    -\jmath\frac{2\pi}{W}
    \left(
    \w  m + \wtt(- n +  k - m' + n' -  k')
    \right)
    \right) \\
    & = 0.
\end{align*}
The zero NLI contributions in the subspace $\mL_0$ for the remaining frequency moments \eqref{Eq:6FreqProduct_2}--\eqref{Eq:6FreqProduct_4} can be derived analogously.

\input{Tex/Tab_Contribution_All}

The remaining subspaces of $\mL$ that lead to nonzero frequency moments are purely energy-dependent ones\footnote{In addition to purely energy correlations, one could also consider amplitude correlations between the I and Q components. Their effects, however, are conjectured to be negligible, as incorporating only energy correlations already yields accurate predictions in the analysis of this paper. A detailed investigation of these terms is left for future work.}, taking the form in \eqref{Eq:SymCorr_Intro}. Table~\ref{Tab:MomBreak} summarizes the energy correlations that contribute to the sixth-order frequency moments in \eqref{Eq:G_SCI}, along with their corresponding time-domain subspaces within $\mL$. In the following, we illustrate the correlation terms listed in Table~\ref{Tab:MomBreak}.

\subsubsection{Identical Time Instance $\mL_{1}$}

The simplest case arises when symbols share an identical time index, forming the subspace
\begin{align*}
   \mL_{1}  \triangleq \left\{ \w \in \mathbb{Z}   :   0 \!\leq\! \w \!\leq\! W\!-\!1, \, \w=\w_1=\w_2=\ldots=\w_6  \right\} .
\end{align*}
The corresponding energy correlations are listed in the first column of Table~\ref{Tab:MomBreak}. The first case $\Exp{|\axwt|^6}$ has been properly incorporated in the EGN model. The latter two cases, $\Ra{\w,\w}{\py \px}=\Exp{|\aywt|^2|\axwt|^4 } $ and $\Ra{\w,\w}{ \px \py}=\Exp{|\axwt|^2|\aywt|^4 }$, represent XP interaction and are neglected in the EGN model.

\subsubsection{Two Different Time Instances $\mLa$}

\input{Tex/Tab_IndexGroups24}

Next, we consider the energy correlations between two groups of distinct time slots, as illustrated in the second and third columns of Table~\ref{Tab:MomBreak}.
The corresponding time subspace, denoted as $\mLa$, is summarized in Table~\ref{Tab:Subset24}.
As an example, $\mLai{1}$ is defined as
\begin{align*}
    &\mLai{1}  \triangleq
    \left\{ (\w,\wtt) \in \mathbb{Z}^2 \,:\,
    0 \leq \w ,\wtt \leq W-1, \,
    \w=\w_1=\w_2, \right. \nonumber\\
    &\left.
    \wtt=\w_3=\w_4=\w_5=\w_6, \,
    \w\neq\wtt
    \right\}. \nonumber
\end{align*}

Note that $\mLa$ can be further categorized into different subspaces according to their associations with the frequency indices $m,n,k,m',n',k'$, as discussed in Appendix~\ref{App:Moments}.

When the symbols with four identical time indices belong to the same polarization, the corresponding correlations are
$\Ra{\w,\wtt}{\px \px}=\Exp{|\axwt|^2 |\axwtt|^4}$, representing SPT interactions, and
$\Ra{\w,\wtt}{\px \py}=\Exp{|\axwt|^2 |\aywtt|^4}$, $\Ra{\w,\wtt}{\py \px}=\Exp{|\aywt|^2 |\axwtt|^4}$, representing XPT interactions.

When the symbols with four identical time indices belong to different polarizations, their correlations involve three distinct symbols, namely
$\Rb{\w,\w,\wtt}{\px \py \px}=\Exp{|\axwt|^2 |\aywt|^2 |\axwtt|^2}$ and
$\Rb{\w,\w,\wtt}{\px \py \py}=\Exp{|\axwt|^2 |\aywt|^2 |\aywtt|^2}$, which both describe XP and XPT interactions.

\subsubsection{Three Different Time Instances $\mLb$}
\input{Tex/Tab_IndexGroups222}

The energy correlations arising from the subspace consisting of three distinct groups of time indices, denoted as $\mLb$, are listed in the last column of Table~\ref{Tab:MomBreak}. As an example, $\mLbi{1}$ is defined as
\begin{align*}
    \mLbi{1}  \triangleq & \left\{ (\w,\wtt,\wttt) \in \mathbb{Z}^3 \,:\,  0 \leq \w ,\wtt,\wttt \leq W-1, \right. \nonumber \\
    & \w=\w_1=\w_4,\;\wtt=\w_2=\w_3,\;\wttt=\w_5=\w_6, \nonumber \\
    & \left. \w\neq\wtt,\;\w\neq\wttt,\;\wtt\neq\wttt \right\}.
\end{align*}
The corresponding correlations are
$\Rb{\w,\wtt,\wttt}{\px \py\px }=\Exp{|\axwt|^2 |\aywtt|^2 |\axwttt|^2}$,
$\Rb{\w,\wtt,\wttt}{\px \px\py }=\Exp{|\axwt|^2 |\axwtt|^2 |\aywttt|^2}$, and
$\Rb{\w,\wtt,\wttt}{\px \py\py }=\Exp{|\axwt|^2 |\aywtt|^2 |\aywttt|^2}$.
These terms involve three symbols across different polarizations and thus represent one SPT interaction and two XPT interactions.

\subsection{Time-Averaged Correlation and Covariance}

So far, the aforementioned correlations are defined in the time domain, forming spaces of size $W^2$ and $W^3$ for two and three groups of time indices, respectively.
Under the cyclo-stationary assumption, one degree of freedom can be reduced.
Since the statistical behavior of the symbol repeats periodically over blocks, it is sufficient to consider the time-averaged correlations to capture their overall influence on the NLI, which are defined as
\begin{subequations}
\begin{align}
    \bRc{ \tau}{\ppol\ppol'}  &\triangleq  \frac{1}{M} \sum_{\w=1}^{M}  \Rc{\w,\w+\tau}{\ppol\ppol'},  \label{Eq:SymCorr1}  \\
    \bRa{\tau}{\ppol\ppol'}  &\triangleq \frac{1}{M} \sum_{\w=1}^{M} \Ra{\w,\w+\tau}{\ppol\ppol'},  \label{Eq:SymCorr2} \\
    \bRb{ \tau,\taup}{\ppol\ppol'\ppol''} &\triangleq  \frac{1}{M} \sum_{\w=1}^{M}  \Rb{\w,\w+\tau,\w+\taup}{\ppol\ppol'\ppol''} .     \label{Eq:SymCorr3}
\end{align}
\end{subequations}

The corresponding time-averaged covariance definitions are given in \eqref{Eq:SymCovAvg1}--\eqref{Eq:SymCovAvg3}. The domain of delays $(\tau,\taup)$ is given by
\begin{align*}
&\mD \triangleq \left\{ (\tau, \taup) \in \mathbb{Z}^2 \,:\, 0 < |\tau| \leq W-1, \right. \nonumber \\ & \left. 0 < |\taup| \leq W-1, \, 0< |\taup-\tau| \leq W-1 \right\} .
\end{align*}
Therefore, the correlation analysis can be simplified by transforming the time indices $ \w,\wtt,\wttt $ in three-dimensional space $\mL$ into the two-dimensional space $\mD$ for delays $ \tau,\taup $.

\subsection{Finite Correlation and Symmetry}\label{App:Symetry}

In this section, we study two key properties of the time-averaged correlations and covariances: \textit{(i)} the finite correlation length $M$ and \textit{(ii)} the symmetry with respect to the delays $\tau$ and $\taup$. Both properties play an important role in simplifying the model derivation. For simplicity, we omit the superscripts denoting the polarizations and use the generic notations for covariance functions $\bKc{\tau}{}$, $\bKa{\tau}{}$, and $\bKb{\tau,\taup}{}$.

\begin{figure}
\centering
\setkeys{Gin}{width=0.24\textwidth}
\subfloat[Finite correlation length.]{
\resizebox{0.7\linewidth}{!}{\subimport{Figures}{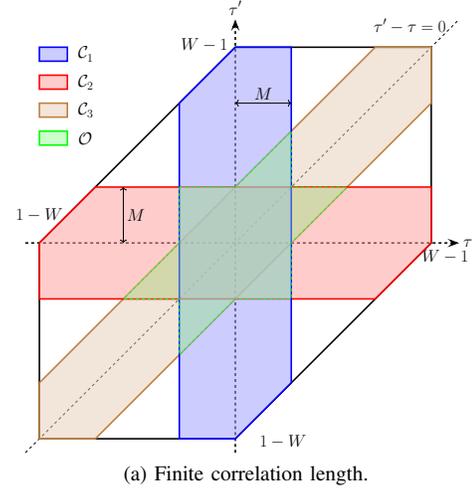}}
}
\hfill
\subfloat[Symmetry for 6 equivalent regions.]{
\resizebox{0.7\linewidth}{!}{\subimport{Figures}{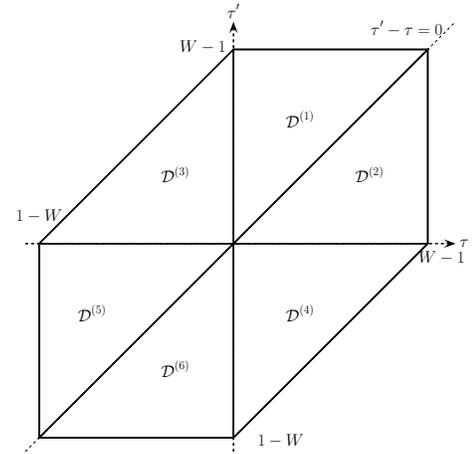}}
}
\caption{The division of the 2-D region for covariance function $\bK(\tau,\taup)$ using the symmetry and finite-correlation properties.}
\label{Fig:CorrProps}
\end{figure}

\begin{figure}
\centering
\setkeys{Gin}{width=0.1\textwidth}
\subfloat[$\bKb{\tau,\taup}{}$ in $\mD$.]{
\includegraphics[width=0.45\linewidth]{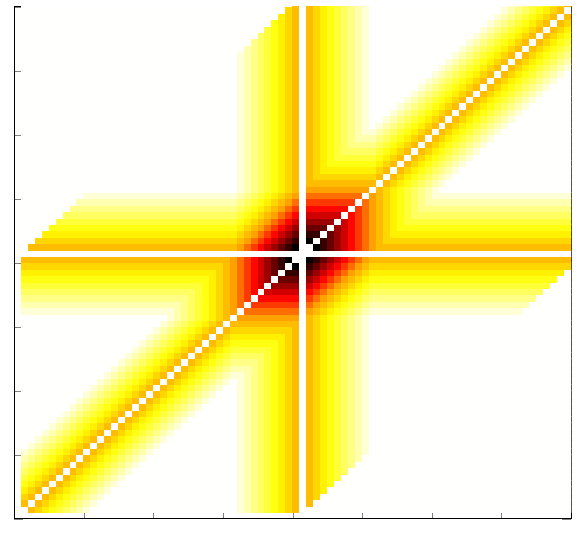}
}
\hfill
\subfloat[$ \Exp{|a|^2}\bKc{\tau}{}$ in $\Ca$.]{
\includegraphics[width=0.45\linewidth]{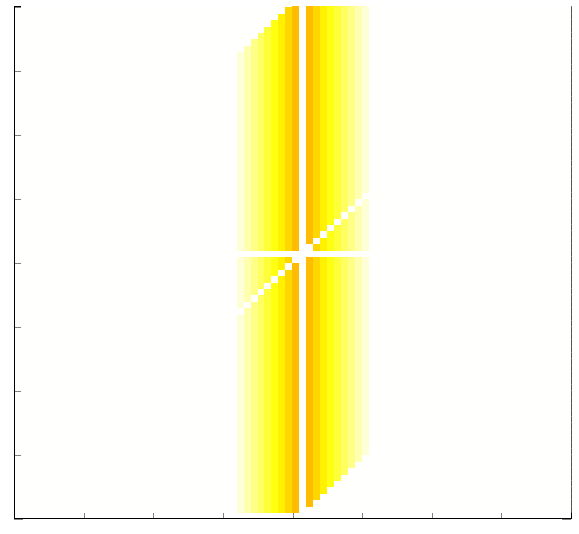}
}
\hfill
\subfloat[$ \Exp{|a|^2}\bKc{\taup}{}$ in $\Cb$.]{
\includegraphics[width=0.45\linewidth]{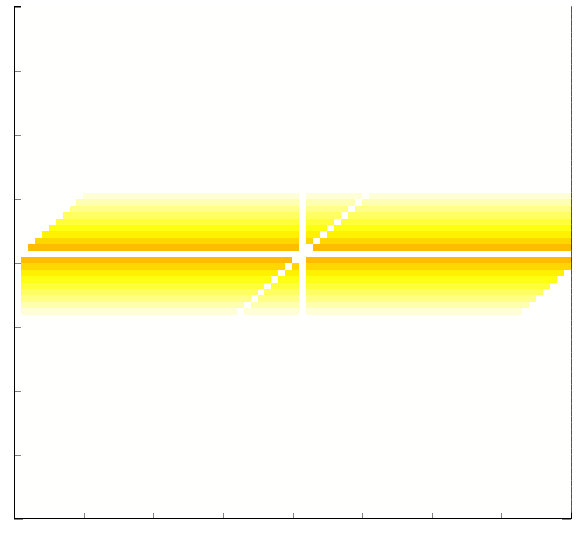}
}
\hfill
\subfloat[$\Exp{|a|^2}\bKc{\taup-\tau}{}$ in $\Cc$.]{
\includegraphics[width=0.45\linewidth]{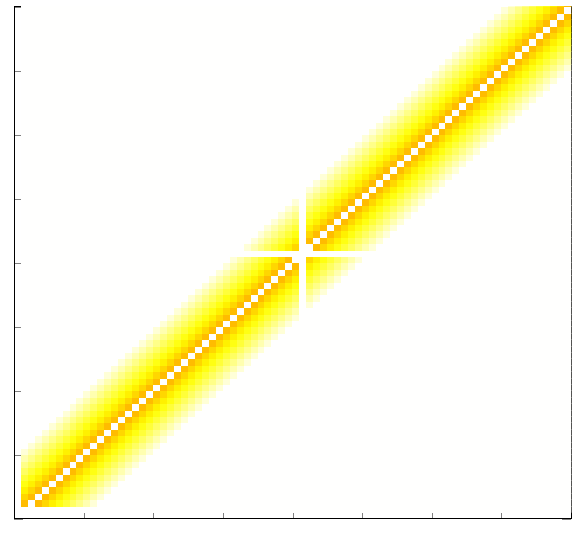}
}
\caption{An example of the divisions of the 2-D region for covariance function $\bKb{\tau,\taup}{}$.}
\label{Fig:CorrPropsCCDM}
\end{figure}

\begin{figure}
    \centering
    \def\SC{0.7}
    \resizebox{\SC\linewidth}{!}{\subimport{Figures}{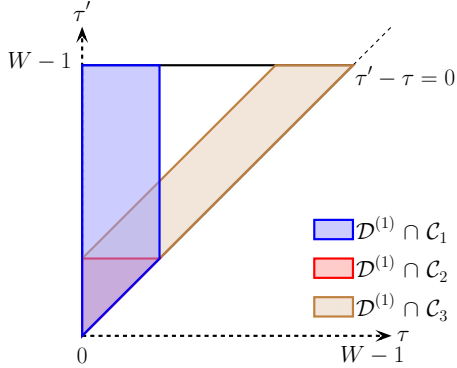}  }
    \caption{The support region in $\mDi{1}$ for covariance function $\bK(\tau,\taup)$. The shaded areas are nonzero. }
    \label{Fig:tau3DComp2}
\end{figure}

With finite correlation length, the correlation vanishes once the delay exceeds the correlation length due to the definite presence of independent symbols. The finite correlation length indicates limited support of the covariances $\bKc{\tau}{}$, $\bKa{\tau}{}$, and $\bKb{\tau,\taup}{}$ that truly contribute to the reduction of NLI.

The symmetry of $\bKc{\tau}{}$ and $\bKa{\tau}{}$ simply indicates that it is an even function
\begin{align}
   \bKc{\tau}{} = \bKc{-\tau}{}, \quad \bKa{\tau}{} = \bKa{-\tau}{}.
\end{align}\label{Eq:corrSymetry1}
and with finite correlation length
\begin{align}
   \bKc{\tau}{}=\bKa{-\tau}{} = 0, \quad \forall |\tau|>M.
\end{align}

These two properties of covariance allow us to consider only the one-sided correlation when $0<\tau\leq M$.

By contrast, $\bKb {\tau,\taup }{}$, as a 2-D function characterized by double delays $(\tau,\taup)\in\mD$, is more complex. The finite correlation further indicates the support region in terms of relative time distances between any two symbols among the symbol triplet. Fig.~\ref{Fig:CorrProps}(a) shows the support regions of $\bKb{\tau,\taup }{}$, which are defined as
\begin{subequations} \label{Eq:corrSymetry3}
 \begin{align}
    \Ca  & \triangleq \left\{ (\tau, \taup) \in \mD \,:\, 0<|\tau| < M \right\}, \\
    \Cb   & \triangleq \left\{ (\tau, \taup) \in \mD \,:\, 0<|\taup| < M \right\}, \\
    \Cc  & \triangleq \left\{ (\tau, \taup) \in \mD \,:\, 0<|\taup-\tau| < M \right\}.
\end{align}
\end{subequations}

In fact, although $\bKb{\tau,\taup }{}$ is defined as a 2-D function of delays $\taup,\tau$, it is only jointly determined by the relative time distances when any two regions of $\Ca,\Cb,\Cc$ are overlapping. This region is defined as
\begin{align*}
      \mO \triangleq (\Ca\cap\Cb)\cup(\Ca\cap\Cc)\cup(\Cb\cap\Cc).
\end{align*}
In all other cases, $\bKb {\tau,\taup }{}$ reduces to a 1-D function, as long as at least two out of three symbols have a time distance larger than $M$. If all time distances are larger than $M$, these three symbols are independent, and $\bKb {\tau,\taup }{}=0$.

By further ignoring the region in $\mO$, $\bKb {\tau,\taup }{}$ can be simplified in the shaded areas $\Ca,\Cb,\Cc$ as visualized in Fig.~\ref{Fig:CorrProps}(a). Hence, $\bKb {\tau,\taup }{}$ can be decomposed as
\begin{equation*}
\setlength{\nulldelimiterspace}{0pt}
 \bKb {\tau,\taup }{} = \left\{\begin{IEEEeqnarraybox}[\relax][c]{l's}
    \bKb {\tau,\taup }{}, & \text{in }$ \mO$\\
    \Exp{|a|^2}\bKc{\tau }{}, & \text{in }$\Ca\setminus(\Cb\cup\Cc)$\\
    \Exp{|a|^2}\bKc{\taup}{}, &\text{in }$\Cb\setminus(\Ca\cup\Cc)$\\
   \Exp{|a|^2}\bKc{ \taup-\tau }{}, &\text{in }$\Cc\setminus(\Ca\cup\Cb)$\\
    0, &\text{in }$\mD\setminus(\Ca\cup\Cb\cup\Cc) $ .
\end{IEEEeqnarraybox}\right.
\end{equation*}
Fig.~\ref{Fig:CorrPropsCCDM} shows an example of the division of the support regions as defined above. Hence, the double summation region in $\mD $ can be partitioned as
\begin{align}\label{Eq:tau3DPartition}
  & \sum_{ \mD } \bKb {\tau,\taup }{}  =\sum_{ \Ca }  \Exp{|a|^2}\bKc{\tau }{} + \sum_{ \Cb }  \Exp{|a|^2}\bKc{\taup}{}\nonumber \\
 &  + \sum_{ \Cc  }  \Exp{|a|^2}\bKc{ \taup-\tau }{} + \sum_{ \mO  } \left[  \bKb {\tau,\taup }{}- \Exp{|a|^2} \right. \nonumber \\
 & \left. \left(\bKc{\tau }{} + \bKc{\taup }{} +\bKc{ \taup-\tau }{} \right)\right].
\end{align}
The idea of this partition is to first approximate $\bKb {\tau,\taup }{}$ as the corresponding 1-D function in $\Ca$, $\Cb$ and $\Cc$. Then, a correction in the region $\mO$ is applied.

Now consider the symmetry of $\bKb {\tau,\taup }{}$, which depends on the SPT and XPT interaction types. For the SPT type $\bKSb{ \tau,\taup}{}$, the regions $\Ca,\Cb  $ and $\Cc $ are equivalent since they describe only the intra-polarization temporal correlation. However, for $\bKXb{ \tau,\taup}{}$, one region in $\Ca,\Cb$ and $\Cc $ must account for SPT, which may not be equivalent to the other two.

For intra-pol temporal symbol-energy correlations $\bKSb{ \tau,\taup}$, or assuming SPT and XPT that have the same strength such that $\bKSb{ \tau,\taup}=\bKXb{ \tau,\taup} $, this 2-D space $\mD$ can be partitioned further into 6 equivalent regions, as shown in Fig.~\ref{Fig:CorrProps}(b), which are
\begin{subequations}
    \begin{align} \label{Eq:corrSymetry2_regions}
    \mDi{1}  \triangleq  &  \left\{ (\tau, \taup) \in \mD \,:\, 0<\tau<\taup \right\} \\
    \mDi{2}  \triangleq &    \left\{ (\tau, \taup) \in \mD \,:\, 0<\taup<\tau \right\}  \\
     \mDi{3}\triangleq  &    \left\{ (\tau, \taup) \in \mD \,:\, \tau<0<\taup \right\} \\
     \mDi{4}\triangleq &    \left\{ (\tau, \taup) \in \mD \,:\, \taup<0<\tau \right\}  \\
     \mDi{5}\triangleq  & \left\{ (\tau, \taup) \in \mD \,:\, \tau<\taup<0 \right\} \\
    \mDi{6}\triangleq  &    \left\{ (\tau, \taup) \in \mD \,:\, \taup<\tau<0 \right\}  .
\end{align}
\end{subequations}

The symmetry of $\bKb{ \tau,\taup}{}$ is expressed as
\begin{equation} \label{Eq:corrSymetry2}
\setlength{\nulldelimiterspace}{0pt}
  \bKb{ \tau,\taup}{} = \left\{\begin{IEEEeqnarraybox}[\relax][c]{l's}
    \bK_0(\tau,\taup), & \text{in }$\mDi{1} $\\
    \bK_0(\taup,\tau), & \text{in }$\mDi{2}  $\\
    \bK_0(-\tau,\taup-\tau), &\text{in }$\mDi{3} $\\
    \bK_0(-\taup,\tau-\taup), &\text{in }$\mDi{4} $\\
    \bK_0(-\taup,-\tau), & \text{in }$\mDi{5} $\\
   \bK_0(-\tau,-\taup), &\text{in }$\mDi{6} $ .
\end{IEEEeqnarraybox}\right.
\end{equation}
Hence, it is sufficient to only consider the covariance function $ \bKb{ \tau,\taup}{}$ in the support of one equivalent region, such as $\mDi{1}$ as shown in Fig.~\ref{Fig:tau3DComp2}.

%% file: Tex/Tab_Contribution_All.tex
\begin{table*}[!t]
\centering
\caption{Breakdown of energy correlations from frequency moments \eqref{Eq:6FreqProduct_1}--\eqref{Eq:6FreqProduct_4} $\w \neq \wtt$, $\w \neq \wttt$, $\wtt \neq \wttt$.}
\label{Tab:MomBreak}
\renewcommand{\arraystretch}{1.55}
\resizebox{\linewidth}{!}{
\begin{tabular}{|c|c|c|c|c|}
\hline
\multirow{2}{*}{\backslashbox[45mm]{Frequency}{Time}} & \multicolumn{4}{c|}{$\Exp{\aw{1}^{\ppol_1}\aw{2}^{\ppol_2*}\aw{3}^{\ppol_3}\aw{4}^{\ppol_4*}\aw{5}^{\ppol_5}\aw{6}^{\ppol_6*}}$ conditioned on different time subspaces} \\ \cline{2-5}

& $\Exp{|\axwt|^6}$ & \multicolumn{2}{c|}{$\Exp{|\axwt|^2 |\axwtt|^4}$} & $\Exp{|\axwt|^2 |\axwtt|^2 |\axwttt|^2}$ \\   \hline

$\Exp{v^{\px}_{m} v^{\px *}_{n} v^{\px}_{k} v^{\px *}_{m'} v^{\px}_{n'} v^{\px *}_{k'} }$   & $\mL_{1}$ & \multicolumn{2}{c|}{$\mLai{1-4}\in \mLa^{a},\mLai{5-8}\in \mLa^{b},\mLai{9}\in \mLa^{c}$} & $\mLbi{1-4}\in \mLb^{a},\mLbi{5-6}\in \mLb^{b}$ \\ \hline\hline

------ & $\Exp{|\aywt|^2|\axwt|^4 }$ & $\Exp{|\axwt|^2 |\aywtt|^4}$ & $\Exp{|\axwt|^2 |\aywt|^2 |\axwtt|^2}$ & $\Exp{|\axwt|^2 |\axwtt|^2 |\aywttt|^2}$ \\ \hline

$\Exp{v^{\px}_{m} v^{\px *}_{n} v^{\px}_{k} v^{\px *}_{m'} v^{\py}_{n'} v^{\py *}_{k'} } $ & \multirow{2}{*}{$\mL_{1}$ } & $\mLai{4} \in \mLa^{a}$  & $\mLai{1,3}\in \mLa^{a},\mLai{5,7}\in \mLa^{b}$ & $\mLbi{1,3}\in \mLb^{a}$ \\ \cline{1-1} \cline{3-5}

$\Exp{v^{\px}_{m} v^{\px *}_{m'} v^{\px}_{n'} v^{\px *}_{k'}  v^{\py *}_{n} v^{\py}_{k} }$  &   & $\mLai{3} \in \mLa^{a}$ & $ \mLai{2,4}\in \mLa^{a},\mLai{5,6}\in \mLa^{b}$ & $\mLbi{1,4}\in \mLb^{a}$ \\   \hline\hline

------ & $\Exp{|\axwt|^2|\aywt|^4 }$ 
& $\Exp{|\aywt|^2 |\axwtt|^4}$ 
& $\Exp{|\axwt|^2 |\aywt|^2 |\aywtt|^2}$ 
& $\Exp{|\axwt|^2 |\aywtt|^2 |\aywttt|^2}$   \\ \hline

 $\Exp{v^{\px}_{m} v^{\px *}_{m'}  v^{\py *}_{n} v^{\py}_{k}  v^{\py}_{n'} v^{\py *}_{k'}}$  
& $\mL_{1}$ 
& $\mLai{5} \in \mLa^{b}$ 
& $\mLai{3,4}\in \mLa^{a},\mLai{8}\in \mLa^{b},\mLai{9}\in \mLa^{c}$ 
& $\mLbi{1}\in\mLb^{a},\mLbi{5}\in\mLb^{b}$ \\ \hline

\end{tabular}
}
\end{table*}

%% file: Tex/Tab_IndexGroups24.tex
\begin{table}[!t]
\centering
\caption{List of all time index subspaces for the \emph{double energy correlations} $\Ra{\w,\wtt }{\ppol \ppol' }$ of $\Exp{\aw{1}\aw{2}^*\aw{3}\aw{4}^*\aw{5}\aw{6}^*}$. The index subgroups identify the corresponding pair and quadruple of indices assuming the same value. } 
\label{Tab:Subset24}
\resizebox{0.8\columnwidth}{!}{
\begin{tabular}{|c|c|c|c|}
\hline 
\multicolumn{2}{|c|}{\backslashbox{Subspace }{Index}} & $\w$ & $\wtt$  \\ \hline
\multirow{4}{*}{$\mLa^{a}$} & $\mLai{1}$ & $\w_1, \w_2$ & $\w_3, \w_4, \w_5, \w_6$ \\ \cline{2-4}  
& $\mLai{2}$ & $\w_5, \w_4$ & $\w_1, \w_2, \w_3, \w_6$ \\ \cline{2-4}  
& $\mLai{3}$ & $\w_3, \w_2$ & $\w_1, \w_4, \w_5, \w_6$ \\ \cline{2-4}  
& $\mLai{4}$ & $\w_5, \w_6$ & $\w_1, \w_2, \w_3, \w_4$ \\ \hline \hline  
\multirow{4}{*}{$\mLa^{b}$} & $\mLai{5}$ & $\w_1, \w_4$ & $\w_2, \w_3, \w_5, \w_6$ \\ \cline{2-4}    
& $\mLai{6}$ & $\w_1, \w_6$ & $\w_2, \w_3, \w_4, \w_5$ \\ \cline{2-4}  
& $\mLai{7}$ & $\w_3, \w_4$ & $\w_1, \w_2, \w_5, \w_6$ \\ \cline{2-4}  
& $\mLai{8}$ & $\w_3, \w_6$ & $\w_1, \w_2, \w_4, \w_5$ \\ \hline   \hline  
$\mLa^{c}$ & $\mLai{9}$ & $\w_5, \w_2$ & $\w_1, \w_3, \w_4, \w_6$ \\ \hline   

\end{tabular}
}
\end{table}

%% file: Tex/Tab_IndexGroups222.tex
\begin{table}[!t]
\centering
\caption{List of all time index subspaces for the \emph{triple energy correlations} $\Rb{\w,\wtt,\wttt}{\ppol \ppol' \ppol''}$ of $\Exp{\aw{1}\aw{2}^*\aw{3}\aw{4}^*\aw{5}\aw{6}^*}$. The index subgroups identify the corresponding pairs of indices assuming the same value. } 
\label{Tab:Subset222}
\resizebox{0.8\columnwidth}{!}{
\begin{tabular}{|c|c|c|c|c|}
\hline
\multicolumn{2}{|c|}{\backslashbox{Subspace  }{Index  }} & $\w$ & $\wtt$ & $\wttt$ \\ \hline
\multirow{4}{*}{$\mLb^{a}$} & $\mLbi{1}$ & $\w_1, \w_4$ & $\w_2, \w_3$ & $\w_5, \w_6$  \\ \cline{2-5}
& $\mLbi{2}$ & $\w_3, \w_6$ & $\w_1, \w_2$ & $\w_4, \w_5$ \\ \cline{2-5}
& $\mLbi{3}$ & $\w_3, \w_4$ & $\w_1, \w_2$ & $\w_5, \w_6$\\ \cline{2-5}
& $\mLbi{4}$ & $\w_1, \w_6$ & $\w_2, \w_3$ & $\w_4, \w_5$ \\ \hline \hline
\multirow{2}{*}{$\mLb^{b}$} & $\mLbi{5}$& $\w_2, \w_5$ & $\w_1, \w_4$  & $\w_3, \w_6$ \\ \cline{2-5}
& $\mLbi{6}$ & $\w_2, \w_5$ & $\w_1, \w_6$  & $\w_3, \w_4$ \\ \hline
\end{tabular}
}
\end{table}

%% file: Tex/Ch_Mom_Deri.tex
\section{Derivation of Frequency Moments} \label{App:Moments}
In this section, we focus our analysis on the single-polarization case $\Exp{v^{\px}_{m} v^{\px *}_{n} v^{\px}_{k} v^{\px *}_{m'} v^{\px}_{n'} v^{\px *}_{k'} }$. As shown in the first row in Table~\ref{Tab:MomBreak}, it leads to temporal symbol-energy correlations $\Rc{\w,\wtt}{\px\px}$, $\Ra{\w,\wtt}{\px\px}$, and $ \Rb{\w,\wtt,\wttt}{\px\px\px}$. For the sake of simplicity, the notation subscript denoting the polarization is hence dropped. The following derivation also applies to the dual-polarization case, which differs only in the types of energy correlations involved, as shown in the second to the last rows in Table~\ref{Tab:MomBreak}.

\subsection{Moments for Identical Time Index}\label{App:A}
We first evaluate the case when the time indices are identical in $\mL_1$, for which we have \cite[Eq.~(102)]{Carena2014}
\begin{align*}
   &  \left.\Exp{v_{m} v^*_{n} v_{k} v^*_{m'} v_{n'} v^*_{k'} } \right|_{\mL_1}   \\
   = &   \Exp{|a|^6} f_0^3  \Pss_{mnkm'n'k'} \sum_{\w=0}^{W-1} e^{-\jmath\frac{2\pi}{W}\w (m-n+k-m'+n'-k')}      \\
   =  &  \Exp{|a|^6} R_s f_0^2 \Pss_{mnkm'n'k'} \times \delta(m-n+k-m'+n'-k') .
\end{align*}
where the delta function is inherently satisfied since $m,n,k$ and $m',n',k'$ satisfy \cite[Eq.~(34)]{Poggiolini2012}
\begin{align}\label{Eq:frqIdxCst}
     if_0 =(m-n+k)f_0 = (m'-n'+k')f_0.
\end{align}

\subsection{Moments for Two Distinct Time Index}\label{App:B}
The subset $\mLa$ for two distinct time indices $\w$ and $\wtt$ is summarized in Table~\ref{Tab:Subset24}. The subset is further categorized into three typical groups, each leading to an equivalent moment expression, because the involved frequency component indices are interchangeable. Hence, considering only one case in each group is sufficient.

\subsubsection{Typical Group $\mLa^{a}$}
For $\mLa^{a}=\{\mLai{1},\mLai{2},\mLai{3},\mLai{4}\}$, we consider the case of $\mLai{1}$, i.e., 

\begin{align}
    &\Exp{v_{m} v^*_{n} v_{k} v^*_{m'} v_{n'} v^*_{k'} }|_{ \mLai{1}}  \nonumber   \\
    = & f_0^3 \Pss_{mnkm'n'k'}\sum_{\w=0}^{W-1} \sum_{\substack{\wtt=0\\\wtt\neq\w}}^{W-1}   \Ra{\w,\wtt}{}  \nonumber  \\
    & e^{-\jmath\frac{2\pi}{W}[\w (m-n)+ \wtt (k-m'+n'-k')]} \nonumber  \\
     = & f_0^3 \Pss_{mnkm'n'k'}\sum_{\w=0}^{W-1} \sum_{\substack{\tau=-\w\\\tau\neq0}}^{W-\w-1} \Ra{\w,\w+\tau}{}  \nonumber  \\
     & e^{-\jmath\frac{2\pi}{W}[\w(m-n+k -m'+n'-k') -\tau (m-n)]}   \nonumber  \\
     = & f_0^3 \Pss_{mnkm'n'k'} \left[\sum_{\tau=1-W}^{-1} \sum_{\w=0}^{W+\tau-1} \Ra{\w,\w+\tau}{}    \right.\nonumber  \\
     & e^{-\jmath\frac{2\pi}{W}[\w(m-n+k -m'+n'-k') -\tau (m-n)]}    \nonumber \\
     & \left.+\sum_{\tau=1}^{W-1} \sum_{\w=\tau}^{W-1} \Ra{\w,\w+\tau}{} \right. \nonumber  \\
     &\left. e^{-\jmath\frac{2\pi}{W}[\w(m-n+k -m'+n'-k') -\tau (m-n)]} \right].\label{Eq:Mom24_0}
\end{align}

Then, we replace the time-dependent correlation $\Ra{\w,\w+\tau}{} $ with the time-averaged correlation $\bRa{\tau}{}$, yielding equivalent results due to the assumptions of $W\to\infty$ and a cyclostationary process on symbol energies. This enables us to discard the summation over time $\w$. Then, after using the symmetry property of the covariance function, \eqref{Eq:Mom24_0} can be rewritten as
\begin{align}
    &\Exp{v_{m} v^*_{n} v_{k} v^*_{m'} v_{n'} v^*_{k'} }|_{\mLai{1}}  \nonumber  \\
    = & f_0^3 \Pss_{mnkm'n'k'} \left[\sum_{\tau=1-W}^{-1}  \bRa{\tau}{}  (W+\tau) \right. \nonumber\\
    & e^{-\jmath\frac{2\pi}{W}[\w(m-n+k -m'+n'-k') -\tau (m-n)]}    \nonumber\\
     & \left. + \sum_{\tau=1}^{W-1}  \bRa{\tau}{}  (W-\tau) e^{-\jmath\frac{2\pi}{W}[\w(m-n+k -m'+n'-k') -\tau (m-n)]}    \right] \nonumber\\
    = & 2 \sum_{\tau=1}^{W-1} \bRa{\tau}{} (W-\tau)  f_0^3 \Pss_{mnkm'n'k'}  \cos{\frac{2\pi\tau(m-n)}{W}} \nonumber \\
    & \times \delta(m-n+k-m'+n'-k') . \label{Eq:Mom24_1}
\end{align} 

Similarly, the moment expressions for the cases of $\mLai{2}$,  $\mLai{3}$, and $\mLai{4}$ can be obtained.

\subsubsection{Typical Group $\mLa^{b}$}

We consider the typical group $\mLa^{b}=\{\mLai{5},\mLai{6},\mLai{7},\mLai{8}\}$ that yield the equivalent moment expressions. In the case of $\mLai{5}$, i.e., 
\begin{align}
     & \Exp{v_{m} v^*_{n} v_{k} v^*_{m'} v_{n'} v^*_{k'} }|_{\mLai{5}} \nonumber \\
     = & 2\sum_{\tau=1}^{W-1}  \bRa{\tau}{}  (W-\tau)f_0^3 \Pss_{mnkm'n'k'}   \cos{\frac{2\pi\tau(m-m')}{W}} \nonumber \\
    & \times \delta(m-n+k-m'+n'-k') .\label{Eq:Mom24_5}
\end{align} Compared to \eqref{Eq:Mom24_1}, its frequency indices in the $\cos(\cdot)$ function are qualitatively different. For the cases of $\mLai{6},\mLai{7},\mLai{8}$, the corresponding moments can be obtained similarly to \eqref{Eq:Mom24_5}.


\subsubsection{Typical Group $\mLa^{c}$}
The remaining typical group has only one case $\mLa^{c}=\{ \mLai{9}\}$. Compared to \eqref{Eq:Mom24_1} and \eqref{Eq:Mom24_1}, its corresponding moment expression only differs in its frequency indices in the $\cos(\cdot)$ function, that is
\begin{align}\label{Eq:Mom24_c}
    &  \Exp{v_{m} v^*_{n} v_{k} v^*_{m'} v_{n'} v^*_{k'} }|_{\mLai{9}} \nonumber   \\
    = & 2  \sum_{\tau=1}^{W-1} \bR(\tau) (W-\tau)  f_0^3 \Pss_{mnkm'n'k'}  \cos{\frac{2\pi\tau(n-n')}{W}}  \nonumber   \\
    & \times \delta(m-n+k-m'+n'-k').
\end{align}

\subsection{Moments for Three Distinct Time Index}
The subset $\mLb$ consisting of three distinct time indices has two typical groups, as shown in Table~\ref{Tab:Subset24}. The moment now involves temporal symbol-energy correlations among three different symbols, and the key to its derivation is to utilize the symmetry properties of $\bRb{\tau,\taup}{} $ as explained in the previous appendix.

Based on the partitions in  \eqref{Eq:tau3DPartition}, the moment for $\mLb$ can be written as
\begin{align}\label{Eq:Mom3Dbreak}
   & \Exp{v_{m} v^*_{n} v_{k} v^*_{m'} v_{n'} v^*_{k'} }|_{ \mLb } \nonumber \\
   = & \hat{\mathbb{E}}  [v_{m} v^*_{n} v_{k} v^*_{m'} v_{n'} v^*_{k'}]  |_{\mLb  \cap \Ca}  +\hat{\mathbb{E}}  [v_{m} v^*_{n} v_{k} v^*_{m'} v_{n'} v^*_{k'}]  |_{ \mLb  \cap\Cb}  \nonumber \\
    & +\hat{\mathbb{E}}  [v_{m} v^*_{n} v_{k} v^*_{m'} v_{n'} v^*_{k'}]  |_{\mLb  \cap \Cc}  + \mathbb{E}^c [v_{m} v^*_{n} v_{k} v^*_{m'} v_{n'} v^*_{k'} ]  |_{\mLb  \cap \mO} .
\end{align}
where the summation of the first three terms approximates the moment, while the last term compensates for the resulting discrepancy. In the following, for each typical group, we show the derivation for these four terms.

\subsubsection{Typical Group $\mLb^{a}$}
The time index groups in the subsets $\mLb^{a}=\{\mLbi{1},\mLbi{2},\mLbi{3},\mLbi{4}\}$ yield the equivalent expressions. In the following, we show the instance of $\mLbi{1}$ 

\begin{align*}
    &\Exp{v_{m} v^*_{n} v_{k} v^*_{m'} v_{n'} v^*_{k'} }|_{ \mLbi{1}}  \nonumber  \\
    =  &\sum_{\w=0}^{W-1} \sum_{\substack{\wtt=0 \\ \wtt\neq\w}}^{W-1} \sum_{\substack{\wttt=0 \\ \wttt\neq\w \\ \wttt\neq\wtt }}^{W-1}   \Rb{\w,\wtt  ,\wttt}{} f_0^3 \Pss_{mnkm'n'k'}  \nonumber\\
    & e^{-\jmath\frac{2\pi}{W}[\w (m-m')+ \wtt (k-n) + \wttt (n'-k')]} .\nonumber
\end{align*}

For the same logic as \eqref{Eq:Mom24_1}, $\Rb{\w,\wtt  ,\wttt}{}$ is replaced with $\bRb{\tau,\taup}{}$. For the first term in \eqref{Eq:tau3DPartition}, which approximate $\bRb{\tau,\taup}{}\approx\Exp{|a|^2}\bRc{\tau}{}$ in the region $\Ca$, we have
\begin{align}\label{Eq:Mom_3Dc1}
    &\hat{\mathbb{E}}  [v_{m} v^*_{n} v_{k} v^*_{m'} v_{n'} v^*_{k'}] |_{\mLbi{1}\cap\Ca}  \nonumber  \\
     = &  f_0^3 \Pss_{mnkm'n'k'} \left[ \sum_{\wttt=0}^{W-1} e^{-\jmath\frac{2\pi}{W} \wttt (n'-k')  }  \right. \nonumber   \\
     & \sum_{\w=0}^{W-1} \sum_{0<|\wtt-\w|<M}^{} \Exp{|a|^2}\bRc{\tau}{} e^{-\jmath\frac{2\pi}{W}[\w (m-m')+ \wtt (k-n)]}  \nonumber \\
     &   - \sum_{\w=0}^{W-1} \sum_{0<|\wtt-\w|<M}^{} \Exp{|a|^2}\bRc{\tau}{} \nonumber \\
     &\left( e^{-\jmath\frac{2\pi}{W}[\w (m-m'+n'-k')+ \wtt (k-n)]}  \right.   \nonumber \\
     & \left.\left. + e^{-\jmath\frac{2\pi}{W}[\w (m-m')+ \wtt (k-n+n'-k')]} \right) \right] \nonumber \\
    = &  f_0^3 \Pss_{mnkm'n'k'} \left[\sum_{\w=0}^{W-1} \sum_{0<|\wtt-\w|<M}^{} W \Exp{|a|^2}\bRc{\tau}{}\right. \nonumber \\
    & e^{-\jmath\frac{2\pi}{W}[\w (m-m')+ \wtt (k-n)]} \delta(n'-k') \nonumber    \\
    &  \left. - \sum_{\w=0}^{W-1} \sum_{0<|\wtt-\w|<M}^{} \Exp{|a|^2}\bRc{\tau}{} \left( e^{-\jmath\frac{2\pi}{W}[ (\wtt-\w) (k-n)]} \right.\right. \nonumber \\
    &\left.\left. +  e^{-\jmath\frac{2\pi}{W}[ (\w-\wtt) (m-m')]}\right)  \delta(m-n+k-m'+n'-k') \right] \nonumber \\
    = & 2\sum_{\tau=1}^{M}   \Exp{|a|^2}\bRc{\tau}{} (W-\tau) f_0^3 \Pss_{mnkm'n'k'}  \nonumber \\
    & \left[ W\cos{\frac{2\pi\tau (k-n)}{W}}  \times \delta(n'-k') \delta(m-m'+k-n) \right. \nonumber   \\
    & \left. - \left( \cos{\frac{2\pi\tau (m-m')}{W}} +\cos{\frac{2\pi\tau (k-n)}{W}} \right) \right. \nonumber \\
    & \left. \times \delta(m-n+k-m'+n'-k') \right]  .
\end{align}

The second term in \eqref{Eq:tau3DPartition} approximate $\bRb{\tau,\taup}{}\approx\Exp{|a|^2}\bRc{\taup}{}$ in the region $\Cb$. We replace the variable notation $\taup\to\tau$ to make the expression as a consistent form as in \eqref{Eq:Mom_3Dc1}, i.e.,
\begin{align}\label{Eq:Mom_3Dc2}
    &\hat{\mathbb{E}}  [v_{m} v^*_{n} v_{k} v^*_{m'} v_{n'} v^*_{k'}]|_{\mLbi{1}\cap\Cb}  \nonumber  \\
     = & 2\sum_{\tau=1}^{M}   \Exp{|a|^2}\bRc{\tau}{}(W-\tau) f_0^3 \Pss_{mnkm'n'k'}  \nonumber \\
    & \left[ W\cos{\frac{2\pi\tau (n'-k')}{W}}  \times \delta(k-n) \delta(m-m'+n'-k') \right. \nonumber   \\
    & \left. - \left( \cos{\frac{2\pi\tau (m-m')}{W}} +\cos{\frac{2\pi\tau (n'-k')}{W}} \right) \right. \nonumber \\
    & \left. \times \delta(m-n+k-m'+n'-k') \right]  .
\end{align}

Note that for delta function $ \delta(n'-k') $ in \eqref{Eq:Mom_3Dc1} and $ \delta(k-n) $ in \eqref{Eq:Mom_3Dc2}, they make the link function to be coupled zero when deriving the corresponding NLI PSD. Therefore, the first term in \eqref{Eq:Mom_3Dc1} and \eqref{Eq:Mom_3Dc2} can be ignored.

The third term in \eqref{Eq:tau3DPartition} approximates $\bRb{\tau,\taup}{}\approx\Exp{|a|^2}\bRc{\taup-\tau}{}$ in the region $\Cc$. After letting notation $\taup-\tau\to\tau$, we have

\begin{align}\label{Eq:Mom_3Dc3}
    &\hat{\mathbb{E}}  [v_{m} v^*_{n} v_{k} v^*_{m'} v_{n'} v^*_{k'}] |_{\mLbi{1}\cap\Cc}  \nonumber  \\
     = & 2\sum_{\tau=1}^{M}  \Exp{|a|^2}\bRc{\tau}{} (W-\tau) f_0^3 \Pss_{mnkm'n'k'}  \nonumber \\
    & \left[ W\cos{\frac{2\pi\tau (k-n)}{W}}  \times \delta(m-m') \delta(k-n+n'-k') \right. \nonumber   \\
    & \left. - \left( \cos{\frac{2\pi\tau (k-n)}{W}} +\cos{\frac{2\pi\tau (n'-k')}{W}} \right) \right. \nonumber \\
    & \left. \times \delta(m-n+k-m'+n'-k') \right]  .
\end{align}

For the last term in \eqref{Eq:Mom3Dbreak}, we use the symmetry properties of \eqref{Eq:corrSymetry2} to divide it into 6 equivalent regions, and use the region $\mO\cap \mDi{1}$ to represent the other regions equivalently.

\begin{align}
    &\mathbb{E}^c \left[{v_{m} v^*_{n} v_{k} v^*_{m'} v_{n'} v^*_{k'} } \right] |_{ \mLbi{1}\cap\mO}  \nonumber  \\
    = &\mathbb{E}^c \left[{v_{m} v^*_{n} v_{k} v^*_{m'} v_{n'} v^*_{k'} } \right]|_{ \mLbi{1}\cap\mO\cap(\mDi{1}\cup\mDi{2}\cup\mDi{3}\cup\mDi{4}\cup\mDi{5}\cup\mDi{6})} \nonumber \\
     = &\sum_{\tau=1}^{M} \sum_{\taup>\tau}^{\tau+M} \left[\bRb{ \tau,\taup }- \Exp{|a|^2} \left(\bRa{ \tau }{} +\bRa{ \taup }{} \right. \right.\nonumber \\
   & \left. \left.\ +\bRa{\taup - \tau }{} \right) \right] (W-\taup) f_0^3 \Pss_{mnkm'n'k'} \nonumber \\
   & \left[ e^{-\jmath\frac{2\pi}{W}[ \tau(k-n)+\taup(n'-k')]} + e^{-\jmath\frac{2\pi}{W}[ \taup(k-n)+\tau(n'-k')]}   \right. \nonumber \\
   &  +   e^{-\jmath\frac{2\pi}{W}[ -\tau(k-n)+(\taup-\tau)(n'-k')]}  \nonumber \\
   & +  e^{-\jmath\frac{2\pi}{W}[ (\taup-\tau)(k-n)-\tau(n'-k')]}   \nonumber \\
  & +  e^{-\jmath\frac{2\pi}{W}[(\tau-\taup)(k-n) -\taup(n'-k')]} \nonumber \\
  &  \left.+  e^{-\jmath\frac{2\pi}{W}[ -\taup(k-n)+(\tau-\taup)(n'-k')]}  \right]  \nonumber \\
  &   \times \delta(m-n+k-m'+n'-k'). \label{Eq:Mom222_a}
\end{align}

\subsubsection{Typical Group $\mLb^{b}$}
For the typical group $\mLb^{b}=\{\mLbi{5},\mLbi{6} \}$, we show the case of $\mLbi{5}=\{\w=\w_2=\w_5,\wtt=\w_1=\w_4,\wttt=\w_3=\w_6 | \w\neq\wtt,\wtt\neq\wttt,\w\neq\wttt\}$.The same approach as in the previous subsection is used and here we give the expressions directly.

For the moment under the constraint of $\Ca$ in \eqref{Eq:Mom3Dbreak}, we have
\begin{align*}
    & \hat{\mathbb{E}}  [v_{m} v^*_{n} v_{k} v^*_{m'} v_{n'} v^*_{k'}] |_{\mLbi{5}\cap\Ca}  \nonumber  \\
     = & 2\sum_{\tau=1}^{M}   \Exp{|a|^2}\bRc{\tau}{} (W-\tau) f_0^3 \Pss_{mnkm'n'k'}  \nonumber \\
    & \left[ W\cos{\frac{2\pi\tau (m-m')}{W}}  \times \delta(k-k') \delta(m-m'+n'-n)  \right. \nonumber   \\
    & \left. - \left( \cos{\frac{2\pi\tau (n'-n)}{W}} +\cos{\frac{2\pi\tau (m-m')}{W}} \right) \right. \nonumber \\
    & \left. \times \delta(m-n+k-m'+n'-k') \right]  .
\end{align*}

Similarly, for the moment under the constraint of $\Cb$ in \eqref{Eq:Mom3Dbreak},
\begin{align*}
    & \hat{\mathbb{E}}  [v_{m} v^*_{n} v_{k} v^*_{m'} v_{n'} v^*_{k'}] |_{\mLbi{5}\cap\Cb}  \nonumber  \\
     = & 2\sum_{\tau=1}^{M}   \Exp{|a|^2}\bRc{\tau}{} (W-\tau) f_0^3 \Pss_{mnkm'n'k'}  \nonumber \\
    & \left[ W\cos{\frac{2\pi\tau (k-k')}{W}}  \times \delta(m-m') \delta(n'-n+k-k') \right. \nonumber   \\
    & \left. - \left( \cos{\frac{2\pi\tau (n'-n)}{W}} +\cos{\frac{2\pi\tau (k-k')}{W}} \right) \right. \nonumber \\
    & \left. \times \delta(m-n+k-m'+n'-k') \right]  .
\end{align*}

The moment approximation for $\Cc$ in \eqref{Eq:Mom3Dbreak} is
\begin{align*}
    & \hat{\mathbb{E}}  [v_{m} v^*_{n} v_{k} v^*_{m'} v_{n'} v^*_{k'}] |_{\mLbi{5}\cap\Cc}  \nonumber  \\
     = & 2\sum_{\tau=1}^{M}   \Exp{|a|^2}\bRc{\tau}{} (W-\tau) f_0^3 \Pss_{mnkm'n'k'}  \nonumber \\
    & \left[ W\cos{\frac{2\pi\tau (m-m')}{W}}  \times \delta(n'-n) \delta(m-m'+k-k')  \right. \nonumber   \\
    & \left. - \left( \cos{\frac{2\pi\tau (m-m')}{W}} +\cos{\frac{2\pi\tau (k-k')}{W}} \right) \right. \nonumber \\
    & \left. \times \delta(m-n+k-m'+n'-k') \right]  .
\end{align*}

For the last term in \eqref{Eq:Mom3Dbreak} that compensates the region $\mO$,
\begin{align}\label{Eq:Mom_I3bO}
    &  \mathbb{E}^c  [v_{m} v^*_{n} v_{k} v^*_{m'} v_{n'} v^*_{k'}] |_{ \mLbi{5}\cap\mO}   \nonumber  \\
     = & \sum_{\tau=1}^{M} \sum_{\taup>\tau}^{\tau+M} \left[\bRb{ \tau,\taup }{}- \Exp{|a|^2} \left(\bRa{ \tau }{}+\bRa{ \taup }{} \right. \right.\nonumber \\
   & \left. \left.\ +\bRa{\taup - \tau }{} \right) \right]  (W-\taup)  f_0^3 \Pss_{mnkm'n'k'} \nonumber \\
     & \left[ e^{-\jmath\frac{2\pi}{W}[ \tau( m-m')+\taup(k-k')]} + e^{-\jmath\frac{2\pi}{W}[ \taup( m-m')+\tau(k-k')]}   \right. \nonumber \\
     &    +  e^{-\jmath\frac{2\pi}{W}[ -\tau( m-m')+(\taup-\tau)(k-k')]}    \nonumber \\
     & +  e^{-\jmath\frac{2\pi}{W}[ (\taup-\tau)( m-m')-\tau(k-k')]}   \nonumber \\
     &  + e^{-\jmath\frac{2\pi}{W}[(\tau-\taup)( m-m') -\taup(k-k')]}   \nonumber \\
     &  \left.+  e^{-\jmath\frac{2\pi}{W}[ -\taup( m-m')+(\tau-\taup)(k-k')]}  \right]  \nonumber \\
      & \times \delta(m-n+k-m'+n'-k') .
\end{align}

%% file: Tex/Ch_PSD_Deri.tex
\section{NLI PSD Derivation From Frequency Moments}\label{App:Kernels_Derivation}
In this appendix, we derive the NLI PSD expression in \eqref{Eq:G_SCI}. For convenience, we first decompose \eqref{Eq:G_SCI} into the contributions associated with the frequency moments in \eqref{Eq:6FreqProduct_1}--\eqref{Eq:6FreqProduct_4}, i.e.,
\begin{align}
    \GSCIx(f) = \GSCIx_1(f) + \GSCIx_2(f) + \GSCIx_3(f) + \GSCIx_4(f).
\end{align}
In the following, without loss of generality, we derive the general form of the NLI PSD expressions without explicitly specifying the underlying frequency moment and thus the corresponding subscripts. We characterize the NLI PSD contributions associated with the time subspaces introduced in the previous section. The channel function corresponding to each time subspace is derived explicitly and referred to as a kernel function. The resulting derivation can subsequently be specialized to the frequency moments in \eqref{Eq:6FreqProduct_1}--\eqref{Eq:6FreqProduct_4}, as discussed in Appendix~\ref{App:DP_NLI_PSD}.

Furthermore, we reformulate the NLI PSD expressions so that they involve only double integrals, thereby easing the numerical evaluation. In the following, we skip the NLI PSD contribution associated with the identical-time-index subspace $\mL_1$, as it is already covered by the EGN model.

\subsection{NLI PSD due to Moments for $\mLa^{a}$}
We consider the NLI PSD induced by the frequency moment belonging to type $\mLa^{a}=\{\mLai{1},\mLai{2},\mLai{3},\mLai{4}\}$. In view of \eqref{Eq:G_SCI} and \eqref{Eq:Mom24_1}, we have $\forall i\in\{1,2,3,4\}$
\begin{align}\label{Eq:G_SC_I2a_deri}
    & \GSCIx_i(f)|_{ \mLa^{a} } \nonumber \\
     = &\frac{64}{81}    f_0^3 e^{ -2\alpha L_s } \sum_{i=-\infty}^{+\infty} \delta(f-if_0) \sum_{m,n,k\in \mathcal{S}_i} \sum_{m',n',k'\in \mathcal{S}_i} \nonumber \\
    & \zeta(k,m,n) \zeta^*(k',m',n')  \Exp{v_{m} v^*_{n} v_{k} v^*_{m'} v_{n'} v^*_{k'} }|_{ \mLa^{a} }  \nonumber \\
    = & \frac{64}{81}   \sum_{\tau=1}^{W-1} \bRa{\tau}{} (R_s-\tau f_0)  f_0^5 e^{ -2\alpha L_s } \sum_{i=-\infty}^{+\infty} \delta(f-if_0) \nonumber \\
     & \sum_{m,n,k\in \mathcal{S}_i} \sum_{m',n',k'\in \mathcal{S}_i} \zeta(k,m,n) \zeta^*(k',m',n')  \Pss_{mnkm'n'k'}  \nonumber \\
     &2 \cos{\frac{2\pi\tau(m-n)}{W}}  \times \delta(m-n+k-m'+n'-k') .
\end{align}

To express \eqref{Eq:G_SC_I2a_deri} in integral form by letting $f_0\to0$, with the constraint \eqref{Eq:frqIdxCst}, the sixth-order summation over the frequency indices $m,n,k,m',n',k'$ leads to a quadruple integral. To further reduce the dimensionality of the integral, we note that $m$ and $k$, or $m'$ and $k'$, are interchangeable. Integrating over $\cos(m-n)$ and $\cos(m'-n')$ is qualitatively equivalent, and thus \eqref{Eq:G_SC_I2a_deri} is

\begin{align}\label{Eq:G_SC_I2a_2}
    & \GSCIx_i(f)|_{ \mLa^{a} } \nonumber \\
    = & \frac{64}{81}   \sum_{\tau=1}^{W-1} \bRa{\tau}{} (R_s-\tau f_0)  f_0^5 e^{ -2\alpha L_s } \sum_{i=-\infty}^{+\infty} \delta(f-if_0) \nonumber \\
     & \sum_{m,n,k\in \mathcal{S}_i} \sum_{m',n',k'\in \mathcal{S}_i} \zeta(k,m,n) \zeta^*(k',m',n')  \Pss_{mnkm'n'k'}  \nonumber \\
     &\left[ \cos{\frac{2\pi\tau(m-n)}{W}} + \cos{\frac{2\pi\tau(n'-m')}{W}} \right]  \nonumber \\
     &  \times \delta(m-n+k-m'+n'-k') \nonumber \\
     = & \frac{128}{81}   \sum_{\tau=1}^{W-1} \bRa{\tau}{} (R_s-\tau f_0)  f_0^5 e^{ -2\alpha L_s } \sum_{i=-\infty}^{+\infty} \delta(f-if_0) \nonumber \\
     & \sum_{m, k } \sum_{m' ,k' } \zeta(k,m,i) \zeta^*(k',m',i)  \Pss_{m(m+k-i)km'(m'+k'-i)k'}  \nonumber \\
      & \left[ \cos^2{\frac{\pi\tau(i-k)}{W}} \cos^2{\frac{\pi\tau(i-k')}{W}}  \nonumber  \right. \\
    & \left. -\sin^2{\frac{\pi\tau(m-n)}{W}} \sin^2{\frac{\pi\tau(m'-n')}{W}} \right]  .
\end{align}

It can be seen in \eqref{Eq:G_SC_I2a_2} that the summation over $m, k $ and the summation over $ m' ,k'$ are conjugates of each other, and this manipulation enables us to obtain a double integral later.

Let $f_0\to 0$, and thanks to the finite correlation length $ \tau\leq M \ll W$, we can approximate
\begin{align}
    \lim_{f_0 \to 0}(R_s-\tau f_0) = R_s.
\end{align}
and let
\begin{align}
    if_0 = f &,&
    mf_0 = f_1 &,&
    nf_0 = f_3 &,&
    kf_0 = f_2 .
\end{align}
The integral form of \eqref{Eq:G_SC_I2a_2} is then
\begin{align}\label{Eq:G_SC_I2a_3}
    & \GSCIx_i(f)|_{ \mLa^{a} } \nonumber \\
    = & \frac{128}{81}   \sum_{\tau=1}^{W-1} \bRa{\tau}{} R_s  e^{ -2\alpha L_s }  \left[     \iint_{\mathcal{B}}  df_1   df_2   S(f_1) \right. \nonumber \\
   & \left. S^*(f_1+f_2-f) S(f_2) \zeta(f_1,f_2,f) \cos^2{ \frac{\pi\tau(f-f_2)}{R_s} }    \right.  \nonumber \\
   & \iint_{\mathcal{B}}  df_1'   df_2'   S^*(f_1') S(f_1'+f_2'-f) S^*(f_2') \nonumber \\
   & \left.  \zeta^*(f_1',f_2',f) \cos^2{ \frac{\pi\tau(f-f_2')}{R_s} }    \right.  \nonumber \\
   & \left. -   \iint_{\mathcal{B}}  df_1 df_2   S(f_1) S^*(f_1+f_2-f) S(f_2) \; \right. \nonumber \\
   &   \zeta(f_1,f_2,f) \sin^2{\frac{\pi\tau(f-f_2)}{R_s} }  \nonumber \\
   &  \iint_{\mathcal{B}}  df_1' df_2'  S^*(f_1') S(f_1'+f_2'-f) S^*(f_2') \; \nonumber \\
   & \left.  \zeta^*(f_1',f_2',f) \sin^2{\frac{\pi\tau(f-f_2')}{R_s} }  \right]  \nonumber \\
     = & \frac{128}{81}   \sum_{\tau=1}^{W-1} \bRa{\tau}{} R_s  e^{ -2\alpha L_s }  \left[   \Bigl| \iint_{\mathcal{B}}  df_1   df_2   S(f_1) \right. \nonumber \\
   & \left. S^*(f_1+f_2-f) S(f_2) \zeta(f_1,f_2,f) \cos^2{ \frac{\pi\tau(f-f_2)}{R_s} } \Bigr|^2  \right.  \nonumber \\
   & \left. -   \Bigl| \iint_{\mathcal{B}}  df_1 df_2   S(f_1) S^*(f_1+f_2-f) S(f_2) \; \right. \nonumber \\
   & \left.  \zeta(f_1,f_2,f) \sin^2{\frac{\pi\tau(f-f_2)}{R_s} } \Bigr|^2 \right] .
\end{align}

By further assuming identical spans of same fiber type with lumped amplifiers exactly compensating for the loss of each span, we can obtain
\begin{align}\label{Eq:G_SC_I2a}
      \GSCIx_i(f) =\sum_{\tau=1}^{W-1} \bRa{\tau}{}  \chi_1(\tau,f),
\end{align}
where $\chi_1(\tau,f)$ is given in \eqref{Eq:Kn_chi1}.

\subsection{NLI PSD due to Moments for $\mLa^{b}$ and $\mLa^{c}$}
Compared to the previous section, obtaining the NLI PSD induced by moments of type $\mLa^{b}$ and $\mLa^{c}$ is similar, except that the manipulation of the trigonometric functions is slightly different.

For $\mLa^{b}$, \eqref{Eq:Mom24_5} can be rewritten as
  \begin{align}
   & \left.\Exp{v_{m} v^*_{n} v_{k} v^*_{m'} v_{n'} v^*_{k'} }\right|_{ \mLa^{b}} \nonumber   \\
   = &  2   \sum_{\tau=1}^{W-1}  \bR(\tau) (W-\tau)  f_0^3 \Pss_{mnkm'n'k'}\nonumber \\
   & \left[ \cos{\frac{2\pi\tau m}{W}} \cos{\frac{2\pi\tau m' }{W}} \right. \left. + \sin{\frac{2\pi\tau m }{W}} \sin{\frac{2\pi\tau m' }{W}} \right]\nonumber \\
   & \times \delta(m-n+k-m'+n'-k').  \label{Eq:Mom24_b}
\end{align}

and for $\mLa^{c}$, \eqref{Eq:Mom24_c} is rewritten as
\begin{align}
    & \left. \Exp{v_{m} v^*_{n} v_{k} v^*_{m'} v_{n'} v^*_{k'} }\right|_{ \mLa^{c}} \nonumber   \\
     = &   2   \sum_{\tau=1}^{W-1} \bR(\tau)(W-\tau) f_0^3 \Pss_{mnkm'n'k'}  \nonumber \\
     &\left[ \cos{\frac{2\pi\tau n}{W}} \cos{\frac{2\pi\tau n' }{W}}   + \sin{\frac{2\pi\tau n }{W}} \sin{\frac{2\pi\tau n' }{W}} \right] \nonumber \\
    & \times \delta(m-n+k-m'+n'-k').
\end{align}

Inserting them into \eqref{Eq:G_SCI}, after the same process as in the previous section, we can obtain their NLI PSD integral form expressions and their kernels $\chi_2(\tau,f)$  in \eqref{Eq:Kn_chi2} and $\chi_3(\tau,f)$ in \eqref{Eq:Kn_chi3}. Hence,
\begin{align}
     \GSCIx_i(f)|_{ \mLa^{b} }
    & =  \sum_{\tau=1}^{W-1} \bRa{\tau}{}  \chi_2(\tau,f), \label{Eq:G_SC_I2b} \\
     \GSCIx_i(f)|_{ \mLa^{c} }
     & =  \sum_{\tau=1}^{W-1} \bRa{\tau}{}  \chi_3(\tau,f). \label{Eq:G_SC_I2c}
\end{align}

\subsection{NLI PSD due to Moments for $\mLb^{a}$}

For the moment type with three distinct times $\mLb $, we analyze the contribution of each term as in \eqref{Eq:Mom3Dbreak}, obtaining
\begin{align} \label{Eq:Gx_break}
     \GSCIx_i(f)|_{ \mLb^{a} }
    = &  \GSCIx_i(f)|_{ \mLb^{a}\cap\Ca} +\GSCIx_i(f)|_{ \mLb^{a}\cap\Cb} \nonumber\\
    & +\GSCIx_i(f)|_{ \mLb^{a}\cap\Cc}+\GSCIx_i(f)|_{ \mLb^{a}\cap\mO}   .
\end{align}

For the first term in \eqref{Eq:Gx_break}, inserting \eqref{Eq:Mom_3Dc1} into \eqref{Eq:G_SCI}, we have
\begin{align}
    & \GSCIx_i(f)|_{ \mLb^{a}\cap\Ca } \nonumber \\
    = & \frac{ 128}{81}   \sum_{\tau=1}^{W-1}   \Exp{|a|^2}\bRc{\tau}{} (R_s-\tau f_0)  f_0^5 e^{ -2\alpha L_s } \sum_{i=-\infty}^{+\infty} \delta(f-if_0) \nonumber \\
     & \sum_{m,n,k\in \mathcal{S}_i} \sum_{m',n',k'\in \mathcal{S}_i} \zeta(k,m,n) \zeta^*(k',m',n')  \Pss_{mnkm'n'k'}  \nonumber \\
     & \left(-\cos{\frac{2\pi\tau (m-m')}{W}} -\cos{\frac{2\pi\tau (k-n)}{W}} \right) \nonumber \\
     &  \times \delta(m-n+k-m'+n'-k') .
\end{align}
where the two $\cos(\cdot)$ functions coincide with those in \eqref{Eq:Mom24_1} and \eqref{Eq:Mom24_5}, meaning that the same kernels are applied. Moreover, the same argument applies to the second term in \eqref{Eq:Mom3Dbreak}. Hence,
\begin{align}\label{Eq:G_SC_I3aC1}
       &  \GSCIx_i(f)|_{ \mLb^{a}\cap\Ca }   =  \GSCIx_i(f)|_{ \mLb^{a}\cap\Cb } \nonumber \\
    = & - \sum_{\tau=1}^{W-1} \Exp{|a|^2}\bRc{\tau}{}  (\chi_1(\tau,f)+\chi_2(\tau,f)) .
\end{align}


For the third term in \eqref{Eq:Gx_break}, inserting \eqref{Eq:Mom_3Dc3} into \eqref{Eq:G_SCI}, and split it into
\begin{align}
        \GSCIx_i (f)|_{ \mLb^{a}\cap\Cc }    =   \GSCIx_{i,1}(f)|_{ \mLb^{a}\cap\Cc } +   \GSCIx_{i,2}(f)|_{ \mLb^{a}\cap\Cc } .\nonumber
\end{align}
 where
\begin{align}
    &  \GSCIx_{i,1}(f)|_{ \mLb^{a}\cap\Cc } \nonumber \\
    = & \frac{ 128}{81}   \sum_{\tau=1}^{W-1}   \Exp{|a|^2}\bRc{\tau}{} R_s(R_s-\tau f_0)  f_0^4 e^{ -2\alpha L_s }\nonumber \\
     &  \sum_{i=-\infty}^{+\infty} \delta(f-if_0) \nonumber  \sum_{m,n,k\in \mathcal{S}_i} \sum_{m',n',k'\in \mathcal{S}_i} \zeta(k,m,n)\nonumber \\
     &  \zeta^*(k',m',n')  \Pss_{mnkm'n'k'}      \cos{\frac{2\pi\tau (k-n)}{W}} \nonumber \\
     &  \times \delta(m-m') \delta(k-n+n'-k') \label{Eq:G_SC_I3aC31}  \\
     = & \frac{ 128}{81}   \sum_{\tau=1}^{W-1}   \Exp{|a|^2}\bRc{\tau}{} R_s(R_s-\tau f_0)  f_0^4 e^{ -2\alpha L_s }\nonumber \\
     &  \sum_{i=-\infty}^{+\infty} \delta(f-if_0) \nonumber  \sum_{m,n,k\in \mathcal{S}_i} \sum_{m',n',k'\in \mathcal{S}_i} \zeta(k,m,n)\nonumber \\
     &  \zeta^*(k',m',n')  \Pss_{mnkm'n'k'}      \cos{\frac{ \pi\tau (k-n)+\pi\tau (k'-n')}{W}}  \nonumber \\
     &  \times \delta(m-m') \delta(k-n+n'-k') \label{Eq:G_SC_I3aC32} \\
      = & \frac{ 128}{81}   \sum_{\tau=1}^{W-1}   \Exp{|a|^2}\bRc{\tau}{} R_s(R_s-\tau f_0)  f_0^4 e^{ -2\alpha L_s }\nonumber \\
     &  \sum_{i=-\infty}^{+\infty} \delta(f-if_0) \nonumber  \sum_{m,n,k\in \mathcal{S}_i} \sum_{m',n',k'\in \mathcal{S}_i} \zeta(k,m,n)\nonumber \\
     & \zeta^*(k',m',n')  \Pss_{mnkm'n'k'}     \left[ \cos {\frac{\pi\tau (k-n)}{W}} \cos {\frac{\pi\tau (k'-n')}{W}} \right. \nonumber  \\
    &\left.- \sin {\frac{\pi\tau (k-n)}{W}} \sin {\frac{\pi\tau (k'-n')}{W}} \right] \nonumber \\
     &  \times \delta(m-m') \delta(k-n+n'-k')  \label{Eq:G_SC_I3aC33}\\
      = & \frac{ 128}{81}   \sum_{\tau=1}^{W-1}   \Exp{|a|^2}\bRc{\tau}{} R_s(R_s-\tau f_0)  f_0^4 e^{ -2\alpha L_s }\nonumber \\
     &  \sum_{i=-\infty}^{+\infty} \delta(f-if_0) \nonumber  \sum_{m } \sum_{k} \sum_{k'} |S(mf_0)|^2 S(kf_0) \nonumber \\
     & S^*(k'f_0)S^*([m+k-i]f_0)S([m+k'-i]f_0)\nonumber \\
     & \zeta(m,k,i) \zeta^*(m,k',i)  \left[ \cos {\frac{\pi\tau (i-m)}{W}} \cos {\frac{\pi\tau (i-m')}{W}} \right. \nonumber  \\
    &\left.- \sin {\frac{\pi\tau (i-m )}{W}} \sin {\frac{\pi\tau (i-m')}{W}} \right]  .\label{Eq:G_SC_I3aC34}
\end{align}
From \eqref{Eq:G_SC_I3aC31} to \eqref{Eq:G_SC_I3aC32}, the Kronecker delta function enables $k-n=k'-n'$. Let $f_0\to 0$, we can factorize the integral form \eqref{Eq:G_SC_I3aC34} as the product of temporal symbol-energy correlations and the kernel $\xi_1(\tau,f) $ in
\eqref{Eq:Kn_xi1}, i.e.
\begin{align}
  \GSCIx_{i,1}(f) |_{ \mLb^{a}\cap\Cc }     = \sum_{\tau=1}^{W-1} \Exp{|a|^2}\bRc{\tau}{}  \xi_1(\tau,f).
\end{align}

For $\GSCIx_{i,2}(f)|_{ \mLb^{a}\cap\Cc }$, it can be written as
\begin{align}
    & \GSCIx_{i,2}(f)|_{ \mLb^{a}\cap\Cc } \nonumber \\
    = & \frac{ 128}{81}   \sum_{\tau=1}^{W-1}   \Exp{|a|^2}\bRc{\tau}{} (R_s-\tau f_0)  f_0^5 e^{ -2\alpha L_s } \sum_{i=-\infty}^{+\infty} \delta(f-if_0) \nonumber \\
     & \sum_{m,n,k\in \mathcal{S}_i} \sum_{m',n',k'\in \mathcal{S}_i} \zeta(k,m,n) \zeta^*(k',m',n')  \Pss_{mnkm'n'k'}  \nonumber \\
    & \left.  \left( -\cos{\frac{2\pi\tau (k-n)}{W}} -\cos{\frac{2\pi\tau (n'-k')}{W}} \right) \right. \nonumber \\
    & \left. \times \delta(m-n+k-m'+n'-k') \right]  .
\end{align}
where the two $\cos(\cdot)$ functions are qualitatively equivalent to the one in \eqref{Eq:Mom24_1}, meaning that the same kernel is applied. Therefore, its integral form on two polarizations can be directly obtained as
\begin{align}
  \GSCIx_{i,2}(f)|_{ \mLb^{a}\cap\Cc }  =  - 2 \sum_{\tau=1}^{W-1} \Exp{|a|^2}\bRc{\tau}{}  \chi_1(\tau,f) .
\end{align}
Therefore,
\begin{align}
        \GSCIx_{i}(f)|_{ \mLb^{a}\cap\Cc }    =    \sum_{\tau=1}^{W-1} \Exp{|a|^2}\bRc{\tau}{}  (\xi_1(\tau,f) - 2\chi_1(\tau,f) ) .\label{Eq:G_SC_I3aC3}
\end{align}

Regarding the NLI PSD under the constraint $ \mLb^{a} \cap \mO$, to obtain double integration later, we expand each exponential function in the frequency moments in \eqref{Eq:Mom222_a} as the product of one exponential function and its conjugate
\begin{align}
      &   \mathbb{E}^c \left[{v_{m} v^*_{n} v_{k} v^*_{m'} v_{n'} v^*_{k'} } \right] |_{ \mLb^{a}\cap\mO}    \nonumber  \\
     = & \sum_{\tau=1}^{M} \sum_{\taup>\tau}^{\tau+M} \bR(\tau,\taup)   (W-\taup) f_0^3 \Pss_{mnkm'n'k'} \nonumber  \\
     & \left[ \left(e^{-\jmath\frac{2\pi}{W}[ \tau(k-n)]} + e^{-\jmath\frac{2\pi}{W}[ \taup(k-n)]} \right) \right. \nonumber \\
    &  \left(e^{\jmath\frac{2\pi}{W}[ \tau(k'-n')]} + e^{ \jmath\frac{2\pi}{W}[ \taup(k'-n')]} \right)\nonumber \\
     &  - \left.    e^{-\jmath\frac{2\pi}{W}[ \tau(k-n)]} e^{\jmath\frac{2\pi}{W}[ \tau(k'-n')]} -  e^{-\jmath\frac{2\pi}{W}[ \taup(k-n)]} e^{\jmath\frac{2\pi}{W}[ \taup(k'-n')]}  \right.\nonumber \\
     &  + \left(e^{-\jmath\frac{2\pi}{W}[ -\tau(k-n)]} + e^{-\jmath\frac{2\pi}{W}[ (\taup-\tau)(k-n)]} \right) \nonumber \\
    & \left(e^{\jmath\frac{2\pi}{W}[ -\tau(k'-n')]} + e^{\jmath\frac{2\pi}{W}[ (\taup-\tau)(k'-n')]} \right)  \nonumber \\
     &   -    e^{-\jmath\frac{2\pi}{W}[ -\tau(k-n)]} e^{\jmath\frac{2\pi}{W}[ -\tau(k'-n')]} \nonumber \\
    & -  e^{-\jmath\frac{2\pi}{W}[ (\taup-\tau)(k-n)]} e^{ \jmath\frac{2\pi}{W}[ (\taup-\tau)(k'-n')]} \nonumber \\
     &  +  \left(e^{-\jmath\frac{2\pi}{W}[ (\tau-\taup)(k-n)]} + e^{-\jmath\frac{2\pi}{W}[ -\taup(k-n)]} \right) \nonumber \\
    & \left(e^{\jmath\frac{2\pi}{W}[ (\tau-\taup)(k'-n')]} + e^{\jmath\frac{2\pi}{W}[ -\taup(k'-n')]} \right)\nonumber \\
     &   -    e^{-\jmath\frac{2\pi}{W}[ (\tau-\taup)(k-n)]} e^{\jmath\frac{2\pi}{W}[ (\tau-\taup)(k'-n')]} \nonumber \\
    & \left. -  e^{-\jmath\frac{2\pi}{W}[ -\taup(k-n)]} e^{\jmath\frac{2\pi}{W}[ -\taup(k'-n')]}  \right]  \nonumber \\
     & \times \delta(m-n+k-m'+n'-k'). \label{Eq:G_SC_I3aO_deri}
\end{align}
Inserting \eqref{Eq:G_SC_I3aO_deri} into \eqref{Eq:G_SCI}, we can obtain their NLI PSD in integral form with kernels $\xi_2(\tau,\taup,f)$ in \eqref{Eq:Kn_xi2}. Hence,
\begin{align}\label{Eq:G_SC_I3aO}
      \GSCIx_{i}(f)|_{ \mLb^{a}\cap\mO }  =   \sum_{\tau=1}^{M} \sum_{\taup=\tau+1}^{\tau+M}  \bRb{\tau,\taup}{}  \xi_2(\tau,\taup,f) .
\end{align}

\subsection{NLI PSD due to Moments for $\mLb^{b} $  }
The derivations for the NLI PSD under the constraints $ \mLb^{b} \cap \Ca$, $ \mLb^{b} \cap \Cb$, and $ \mLb^{b} \cap \Cc$ are similar to those in the previous section.
\begin{align}
     \GSCIx_{i}(f)|_{ \mLb^{b} }
    = &  \GSCIx_{i}(f)|_{ \mLb^{b}\cap\Ca} +\GSCIx_{i}(f)|_{ \mLb^{b}\cap\Cb} \nonumber\\
    & +\GSCIx_{i}(f)|_{ \mLb^{b}\cap\Cc}+\GSCIx_{i}(f)|_{ \mLb^{b}\cap\mO}   .
\end{align}

Here we give the results directly:
\begin{align}
       &\GSCIx_{i}(f)|_{ \mLb^{b}\cap\Ca }  = \GSCIx_{i}(f)|_{ \mLb^{b}\cap\Cb }   \nonumber  \\
       =  &  \sum_{\tau=1}^{W-1} \Exp{|a|^2}\bRc{\tau}{}  (\psi_1(\tau,f) -  \chi_2(\tau,f) - \chi_3(\tau,f) ). \label{Eq:G_SC_I3bC12}
\end{align}
and
\begin{align}
       \GSCIx_{i}(f)|_{ \mLb^{b}\cap\Cc }     =  \sum_{\tau=1}^{W-1} \Exp{|a|^2}\bRc{\tau}{}  (\psi_2(\tau,f) - 2 \chi_2(\tau,f) ).\label{Eq:G_SC_I3bC13}
\end{align}
where $\psi_1(\tau,f)$ and $\psi_2(\tau,f)$ are given in \eqref{Eq:Kn_psi1} and \eqref{Eq:Kn_psi2}, respectively.

Regarding the NLI PSD under the constraint $ \mLb^{b} \cap \mO$, we also first expand the frequency moments in \eqref{Eq:Mom_I3bO} each exponential function as the product of one exponential function and its conjugate version
\begin{align}
    &\Exp{v_{m} v^*_{n} v_{k} v^*_{m'} v_{n'} v^*_{k'} }|_{ \mLb^{b}\cap \mO }   \nonumber  \\
    = &  \sum_{\tau=1}^{M} \sum_{\taup>\tau}^{\tau+M} (W-\taup)\bR(\tau,\taup) f_0^3 \Pss_{mnkm'n'k' }\nonumber \\
     & \left[ e^{-\jmath\frac{2\pi}{W} (\tau m +\taup k)} e^{\jmath\frac{2\pi}{W} (\tau m' +\taup k')}\right. \nonumber \\
    &  + e^{-\jmath\frac{2\pi}{W} (\taup m +\tau k)} e^{\jmath\frac{2\pi}{W} ( \taup m' +\tau k')}    \nonumber \\
    &  +e^{-\jmath\frac{2\pi}{W} [-\tau m +(\taup-\tau) k]} e^{\jmath\frac{2\pi}{W} [-\tau m' +(\taup-\tau) k']} \nonumber \\
    &  + e^{-\jmath\frac{2\pi}{W} [(\taup-\tau) m -\tau k]} e^{ \jmath\frac{2\pi}{W}  [(\taup-\tau) m' -\tau k']}    \nonumber \\
     & +e^{-\jmath\frac{2\pi}{W} [(\tau-\taup) m -\taup k]} e^{\jmath\frac{2\pi}{W} [(\tau-\taup) m' -\taup k']} \nonumber \\
    &  \left. + e^{-\jmath\frac{2\pi}{W} [-\taup m +(\tau-\taup) k]} e^{\jmath\frac{2\pi}{W}  [-\taup m' +(\tau-\taup) k']} \right]  \nonumber \\
      & \times \delta(m-n+k-m'+n'-k').  \label{Eq:Mom222_b}
\end{align}
Inserting \eqref{Eq:Mom222_b} into \eqref{Eq:G_SCI}, we can obtain their NLI PSD integral form expressions and their kernels $\psi_3(\tau,\taup,f)$  in \eqref{Eq:Kn_psi3}.
\begin{align}\label{Eq:G_SC_I3bO}
     \GSCIx_{i}(f)|_{ \mLb^{b}\cap\mO }  =   \sum_{\tau=1}^{M} \sum_{\taup=\tau+1}^{\tau+M}  \bRb{\tau,\taup}{}  \psi_3(\tau,\taup,f) .
\end{align}

\section{Dual-Polarization Extension of the NLI PSD}\label{App:DP_NLI_PSD}


In the previous section, we derived the NLI PSD contributions associated with the different time subspaces in \eqref{Eq:G_SC_I2a}\eqref{Eq:G_SC_I2b}\eqref{Eq:G_SC_I2c}\eqref{Eq:G_SC_I3aC1}\eqref{Eq:G_SC_I3aC3}\eqref{Eq:G_SC_I3aO}\eqref{Eq:G_SC_I3bC12}\eqref{Eq:G_SC_I3bC13}\eqref{Eq:G_SC_I3bO}. For each subspace, the frequency moments in \eqref{Eq:6FreqProduct_1}--\eqref{Eq:6FreqProduct_4} induce different types of temporal symbol-energy correlations, as summarized in Table.~\ref{Tab:MomBreak}. The complete NLI PSD is therefore obtained by summing the corresponding contributions across the time and polarization dimensions.

\subsection{NLI PSD due to Single-Polarization Frequency Moments}
Consider the first row of Table~\ref{Tab:MomBreak}, where the SCI is contributed by the moments $\Exp{v^{\px}_{m} v^{\px *}_{n} v^{\px}_{k} v^{\px *}_{m'} v^{\px}_{n'} v^{\px *}_{k'} }$,
and only SPT energy interactions are involved. In view of Table~\ref{Tab:MomBreak}, we count the number of time-index types involved
that corresponds to the coefficients for each kernels. Hence, we have

\begin{align} \label{Eq:G1_final}
    &\GSCIx_{1}(f) \nonumber \\
    =& 4G_{1}(f)|_{\mLa^{a} } + 4G_{1}(f)|_{\mLa^{b} }  + G_{1}(f)|_{\mLa^{c} } \nonumber \\
    & + 4G_{1}(f)|_{\mLb^{a} } + 2G_{1}(f)|_{\mLb^{b} } \nonumber \\
    = & \sum_{\tau=1}^{M}  \Exp{|a|^2}\bKSc{\tau}    [ 4\xi_1(\tau,f) + 4\psi_1(\tau,f)] \nonumber  \\
     &  + 2\psi_2(\tau,f)  -16\chi_1(\tau,f) -16\chi_2(\tau,f) -4  \chi_3(\tau,f) \nonumber  \\
   & + \sum_{\tau=1}^{M}  \bKSa{ \tau} \left[ 4\chi_1(\tau,f) + 4\chi_2(\tau,f)  +  \chi_3(\tau,f) \right]\nonumber  \\
  & + \sum_{\tau=1}^{M} \sum_{\taup=\tau+1}^{\tau+M} \left[ \bKSb{\tau,\taup}   -\Exp{|a|^2}\right. \left. \left( \bKSc{\tau} +\bKSc{\taup}  \right. \right. \nonumber \\
    &  \left.  \left. +\bKSc{\taup-\tau}\right) \right] \left[4\xi_2(\tau,\taup,f) + 2\psi_3(\tau,\taup,f) \right]  .
\end{align}

\subsection{NLI PSD due to Dual-Polarization Frequency Moments}
For the last three rows of Table~\ref{Tab:MomBreak}, corresponding to the dual-polarization frequency moments, we similarly determine the temporal symbol-energy correlations involved, as defined in \eqref{Ch:Cov_XPT_3}, and the coefficients by counting the number of associated time-subspace types. Hence, we have

\begin{align} \label{Eq:G234_final}
    &\GSCIx_{2}(f)+\GSCIx_{3}(f)+\GSCIx_{4}(f)  \nonumber \\
    =& 3 \bKXa{ 0} \phi_4(f) \nonumber  \\
     &  +\Exp{|a|^2} \bKXc{ 0} (5\phi_2(f) + \phi_3(f)- 12 \phi_4(f))\nonumber  \\
     &   \sum_{\tau=1}^{M}  \Exp{|a|^2}\bKSc{\tau}    [ \xi_1(\tau,f) + \psi_1(\tau,f)  \nonumber  \\
     & -6\chi_1(\tau,f) -5\chi_2(\tau,f) -  \chi_3(\tau,f) ]  \nonumber   \\
     & + \sum_{\tau=1}^{M}  \Exp{|a|^2}\bKXc{\tau}    [ 4\xi_1(\tau,f) + \psi_1(\tau,f) +  \psi_2(\tau,f)  \nonumber  \\
     &   -14\chi_1(\tau,f) -9\chi_2(\tau,f) - \chi_3(\tau,f) ]   \nonumber  \\
   & + \sum_{\tau=1}^{M}  \bKXa{ \tau} \left[ 2\chi_1(\tau,f) +  \chi_2(\tau,f)    \right]  \nonumber   \\
    & + \sum_{\tau=1}^{M}  \left[\bKXb{0,\tau}-\Exp{|a|^2} \bKXc{ 0} \right] \nonumber  \\
    &   \left[ 6\chi_1(\tau,f) +  5\chi_2(\tau,f) +\chi_3(\tau,f)   \right]  \nonumber  \\
  & + \sum_{\tau=1}^{M} \sum_{\taup=\tau+1}^{\tau+M} \left[ \bKXb{\tau,\taup}   -\Exp{|a|^2}\right. \left. \left( \bKXc{\tau} +\bKXc{\taup}  \right. \right.  \nonumber \\
    &  \left.  \left. +\bKSc{\taup-\tau}\right) \right] \left[5\xi_2(\tau,\taup,f) + \psi_3(\tau,\taup,f) \right]    .
\end{align}


Note that for the special case of independent polarizations, we only keep the terms involving SPT interactions.

Finally, we sum up all the contributions from these four frequency moments, i.e.,
\begin{align}\label{Eq:G_final}
    G (f) = \sum_{i=1}^4 \GSCIx_{i}(f) + \GSCIy_{i}(f),
\end{align}
where $\GSCIy_{i}(f)$ contributes to the NLI equally as $\GSCIx_{i}(f)$ under symmetric polarization assumptions. Next, after inserting \eqref{Eq:G1_final} and \eqref{Eq:G234_final} into \eqref{Eq:G_final}, we group the terms according to the SPT, XP, and XPT types, and assemble the corresponding kernels into the channel functions for each interaction type. Since our objective is to derive the correction $G_C(f)$ in \eqref{Eq:FEGN_Model_0} to the NLI PSD relative to the EGN model, we replace the correlations with their covariance functions. This leads to the full MEGN model formula presented in Theorem~\ref{Thm:MEGN_Full} of Sec.~\ref{Sec:SXPT}.